**Kurdistan Regional Government**
**Ministry of Higher Education & Scientific Research**
**University of Sulaimani**
**College of Agricultural Engineering Sciences**

# COMPATIBILITY STUDIES OF LOQUAT SCIONS WITH LOQUAT AND QUINCE ROOTSTOCKS

**A Dissertation**

**Submitted to the Council of the College of Agricultural Engineering Sciences at the University of Sulaimani in Partial Fulfillment of the Requirements for the Degree of Doctor of Philosophy**

**in**

**Horticulture**

**Evergreen Fruit Production**

By

**Rasul Rafiq Aziz**

B.Sc. Horticulture (2004), College of Agriculture, University of Sulaimani
M.Sc. Pomology (2011), College of Agriculture, University of Sulaimani

Supervised by:

**Dr. Fakhraddin Mustafa Hama Salih** — Assistant Professor

**Dr. Ibrahim Maaroof Noori** — Assistant Professor

2724 K. 2024 A. D.

بسم الله الرحمن الرحيم

# قالوا سبحانك لا علم لنا إلا ما علمتنا إنك أنت العليم الحكيم

صدق الله العظيم

البقرة (32)

# SUMMARY

This research was conducted from 2021 to 2023 at the College of Agricultural Engineering Sciences, University of Sulaimani, Kurdistan Region-Iraq, to investigate the grafting and rooting percentages of loquat/quince and loquat/loquat scion-rootstock combinations. The study considered various factors, including different indole-3-butyric acid (IBA) concentrations, grafting dates, stocks, and rooting substrates, across two locations. The location 1 situated at the College, where four experiments were conducted in the lathhouse over three growing seasons. However, the location 2, involved grafting loquat onto quince trees grown in Kani Waysa village and loquat trees grown in the College. Cleft grafting was used for all experiments. The experiments were laid out in a Factorial Randomized Complete Block Design with three replications, and the data were analyzed using ANOVA, and the means were compared using Duncan's multiple range tests ($P \leq 0.05$).

**Experiment 1. Rooting of Hardwood Cuttings of Quince *Cydonia oblonga* Mill. as Influenced by IBA and Rooting Substrate:**

This experiment took place in the lathhouse at the college site, during the growing season, from February to July, 2021, and aimed to investigate the impact of various IBA concentrations (0, 1000, 2000, and 3000 $mg.L^{-}$), and different rooting substrates; river sand, river sand + peat moss (1:1 v/v), and perlite + peat moss (1:1 v/v) on the rooting success of hardwood cuttings of quince (*Cydonia oblonga* Mill.). The parameters, included root characters, shoot characters, and leaf chlorophyll content. The effect of individual factors showed that rooting and other traits of rooted cuttings were independent of the IBA effect. The highest rooting percentage (62.50%) was achieved in the control cuttings, along with improved other traits. Additionally, the best rooting (64.58%) was found in the cuttings planted in river sand medium. Interaction effects of the two factors showed that control cuttings planted in river sand gave the highest rooting (70.83%) and the highest other root and shoot traits. Both river sand and river sand + peat moss were outstanding for the quince cuttings, but IBA was not needed at the concentrations used in this study.

**Experiment 2. The Impact of Grafting Dates, The Cutting Types, and The IBA Concentrations on Grafting Success of Bench Grafted Loquats:**

This experiment was conducted from February to June 2022 in the lathhouse at the college site, which is to evaluate the grafting of loquat onto two cutting stocks; loquat and quince, performed on various dates (February 10, February 20, March 2, and March 12), and IBA concentrations (0, 1000, 2000, 3000 and 4000 mg.L$^{-}$). Graft bud sprout percentages were recorded on March 30, April 30, and May 30, along with rooting percentages. The results for loquat stock cuttings showed that the highest graft bud sprout percentage (37.50%) for loquat cutting, recorded on May 30, was linked to February 20 grafting and treatment with 1000 mg.L$^{-}$ IBA. Conversely, the lowest sprout percentage (0.00%) occurred when grafting was done on March 2 and 12, notably with 2000 and 3000 mg.L$^{-}$ IBA. In contrast, quince stock cuttings exhibited a peak sprout percentage of (41.67%) on March 30 when grafted on March 2 without IBA. Another notable peak of (41.67%) was observed on April 30 and May 30 with quince cuttings grafted on February 10 and treated with 3000 mg.L$^{-}$ IBA. February 20 grafting with loquat stock showed the highest interaction success on May 30 (37.50%), while the lowest percentages were observed at (0.00% for both stocks across different dates and IBA concentrations. Rooting consistently yielded (0.00%), emphasizing the need for further exploration in loquat bench grafting.

**Experiment 3. The Performance of Grafting Loquats Combined onto Loquat and Quince Rootstocks on Different Dates:**

This study was conducted in the lathhouse at the college site from February 20 to July 1, 2023, aiming to assess the impact of rootstock types (loquat and quince) and grafting dates (February 20, March 10, and March 30) in loquat grafting success. Grafting percentage, shoot traits, and chlorophyll content were recorded. The results indicated a significantly higher grafting success rate (97.78%) for loquat rootstock compared to quince rootstock (84.44%). The optimal grafting success of (96.67%) occurred on February 20 when both loquat and quince rootstocks were used. Conversely, the least favorable outcome (86.67%) was observed on March 30 using the same rootstocks for grafting. Remarkably, combining loquat rootstock with two grafting dates (February 20 and March 30) resulted in the highest significant grafting success percentage (100%).

**Experiment 4. The Effect of Grafting Dates and Stock Types on Grafting Success of Loquat and Quince Tree Stocks:**

This experiment was conducted at two locations: College site and Kani Waysa village, and aimed to evaluate the impact of tree stock types (loquat and quince) and grafting dates (February 20, March 10, and March 30) on loquat grafting success. Grafting percentage, shoot traits, and chlorophyll content, were measured. The findings revealed no significant differences in grafting success between the two tree stock types. However, the loquat tree stock demonstrated a higher graft success percentage (73.33%) compared to quince (66.67%). The grafting success percentages for all three dates were uniform at (70%). The interaction of loquat tree stock with the three grafting dates yielded the highest success (73.33%), while the interaction of quince tree stock with the same dates resulted in the lowest grafting success (66.67%).

**Experiment 5. The Impact of Cutting Types and Grafting Dates on Graft Bud Sprout and Rooting Percentage in Loquat Bench Grafting:**

This experiment was conducted at the college site in the lathhouse, to investigate the effect of loquat bench grafting on graft bud sprout percentage, leaf development, and biochemical parameters, utilizing two stock cutting types (loquat and quince) across different grafting dates (February 20, March 10, and March 30). Graft bud sprout percentages varied significantly, reaching the highest (86.11%) on loquat stock cuttings at 60 days after grafting (Dag), while quince stock cuttings recorded the lowest (59.72%). Graft bud sprout percentage consistently increased from February 20 to March 30. Leaf numbers favored loquat (2.70), with optimal development on February 20 (2.78). Biochemical parameters revealed distinct variations and showed higher carbohydrate content (7.44%) and lower total phenols level (825.67 %) in loquat. Emphasizing the critical role of grafting date selection, February 20 and March 30 yielded the highest carbohydrate percentages (6.58%).

## List of Contents

# List of Tables

# List of Figures

## List of Appendixes

# CHAPTER ONE

# INTRODUCTION

Loquat (*Eriobotrya japonica* Lindl.) is an important sub-tropical fruit tree that belongs to the family Rosaceae (Abbasi *et al.*, 2014). The Dadu River Valley of China is believed to be the original home of the genus *Eriobotrya* (Hussain, 2009). It is cultivated in many countries. China is the largest loquat-producing country in the world, with a production of 650,000 tons (Zheng *et al.*, 2019), followed by Spain, Pakistan, and Turkey (Tepe and Koyuncu, 2018). So, quince belongs to the Rosaceae family. It has about 30 genera and 1,000 species, including the genus *Cydonia*, which has only one species; *Cydonia oblonga* Mill., but with a larger number of varieties (Kole, 2011).

The central regions of quince (*Cydonia oblonga* Mill.) are located between Dagestan and Talysh in Trans-Caucasia and north of Iran. Historic documents present evidence of quince domestication in Mesopotamia, between 5000 and 4000 B, by the end of the nineteenth century (Abdollahi, 2019).

Historically, the loquat was known as an ornamental tree with small fruits. Later, in the nineteenth century, since the selection of cultivars with larger fruits, the fruits were used by consumers. Presently, loquat fruits are consumed largely as fresh fruit although small amounts are used in jams, jellies, syrups, and pies (Polat and Caliskan, 2007). On the other hand, quince is very interesting for growing, not only in gardens but in commercial orchards as well. It is characterized by high-quality and aromatic fruits that are appropriate for different purposes, specifically for processing into juice, compote, jam, jelly, and more recently a brandy that is highly appreciated due to its particular aroma (Rop *et al.*, 2011).

Loquat has formed a variety of ecological types in different zones during its cultivation and acclimatization. Generally, it can be found between latitudes 20 and 35 north and south (Vilanova *et al.*, 2001), but can be grown up to latitude 45 (Polat and Caliskan, 2007). Loquat has adapted well to the Mediterranean climate and is produced in the same areas where citrus is cultivated (Badenes *et al.*, 2003). However, it has more specific environmental requirements than citrus (Caballero and Fernandez, 2003 and Durgac *et al*., 2006). Commonly, the loquat tree is well adapted to almost all soils that have good drainage and hence grows equally well in acidic as well as in alkaline soils. The tree is cold tolerant to -10 °C but the fruits freeze at a low temperature of about -3 °C (Häkkinen, 2007). To establish a loquat orchard, the winter temperature should be higher than -3 °C and the summer temperature not above 35 °C (Lin, 2007). However, quince is cultivated on all continents in warm-temperate and temperate climates. It requires a cooler period of the year, with temperatures under 7 °C, to flower properly.

The loquat flowers are abundantly designed on panicles during autumn, with each panicle holding over 100 flowers. Each flower consists of 5 petals, 5 sepals, 5 stigmas, approximately 20 stamens, and 5 carpels, with each carpel containing 2 ovules. The loquat fruit begins to develop during winter and ripens in early spring. Due to its unique phenology, it reaches the market earlier than any other fruit during the spring season (Cuevas *et al.*, 2007). The fruit usually contains 1 to 5 large seeds, although up to 10 seeds may develop. While the loquat is primarily pollinated by bees, certain cultivars are self-infertile, and others are only partially self-fertile. Flowers from the early and late flushes tend to have abnormal stamens and very little viable pollen (Merino and Nogueras, 2003). In Kurdistan, loquat fruit becomes available in April-May when no other fresh fruit is present in the market, thereby filling a gap in the availability of seasonal fruits.

Loquat trees can be propagated through seed but do not ensure the protection of parental traits. Also, they need 6 to 8 years to grow in the juvenile period before flowering and fruiting. Therefore, vegetative propagation is one of the best options, to ensure the true genotype of loquat cultivars, and also bearing needs 1 to 2 years after planting Therefore, budding and grafting are the optimal methods in vegetative propagation for getting true-to-type plants but sometimes are difficult, because vegetatively propagated plants primarily depend upon proper time, and season (Hartmann *et al*., 2014). On the other hand, endogenous and exogenous factors such as growth substances, the anatomical structure of cutting, date of cutting, rooting media, humidity air and light conditions, and dates of budding and grafting, are always required to obtain satisfactory propagation success (Abbasi *et al.*, 2014). In contrast, quince trees are normally propagated by cuttings taken during the pruning season at the end of hibernation. The success in stem cutting multiplication of fruit crops depends upon some factors such as the condition of the mother plant, part of the tree, age of the tree from where the cuttings are made, date of planting, rainfall, humidity, and rooting media. Some researchers state that when quince is used as rootstock for loquats, the trees show dwarf growth, fruiting early, ripening earlier, fruits are larger and of higher quality (Polat and Caliskan, 2007). In some studies, it has been determined that quince rootstock provides 20-25% dwarfing in loquat varieties compared to loquat seedling rootstock (Polat and Akkus, 2022).

Plant growth regulators, especially indole-3-butyric acid (IBA), are important for helping fruit crops grow roots when using methods like cuttings, layering, or stooling. The success of these growth regulators, especially IBA, depends on factors like how much of the regulator is used, the type of the plant, and the time of the cuttings collection. IBA and similar growth regulators positively affect different aspects of cuttings, like moving nutrients and encouraging cells division. Understanding these processes is crucial for making propagation methods work better in various fruit crops. This understanding helps improve the effectiveness of propagation methods and ensures strong root development (Taiz *et al.*, 2014).

Media is a substrate that helps to provide moisture, support, nutrients, and aeration to the growing plant and helps in the proper growth and development of plant. The type of rooting medium plays a crucial role in increasing the rooting percentage, so the selection and the combination of medium components are important in the rooting success due to providing adequate aeration and drainage to ensure faster and better-quality root development. Soil texture is also an important physical property that plays a key role in seed germination and rooting of cutting (Raviv, 2005).

**The Aims of the Study:**

The aims of the study could be summarized as follows:

1. To determine the most effective propagation method for loquat and quince, evaluating the use of either seeds or cuttings for rootstock production.
2. To determine the optimal rooting medium for quince cuttings, and to compare the effectiveness of river sand, river sand with peat moss, and peat moss with perlite.
3. To identify the most suitable IBA (Indole-3-butyric acid) concentrations for promoting root formation in quince and loquat cuttings.
4. To assess the compatibility between produced rootstocks (both loquat and quince) and loquat scion cultivars to determine the best grafting combinations.
5. To determine the ideal timing for the vegetative propagation of loquat and quince through the use of cuttings and grafting techniques.
6. To produce dwarf rootstock and expedite early fruit production for loquat grafting.

# CHAPTER TWO
# LITERATURE REVIEW

## 2.1 Plant Propagation

The foremost aim of propagation is to produce more plants. The fruit plants are propagated both by sexual and asexual means. In sexual methods, plants are raised through seeds, and such plants are called seedlings. However, in asexual or vegetative propagation, a part of the plant viz. leaves, stems, branches and roots are used to develop new plants which are called saplings (Singh, 2013). Various propagation methods, including budding, grafting, cuttings, and sucker removal are commonly employed in this process. Each method carries its distinct advantages and limitations.

Grafting: The process includes attaching a segment of the desired tree (the scion) to the rootstock of a different tree. While it allows precise control over the characteristics of the resulting tree, it requires specialized skills and knowledge for effective implementation.

Hardwood cuttings, prepared during the dormant season from one to two-year-old shoots of the previous season's growth, are derived from mature, dormant, and firm wood after leaf abscission. This method is considered as one of the simplest and most economical techniques of vegetative propagation (Abbasi *et al.*, 2014). The potential for root development in stem cuttings varies depending on several factors, including species and specific cultivar requirements, stock plant condition and age, the position and type of cuttings taken, the removal of leaves, and the use of etiolation and girdling techniques. Other influential factors include the timing of cutting, as well as growing conditions such as the choice of growing media, misting, bottom heat, hormone application, fertilization, and supplementary lighting (Hartmann and Kester, 1990). Studies conducted by Mohammed *et al.* (2020) emphasized the successful propagation of quince through hardwood cuttings from one-year-old wood, with the recommended length of cuttings being around 20 cm. Hartmann *et al.* (2014) further highlighted that the basal cut is typically made just below the basal node of the cuttings. In horticulture, cuttings are commonly taken just below a node, because the nodes existed on the stem are known sites for the formation of adventitious roots.

## 2.2 Plant Growth Regulators (PGRs)

Plant growth regulators are a significant factor in the successful enhancement of root formation in fruit tree cuttings. Researchers have demonstrated that various physiological processes occur during the rooting of cuttings, including the influence of growth regulators, the role of vitamins, the presence of buds and leaves, the rooting co-factors, the nutritional factors, and the endogenous rooting inhibitors. These natural occurrences may act as promoters and inhibitors in the process of root initiation. Additionally, various classes of growth regulators, such as auxins, cytokinins, gibberellins, abscisic acid, and ethylene influence the rooting of cuttings. However, as the most common growth regulator used in rooting cuttings is indole-3-butyric acid (IBA), this part will focus on the role of this PGR, which is considered to have the greatest effect on root formation in cuttings. The required concentration of IBA varies depending on the plant and the type of cuttings used. Rooting can be further enhanced by the use of growth regulators such as IBA and different rooting media. Synthetic auxins, including IBA, are extensively used for inducing rooting. Due to its ability to increase rooting and induce a fibrous root system, IBA is widely preferred (Singh *et al*., 2019a). IBA is an organic compound that influences plant growth by promoting cell enlargement, elongation, and root initiation. It is commonly utilized in tissue culture procedures to stimulate root initiation in explants or calluses ( Williamson and Jackson, 1994).

IBA concentration of 1000 mg.L$^-$ significantly influenced vegetative growth and multiplication rate of stem cuttings of quince rootstocks (Mehta *et al*., 2016). Bhusal (2001) examined the effect of indole butyric acid at concentrations of 4000 mg.L$^-$ on stem cuttings of *Citrus* and observed that IBA-treated cuttings exhibited an increase in both the number and the length of roots compared to non-treated cuttings. However, the final rooting percentage did not show significant differences between the various IBA treatments and the control group. Ak *et al.* (2021) conducted a study under Isparta, Turkey conditions to examine the impact of cutting period and the rooting medium on the rooting rate and root quality in black mulberry and white mulberry cuttings. His study revealed that the average rooting rate ranged from (2.22% to 71%) in black mulberry cuttings and from (3.33% to 50%) in white mulberry cuttings when treated with a dose of 5000 mg.L$^-$ IBA hormone. In another study, the same researcher used various growth regulators, including 10-100 mg.L$^-$ IAA, IBA, and NAA were applied to white mulberry. The research findings indicated that these applications effectively stimulated callus and root formation under all tested conditions. Moreover, the study highlighted that the application of these growth regulators encouraged rooting even when rooting did not occur under standard conditions. Rahman *et al*. (2002) noted that the cutting of the olive cultivar Coratina when treated with 3000 mg.L$^-$ IBA, exhibited the best results in terms of shoot length. Also, Tsipouridis *et al.* (2003) found that the rooting percentage in apple cuttings reached (31.4%) at an IBA concentration of 3000 mg.L$^-$. Additionally, Dvin *et al.* (2011) found that the rooting percentage for M.26 and MM106 rootstock cuttings were (7.2% and 25%), respectively, when using 2000 mg.L$^{-}$. Also, Kauppinen *et al.* (2003) found that IBA promoted rooting in base cuttings of Japanese quince. The ones treated with 100 mg.L$^{-}$ or 200 mg.L$^{-}$ IBA resulted in the highest percentage of rooted cuttings. According to Štefančič *et al*. (2005), the process of adventitious root formation is influenced by various internal and external factors. Among these factors, phytohormones, particularly auxins, are attributed to the most crucial role. It is widely acknowledged that auxins play a significant role in initiating rooting and thus governing the growth and development of plants, including lateral root initiation, root gravity response, and other parameters of vegetative growth, such as seedling height, vegetative fresh weight, vegetative dry weight, the number of leaves per plant, and the leaf area. Arpaia (2005) noted that IBA proved to be the most suitable plant growth regulator for improving success, along with other rooting and vegetative characteristics of the Shan-i-Punjab peach. Alam *et al*. (2007) observed that IBA enhanced the percent plant survival, the number of roots per plant, root length, root weight, root diameter, number of leaves, and shoot diameter in Kiwi cuttings. Pelicano *et al*. (2007) found that the application of IBA to mulberry (*Morus alba*) resulted in the best responses in terms of root growth. Their study revealed that young cuttings treated with 3500 mg.L$^{-}$ of IBA exhibited a high percentage of root growth, reaching (80%) in the first month and (83%) in the second month.

## 2.3 Rooting Media

A rooting medium refers to any substrate that fosters the growth of roots, typically composed of various organic and inorganic components. The choice of the best rooting medium depends on the grower's available materials and the specific plant species. Various types of rooting media, including sand, soil, peat moss, coconut husk, vermiculite, and perlite, are utilized for the growth of new seedlings or cuttings. However, the success of a rooting medium varies based on factors such as location, materials used, and the chosen propagation method. The selection of a suitable growing medium plays a crucial role in the root proliferation and further growth of plants propagated by stem cutting. It is imperative to consider the rooting medium as an integral part of the propagation system. An ideal growing medium should provide adequate anchorage or support for the plant, act as a reservoir for nutrients and water, and facilitate oxygen diffusion to the roots, as well as gaseous exchange between the roots and the external atmosphere (Abad *et al.*, 2002).

The selection of a growing medium is critical, as it should possess certain key characteristics, including porosity, uniform texture, the capacity to retain sufficient moisture, and effective drainage (Esringü *et al.*, 2022). Bhardwaj (2013) emphasized that the medium provides the necessary physical support, aeration, and water for the plants. Notably, planting loquat cuttings in a mixture of sand and peat moss after the application of IBA under a mist propagation unit has demonstrated a (50%) rooting rate in the greenhouse (Abbasi *et al.*, 2014).

Observations gained by Al-Zebari and Al-Brifkany (2015) suggest that a medium composed of 1 part peat moss and 2 parts sand effectively enhances the rooting percentage in stem cuttings of the *Citrus medica* L. Corsian cultivar. Singh *et al*. (2018) recommend the use of peat or a mixture of peat, perlite, and cocopeat in a 2:2:1 ratio to increase plant height and leaf number in citrus cuttings. Furthermore, mixtures such as perlite plus peat, coconut fiber, or vermiculite have also yielded favorable outcomes.

## 2.4 Rootstock

The choice of a proper rootstock is an essential stage in successful fruit production. The selection of a rootstock in any situation depends on the choice of the training system, spacing, site, tree vigor, scion variety, plant growth and development stage, and soil (Baron *et al.*, 2019). Rahman *et al.* (2017) reported that rootstocks play a vital role in modern fruit production, adapting to specific cultivars in diverse environmental conditions. Rootstocks influence tree growth, precocity, production, nutritional elements of the tree (Milošević and Milošević, 2015), alternate bearing (Reig *et al.*, 2018), chilling and frost tolerance (Robinson, 2007), drought tolerance (Tworkoski *et al*., 2016), physical and chemical composition of the fruit (Kviklys *et al.*, 2017). From these aspects, Polat and Caliskan (2007) recommended grafting loquat onto quince rootstock, resulting in a significantly reduced size of 'Magdal' loquat in Spain. The grafted trees, planted at (2.5 × 1.7) meter spacing (2,353 trees/hectare), reached a maximum height of (1.87) meters. This size reduction led to cost savings in labor, especially during blossom, fruit thinning, and harvest. Additionally, earliness of production and higher yield have been achieved with the use of this rootstock under protected cultivation (Hueso *et al.*, 2007). It has been reported that dwarfing rootstocks reduce the concentration of elements such as nitrogen and potassium in the leaf of the scion compared to vigorous rootstocks (Sotiropoulos, 2008). Bermede and Polat (2011) indicated that the integration of dwarfing rootstocks in loquat production results in a reduction in tree size, enabling the planting of a higher number of trees per unit area. This practice not only leads to an increase in early yield, facilitating a streamlined harvest and reducing costs but also enhances overall orchard management. In the context of loquat cultivation, the implementation of quince as a rootstock effectively decelerates scion growth, leading to a notable decrease in tree size by approximately (20 to 25%) compared to loquat seedlings. This process simultaneously promotes earlier fruiting and contributes to an improvement in both fruit quality and size. However, the successful grafting of loquat on quince encounters significant challenges due to a notably low success rate, ultimately hindering the commercial production of dwarf loquat trees. Rabi *et al*. (2014) reported that grafting five apple cultivars (Royal Gala, Mondial Gala, Treco Gala, Gala Must and Spartan) by bench grafting on rootstocks MM-111,106, M-9, 26 and crab apple increased graft take success (91.10%). In addition, plant height (107.97cm), scion diameter (9.38 mm), the number of leaves per plant (116.5) were noted in cultivar Gala Must with crab apple rootstock. The focus of our comprehensive two-year study was to evaluate diverse methods aimed at facilitating the rooting and grafting success of quince as a dwarfing rootstock for loquat trees. When selecting an appropriate rootstock for loquat trees, several crucial factors come into play, including the success rate, and compatibility with the scion, as well as the prevailing soil and environmental conditions. The concentration of leaf nutrient elements varies with the rootstock, with dwarfing rootstocks absorbing fewer nutritional elements than vigorous rootstocks (Ikinci *et al.*, 2014). Quince rootstocks have been widely used due to some beneficial characteristics such as tree size reduction, yield precocity, and improvements in fruit size and quality. The rootstocks have a significant effect on the growth characteristics of fruit trees, with their effect accounting for up to (50%) of the economic outcome in some fruit tree species (Mezey and Leško, 2014). Dogra *et al*. (2018) observed that endogenous plant hormones are thought to be involved in regulating the complex relationships between rootstock and scion, emphasizing the importance of selecting a good rootstock and a suitable variety to ensure a healthy relationship and avoid economic losses for producers in the future.

### 2.5 Graft Union

Grafting involves bringing together two genetically distinct but similar plant parts that, under suitable conditions, can form a composite plant. For most grafting purposes, the two plant parts used are the rootstock and the scion. The rootstock comprises the root system and an above-ground stem portion. Rootstocks are often classified based on tree vigor, such as dwarfing, semi-dwarfing, semi-vigorous, and vigorous. Rootstocks also vary in their resistance to pests and diseases, as well as their influence on fruiting. On the other hand, the scion refers to the specific cultivar or variety that forms the top part of the plant. In the case of apple trees, the scion can be chosen from the numerous apple varieties available, depending on the preferences of the grower.

There are two main types of grafting commonly used in the industry and in the research: bench grafting and grafting on rootstock. Bench grafting involves using scion tissue containing several viable buds and is typically performed in late winter and early spring when the rootstock and scion tissue are dormant. The grafted trees are then transplanted from the nursery in the spring before bud break (known as a 'sleeping eye') or in the fall after a season of growth. Hartmann *et al.* (2014) provided an excellent review of several commonly used methods of grafting and budding. The sequence of events in graft formation has been thoroughly examined and documented for both herbaceous and woody plants (Pina and Errea, 2005). Regardless of how each event is presented, all descriptions of graft formation share four common characteristics: initial necrotic/isolation layer deposition, callus bridge formation, cambial layer differentiation, and secondary xylem and phloem development. Budding is done during the active growth stage, while grafting is done during dormancy in winter or early spring. Some important methods of grafting apples are whip or tongue grafting and cleft grafting (Chaudhry, 1996).

Although there is no common method currently used to successfully graft genetically incompatible plants, cultural practices can help improve graft development and plant survival of compatible combinations. Hartmann *et al.* (2014) identified four conditions necessary for successful grafting, which are separate from the issue of compatibility:

- The vascular cambium of the rootstock and the scion should be placed in direct contact and held tightly together. Similarly sized graft partners are desired when doing bench grafting.
- Grafting should take place when the scion and rootstock are in the appropriate physiological condition, which depends on the grafting method.
- All of the cut surfaces need to be protected from desiccation by wrapping them with tape, parafilm tape, and budding tape, covered with moist sawdust, and/or kept in humid conditions.
- The removal of suckers and staking of new growth to direct energy to where it is needed and to prevent the breaking of new growth when the graft is weak.

Several studies have sought to investigate practical approaches to provide the above conditions. Singh *et al.* (2019a) found that "V" grafting of two apple cultivars on M.26 and three pear cultivars on quince resulted in increased callus formation, graft survival, growth, and branching when compared to "Omega" grafting. Additionally, Mng'Omba and du Toit (2013) investigated the effect of the length of the cut surface on spliced graft mango, avocado, and peach trees.

They found that for mango and peach, the increased length of the cut surface improved graft success, while it did not affect avocado. Also, they found that 40-millimeter cuts had good graft success for all three species. From these studies, success may be improved by various grafting methods and techniques.

## 2.6 Compatibility and Incompatibility in Grafting

Graft incompatibility refers to the failure of a rootstock and scion to form a successful graft union, which can be caused by various factors such as anatomical mismatching, poor graftsmanship, adverse environmental conditions, diseases, and physiological responses between the grafting partners. It can also result from the transmission of viruses or phytoplasma and anatomical abnormalities of vascular tissues in the callus bridge. Distinguishing between compatible and incompatible graft unions may not always be straightforward (Dogra *et al.*, 2018).

Despite the challenges, grafting remains a widely used technique in contemporary fruit-growing practices, facilitating the successful union and development of two distinct plants into a single composite plant. It is employed for the reproduction of various fruit species and varieties, allowing grafting within the same variety, cultivar, species, or genus, as well as between different cultivars, species, or genera (Dogra *et al.*, 2018). The performance of both the scion and the rootstock hinges on the compatibility of the two components. Generally, closely related cultivars and species are compatible, while distantly related plants are often graft-incompatible. "Compatible" here refers to a satisfactory union that considers the physical connection at the graft union and the physiological harmony of the genetic system. For instance, Bartlett's pear on most quince rootstocks is weak and prone to breakage at the union. Bartlett on Oriental sand pear (*Pyrus pyrifolia*) root, although physically strong at the union, can lead to severe fruit disorders, such as blackening of the calyx end (known as 'blackened'). Some incompatibilities may manifest after a significant delay, while most incompatibilities in orchard trees become evident at an early stage. Graft unions after several years of growth can be of three types: a) equal size scion and stock, b) scion overgrowth, and c) stock overgrowth. Unequal size between the scion and stock does not necessarily indicate incompatibility; many unions with scion overgrowth are compatible and reflect the genetic tendency of the scion to undergo increased wood growth and cambial activity. For example, Comic pear overgrows all types of rootstocks, even though it is completely compatible (Chalise *et al.*, 2013).

The symptoms mentioned by Hartmann *et al.* (1997) are useful in identifying potential cases of graft incompatibility. These indicators can help distinguish between successful and unsuccessful graft unions, enabling growers to take appropriate measures in managing the grafting process. The symptoms include:

- Failure to form a successful graft or bud union in a high percentage of cases: An indication that the grafting process has not resulted in a robust and durable union between the scion and rootstock.
- Yellowing foliage in the later part of the growing season followed by early defoliation, a decline in vegetative growth, appearance of shoot die-back, and general ill health of the tree: Signs of physiological stress or poor compatibility between the grafted components lead to compromised health and vitality.
- The premature death of the trees which may live for only a year or two in the nursery: A severe consequence of graft incompatibility, resulting in the early demise of the grafted plants.
- Marked differences in growth rate or vigor of the scion and rootstock: Discrepancies in the growth patterns and overall vigor between the grafted components, suggest an imbalance or lack of synchronization between the scion and rootstock.
- Differences between scion and rootstock in the time at which vegetative growth for the season begins or ends: Disparities in the seasonal growth behavior of the scion and rootstock indicate a lack of harmonious physiological response between the grafted components.
- Overgrowths at or above the graft union: Excessive growth occurring either above or below the graft union, signify a lack of proper fusion and integration between the scion and rootstock.
- Suckering of rootstock: Emergence of shoots or growth from the rootstock below the graft union, which can lead to a competitive relationship with the scion and affect overall plant development.
- Graft components breaking apart cleanly at the graft union: Weak graft union leads to the separation of the scion and rootstock, often resulting from poor integration or compatibility issues between the two components.

## 2.7 Grafting Success

Grafting can be understood as the practice of connecting two sections of living plant tissue in a manner that fosters their fusion and subsequent growth and development as a unified plant entity (Chalise *et al.*, 2013). A successful grafting process relies on the proper attachment of the scion and rootstock, facilitated by the formation of callus and subsequent tissue differentiation. The pivotal factor in grafting success is the fusion of the cambium of the stock and the scion, as no graft union can occur without this critical step. However, it is important to note that there exists considerable variation in the ability of certain plants to successfully unite with others. Some plants readily amalgamate, making the grafting process relatively straightforward, while others present more challenging prospects, and certain plant combinations do not unite at all. This diversity underscores the significance of understanding the compatibility between different plant species and selecting appropriate grafting techniques accordingly.

### 2.7.1 Conditions for successful grafting

In all grafting procedures, the precise alignment of the cambial layer of the stock with that of the scion is of paramount importance. To achieve this, both the stock and the scion must be of the same thickness to ensure a proper match. Additionally, the cut should be uniform and smooth, facilitating a firm and secure connection between the two components, thereby preventing any open gaps. Poor-fitting between the scion and the stock may result in the need for an increased amount of callus formation, potentially leading to a weak or incomplete union. Karadeniz (2003) studied the anatomical and histological developments of chip budding and tongue grafts on loquat under controlled conditions, 25±2 C° temperature, and 75-80% humidity. The graft union and the development were studied at the cross and longitudinal graft sections after 15, 22, 30, 45, 60, and 120 days from grafting. demonstrated that callus bridges can be observed 22 days after grafting in both grafting techniques. also confirmed that the graft union and the growth of the loquat were excellent, consequently grafting of the loquat was similar to the other fruit species.

The following basic prerequisites are crucial for the successful formation of a graft union (Chalise *et al.*, 2013):

- Grafting should be performed when the stocks are in an active growth stage and there is a sufficient flow of sap.
- Ideally, the thickness of the stock and the scion should be the same, with the scion being of a pencil size and containing at least two buds.
- Exclusion of air from the point of union is essential, necessitating close contact between the two graft components.
- Protection from direct sunlight and rain is necessary until the union is complete.
- It is crucial to use the appropriate age of the scion and stock, avoiding the use of excessively old stock.
- Regulation of moisture and temperature is necessary to maintain the plant in an active growth stage.
- The tying materials should only be removed after a complete union has been achieved.

### 2.7.2 Factors affecting graft union success

The factors listed by Hartmann *et al*. (2014) are essential considerations when assessing the success of graft unions. They underscore the complex interplay between various biological and environmental factors that can influence the outcome of the grafting process. These factors include:

- Incompatibility: This refers to the degree of suitability between the scion and the rootstock, encompassing physiological and genetic aspects that influence the formation and strength of the graft union.
- Plant species and type of graft: Different plant species and graft types may exhibit varying degrees of compatibility and success rates, impacting the overall efficacy of the grafting process.

  The results of Hussain *et al*. (2017) verified that wedge or (cleft) and side grafting had the highest graft take with Kinnow mandarin (90.00% and 86.67%, respectively): tongue grafting had the highest graft take with the Jaffa sweet orange (76.67%): and the side grafting alone had the highest graft take with the Succari sweet orange. The shoot length was significantly higher in Mandarin cv. Kinnow (16.00 cm) followed by sweet orange cv. Succari (11.67cm) as compared to Jaffa (11.00 cm). In general, the results showed that side grafting was the most effective method of propagation for all evaluated cultivars.

- Environmental conditions during and following grafting: Factors such as temperature, moisture levels, and plant water relations play a crucial role in determining the success of graft unions, as they directly affect the physiological processes involved in graft formation and subsequent growth.
- Growth activity of the rootstock: The growth stage and activity of the rootstock can significantly impact the success of the graft, as it influences the physiological response of the plant to the grafting process.

  The apple cultivar 'Custard' grafting study was carried out at different times ($3^{rd}$ week of January, $1^{st}$ week of February, $3^{rd}$ week of February, and $1^{st}$ week of March), having different ages of rootstock i.e. (seven months, eight months, nine months and ten months). The results from three consecutive years of investigation revealed that rootstock age and grafting time significantly influence the grafting success percentage. Seven-month-old rootstock grafting during $3^{rd}$ week of January showed the highest grafting success (95.11%), the maximum height of bud shoot (64.22 cm), the highest number of leaves (17.89), and diameter (1.06 cm) obtained on ten-month-old rootstock grafted at $1^{st}$ week of March (Dharmik *et al*. 2022).
- Virus contamination, insects, and diseases: Pathogens, pests, and diseases can affect the health and vigor of the plant, potentially compromising the success of the graft union and overall plant development.
- Plant growth regulators: The use of growth regulators can influence the physiological processes of the grafted plants, affecting their growth, development, and the formation of the graft union.
- Post-graftage bud forcing methods: Techniques employed after grafting, such as bud forcing, can have a significant impact on the subsequent growth and development of the grafted plant, influencing the overall success of the graft union.

## 2.8 The Effect of Scion Treatments with Plant Growth Regulators

Plant growth regulators (PGRs), including both endogenous plant hormones and exogenous substances, have long been recognized for their crucial roles in various metabolic processes, such as reproduction, growth, and development. These substances, such as auxins, gibberellins, cytokinin, abscisic acid, and ethylene, are currently subjects of extensive research in diverse fields.

In a study by (Simkhada, 2007), plant growth regulators, particularly auxins, applied to tree wounds or graft unions, were found to yield varying results in terms of wound response and graft union formation. For instance, auxins such as (IAA and NAA) and cytokinin such as (BA) were observed to enhance graft success in *Picea* scions, while dikegulac stimulated scion growth by retarding rootstock development. Additionally, cytokinin was found to enhance the patch budding of Persian Walnuts (*Juglans regia* L.). Nonetheless, unlike their usage in cutting propagation, plant growth regulators are not commonly used in commercial grafting and budding systems. It is noted that these regulators do not uniformly enhance grafting and may not overcome graft incompatibility.

Shaban (2010) examined the endogenous hormones and phenols of seedling trees from various polyembryonic mango cultivars commonly used as rootstocks, the study their effects on the scion vigor of cv. Alphonso. The study revealed that the hormonal levels in the rootstocks correlated with the growth and development of the scion. Moreover, Sorce *et al.* (2002) conducted a study on the hormonal relationships in the xylem sap of grafted and ungrafted *Prunus* rootstocks. Their research demonstrated that the growth potential was influenced by the transport rates of IAA and zeatin riboside in the xylem sap.

Furthermore, Hama Salieh (2004) investigated the impact of plant growth regulators, specifically indole-3-acetic acid (IAA) and kinetin, on the success of T-budding in pistachio (*Pistacia vera* L.). The findings of the study demonstrated significant effects attributed to the use of various plant growth regulators. Sorce *et al.* (2007) focused on the hormonal factors influencing the control of vigor in grafted peach seedlings (*Prunus persica* L. Batsch). Their findings indicated that the flow of IAA and cytokinin sap was positively correlated with tree vigor, suggesting that the supply of these hormones through the xylem enhances shoot development. Moreover, the growth of the crown of peach scions was found to be lower than that of the rootstocks, emphasizing the role of hormonal modulation in scion growth.

## 2.9 The Effect of Time on Grafting Success

Grafting, an ancient horticultural method dating back thousands of years, involves uniting the scion and rootstock for vegetative propagation, especially in fruit crops like loquat. Rootstocks, sourced from seedlings, cuttings, or layered plants, significantly influence characteristics such as tree size, growth habits, yield, and fruit maturity time (Beshir *et al.*, 2019). This technique is employed for challenging cultivars where sexual reproduction or vegetative methods pose difficulties, capitalizing on the advantageous traits of specific rootstocks (Kako *et al.*, 2012).

Successful graft unions depend on the wound response sequence, fostering a continuous cambium and vascular system between the scion and rootstock (Rasool *et al.*, 2020). Grafting, serving as an asexual propagation method, ensures the production of true-to-type plants by merging two living plant tissues into a single entity. Compatibility allows for successful unions, extensively used in reproducing various fruit species and varieties, even within different varieties, cultivars, species, or genera (Hartmann *et al.*, 2014 and Dogra *et al.*, 2018). While cleft-grafted plants often exhibit superior architecture compared to trees propagated by other methods, such as whip grafting (Rasool *et al.*, 2020). The timing of grafting plays a crucial role in success. Grafting time impacts parameters like bud take, shoot length, and overall success (Vural *et al*., 2008). Additionally, Gautam *et al*. (2001) documented higher success rates in mandarin grafting on January 31st and January 16th. Adhikari (2006) noted the highest success rate (79.73%) in acid lime grafting on January 16th. Ghojage *et al*. (2011) reported the highest success rate in February (81.66%), comparable to October (80.00%). Moreover, Mir and Kumar (2011) reported that conducting grafting in the third week of February resulted in the highest scion length and leaf count. They observed earlier bud burst occurring at 29 and 32 days when grafting was performed in both the third and fourth weeks of February. Aminzadeh *et al.* (2013) from Iran emphasized the significant impact of grafting time on both graft take and survival. Wani *et al*. (2017) reported increased scion length, leaf count, and scion thickness when grafting took place on January 20th and 30th in Kashmir. Mehta *et al*. (2018) reported the shortest time to bud burst (25 days) in tongue grafting during the first week of March. They also observed an increase in scion length. Grafting time was identified as a crucial factor for success (Yetgin, 2010).

Selecting appropriate rootstocks and timing the grafting process according to optimal growing conditions can enhance success rates (Mng'omba *et al.*, 2010). The researcher Hama Salieh (2004) found that budding on August 15 resulted in the highest values for bud shoot height, bud shoot diameter, and budding success for pistachio. These studies indicate that the timing of the grafting process significantly influences the success and growth parameters of the grafted plants. Pear scion cultivars (Carmen, Abate Fetel, William Bartlett, and Chinese Sandy Pear) were bench grafted (cleft grafting) on five rootstocks (Quince C, BA-29, Quince seedling, Kainth seedling and Pear suckers) during Mid-March. The earliest bud burst at 14th April was recorded in Chinese sandy Pear/Quince and the highest grafting success (96.67%) was recorded in Carmen/Quince C, Chinese Sandy Pear/Quince C, Abate Fetel/BA29 and Chinese Sandy pear/Quince (Rathore *et al.*, 2023).

**Table 2.1 The developmental stages and time intervals of graft union formation in T-budded Citrus.**

| Stage of development | Approximate time after budding |
|---|---|
| 1. First cell division | 24 hours |
| 2. First callus bridge | 5 days |
| 3. Differentiation of callus | |
| a) In the callus of the bark flaps (rootstock) | 10 days |
| b) In the callus of the shield bud (scion) | 15 days |
| 4- First occurrence of xylem tracheids | |
| a) In the callus of the bark flaps | 15 days |
| b) In the callus of the shield | 20 days |
| 5. Lignification of the callus completed | |
| a) In the bark flaps | 25 to 30 days |
| b) Under the shield | 30 to 45 days |

## 2.10 Post–Graftage Bud Forcing

Hartmann *et al.* (1997) mentioned that after graft union formation has occurred in grafting or budding, it is often necessary to force out the scion or the scion bud. The axillary buds of the rootstock, which develop into photosynthesizing branches, are initially important for the growth of the composite plant. However, they can inhibit the growth of the scion through apical dominance, which is an auxin response. By crippling (cutting halfway through the rootstock shoot above the bud union and breaking the shoot over the rootstock stem), girdling, or removing the rootstock above the scion bud union, apical dominance is broken, and the scion bud rapidly elongates.

# CHAPTER THREE
# MATERIALS AND METHODS

## 3.1 Locations

### 3.1.1 Location 1

The experiments were carried out during 2021-2023 in the Department of Horticulture, College of Agricultural Engineering Sciences, University of Sulaimani, located in Bakrajo 15 km southwestern of Sulaymaniyah City, Kurdistan Region-Iraq (Figure 3.1). The location coordinates by GPS were (35°32′ 15″ N, 45°21′ 52″ E), and 730 meters above mean sea level, the meteorological data of the Sulaymaniyah location during the study period are shown in (Table 3.1).

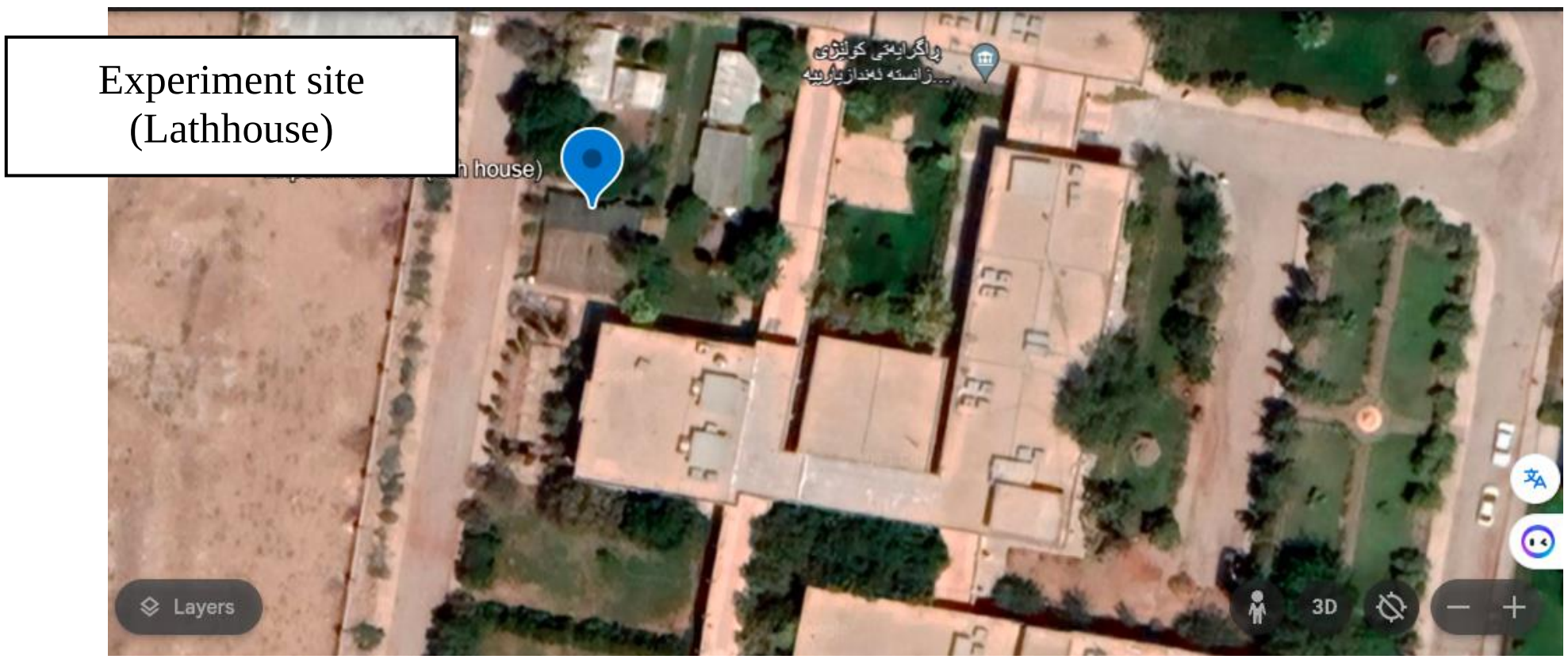

**Figure 3.1 The study site for the experiment was the College of Agricultural Engineering Sciences.**

**Table 3.1 Meteorological data of Suliamaniyah city during the study period.**

| Year | Month | Min. Temp.(°C) | Max. Temp.(°C) | Avg. Temp.(°C) | Relative Humidity (%) |
|---|---|---|---|---|---|
| 2021 | February | 4.6 | 17.5 | 11.0 | 63 |
| | March | 7.7 | 20.2 | 14.0 | 57.3 |
| | April | 13.9 | 29.4 | 21.7 | 42.6 |
| | May | 20.3 | 35.4 | 27.9 | 33.2 |
| | June | 22.6 | 40.0 | 31.3 | 27.5 |
| 2022 | February | -2.13 | 19.92 | 8.90 | 64.81 |
| | March | - 4.56 | 22.69 | 9.07 | 64.31 |
| | April | 5.12 | 29.98 | 17.55 | 46.12 |
| | May | 8.48 | 37.22 | 22.85 | 40.0 |
| | June | 18.01 | 42.58 | 30.3 | 20.56 |
| 2023 | February | 2.51 | 11.37 | 6.94 | 62.66 |
| | March | 5.23 | 15.72 | 10.48 | 59.94 |
| | April | 9.19 | 21.58 | 15.39 | 52.7 |
| | May | 14.87 | 28.26 | 21.57 | 38.51 |
| | June | 20.58 | 35.58 | 28.08 | 21.04 |

*Data obtained from; Nasa https://power.larc.nasa.gov/

### 3.1.2 Location 2

Part of the experiment was carried out during the growing season 2023 on the local quince trees, 4-5 years old, in a private orchard cultivated at Kani Waysa village, Sitak, 25 km northeast of Sulaymaniyah city (Figure 3.2) at 1057 m above sea level with the coordinates 35˚40′ 02″ N and 45˚34′30″ E. The meteorological data of the location during the study period are shown in (Table 3.2).

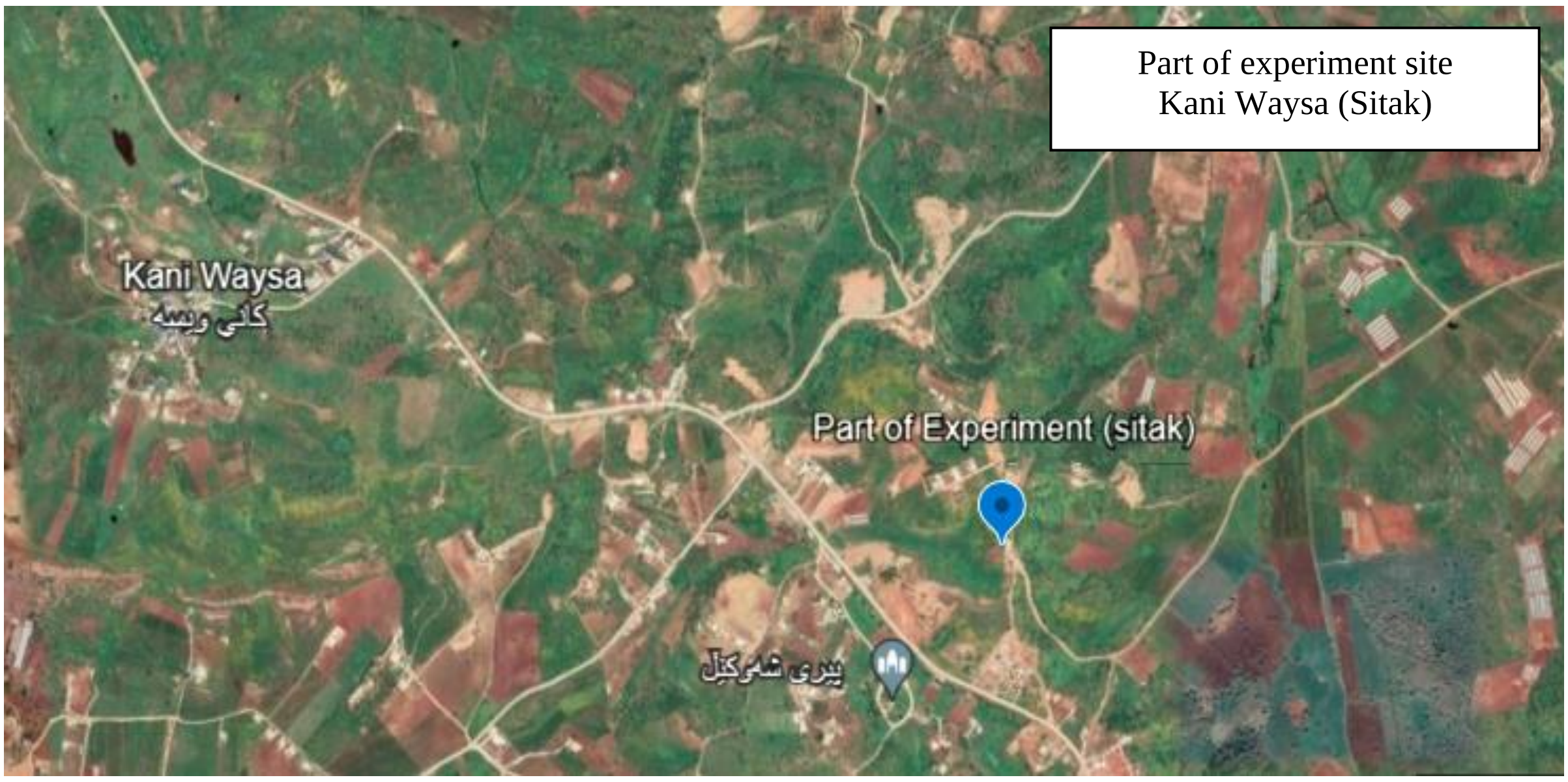

**Figure 3.2 The sitemap for location2 of the experiment.**

**Table 3.2 Meteorological data of Kani Waysa during the study period.**

| Year | Month | Temperature (°C) * | | | Relative Humidity (%) |
|---|---|---|---|---|---|
| | | Max. | Min. | Avg. | |
| 2023 | February | 9.75 | -1.30 | 4.23 | 65.93 |
| | March | 16.02 | 5.18 | 10.6 | 71.07 |
| | April | 21.21 | 8.50 | 14.86 | 62.72 |
| | May | 28.05 | 13.82 | 20.94 | 49.95 |
| | June | 35.88 | 19.45 | 27.67 | 31.21 |

*Data obtained from; Nasa https://power.larc.nasa.gov/

## 3.2 The Preparation of Plant Growth Regulator Solutions

### 3.2.1 IBA (Indole-3-butyric acid) solutions

IBA solutions were prepared at concentrations of (1000, 2000, 3000, and 4000 $mg.L^{-}$) by dissolving 4 g of IBA in 500 mL ethanol with 96% purity, and then completing the volume to 1 L for preparing the 4000 $mg.L^{-}$, followed by a dilution series with distilled water for the lower concentrations, based on the method outlined by (Evert and Smittle, 1990).

### 3.2.2 BA (Benzyl adenine) solutions

Cytokinin such as (BA) was used in grafting, 10 $mg.L^{-}$ benzyl adenine (BA) solution was prepared by dissolving (10) milligrams of BA in 10 ml of (NaOH 0.1N), and the volume was completed to 1 L with distilled water.

## 3.3. The Preparation of Mother Trees for Cutting Collection

### 3.3.1. Selection of loquat mother tree stocks

Loquat cuttings were taken from the loquat tree stocks of 10-12 years old, known for their abundant and high-quality fruit characteristics, to be used in this study. The trees were healthy and pest and disease-free, to which cuttings of uniform diameters were selected for the experiment. The cuttings were obtained from the middle part of 2-year-old shoots, measuring 25 cm in length and 10±2 mm in diameter, at various times according to the experiment's method. These cuttings were taken in the Qularaisy region, located 5 kilometers northwest of Sulaymaniyah city, with coordinates 35˚36′10″ N and 45˚21′47″ E, at an elevation of 820 meters (Figure 3.3). After obtaining the cuttings at different times, they were transferred to the experimental site. The physical and chemical analyses of the soil of the experimental site are shown in (Table 3.3).

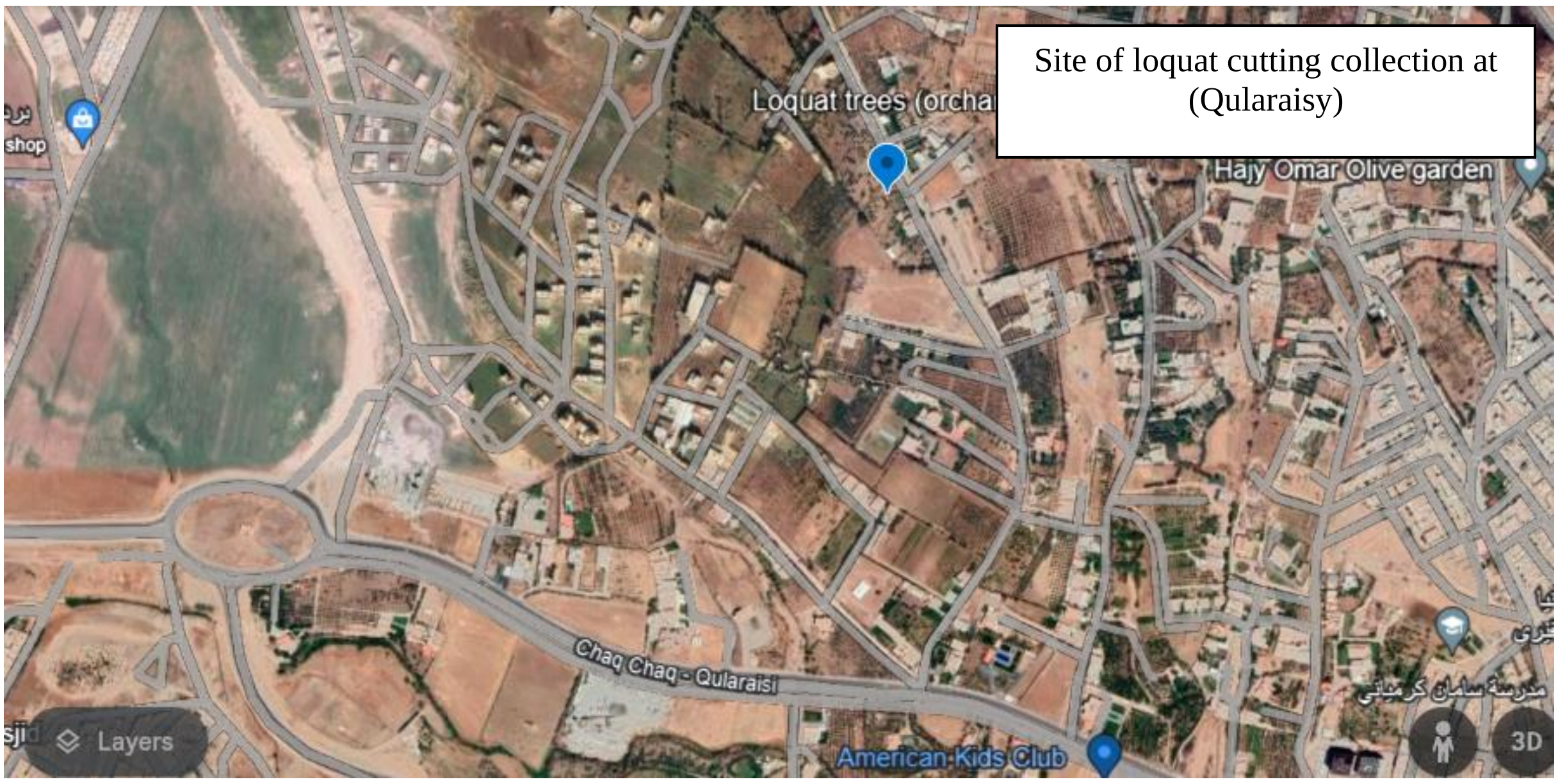

**Figure 3.3 The sitemap for the loquat cutting collection at Qularaisy.**

**Table 3.3 Some physical and chemical characteristics of the Qularaisy soil.**

| Soil components | Quantities | Unit |
|---|---|---|
| Sand | 155 | |
| Silt | 351 | g kg$^{-}$ |
| Clay | 494 | |
| Textured name | Clay | |
| pH | 7.6 | |
| EC | 0.7 | dS m$^{-}$ |
| Organic Matter | 17.5 | g.kg$^{-}$ |
| CEC | 40.90 | cmolc kg$^{-}$ |
| Calcium (Ca++) | 7.7 | meq L$^{-}$ |

## 3.3.2 Selection of quince mother tree stocks

Cuttings were taken from the quince trees and used in this study. The cuttings were carefully chosen for their health, free from pests and diseases, and uniformity in diameter, making them ideal candidates for the experiment. The selection of the mother plant plays a crucial role in the successful propagation of fruit plants through cuttings. Optimal growing conditions and the quality of the cutting taken from mother plants are essential factors that contribute to a higher success rate in rooting percentage. Plant material of the local variety of quince was taken from 3-year-old quince trees at Kani Panka Nursery Station, located 40 kilometers east of Sulaymaniyah city, with coordinates 35˚22′44″ N and 45˚43′18″ E, at an elevation of 548 meters above sea level (Figure 3.4). The material was obtained from the middle part of one-year-old branches, measuring 25±1 cm in length and 10±2 mm in diameter. After obtaining the cuttings at different times, they were transferred to the experimental site. The physical and chemical analyses of the soil of Kani Panka are shown in (Table 3.4).

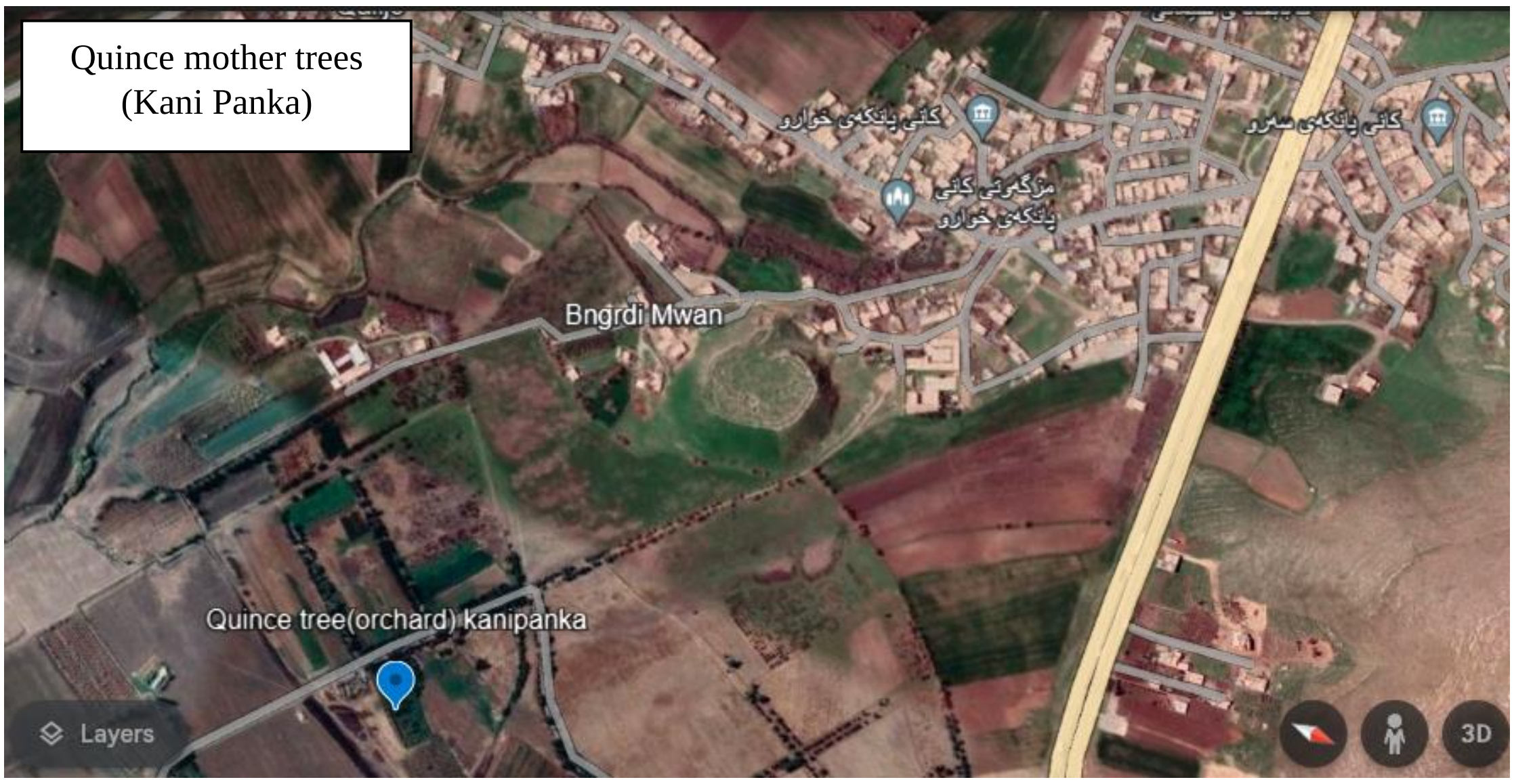

**Figure 3.4 The sitemap of the location for quince mother trees at the experiment site of Kani Panka.**

**Table 3.4 Some physical and chemical properties of the soil at Kani Panka location.**

| Soil components | Quantities | Unit |
|---|---|---|
| Sand | 306.0 | g kg$^{-}$ |
| Silt | 344.0 | |
| Clay | 350.0 | |
| Textured name | Clay Loam | |
| pH | 7.8 | |
| EC | 0.40 | dS m$^{-}$ |
| Organic Matter | 8.7 | g.kg$^{-}$ |
| CEC | 33.0 | cmolc.kg$^{-}$ |
| Calcium (Ca++) | 2.28 | meq.L$^{-}$ |

## 3.4 Selection of Loquat Mother Trees for Scion Collection

The experiment involved the selection of twenty trees situated in various locations of Sulaymaniyah city, including Ibrahim Pasha, Ibrahim Ahmad, Qrga, Wuluba, Iskan, Aqari, Chwarchra, Hawari Shar, Rapareen, and Bakarjo. From each of these locations, two trees were carefully chosen (Table 3.5), and their fruits were harvested at the time of maturity. This process was done with three replications, in which ten fruits were collected from each replicate. After that, various physical characteristics of the fruits were measured. Subsequently, the sensory evaluations were carried out for fruits from each replicate, and the resulting data are presented in (Tables 3.6 and 3.7). Following the collection of data on the physical characteristics and sensory evaluations of the fruits, the next step involved conducting statistical analyses to gain a deeper insight into the experimental outcomes. The statistical analysis utilized the analysis of variance (ANOVA), and any significant differences were further explored using Duncan's multiple comparison tests ($P \leq 0.05$). For this analysis, the XLSTAT software was employed (https://www.xlstat.com). Based on the significant findings about specific physical attributes and sensory evaluations, a particular tree from the Iskan location (T9) at the center of Sulaymaniyah city (Figure 3.5) at 909 m above sea level with 35˚34′ 16″ N and 45˚26′59″ E. was singled out for a taste assessment under various experimental conditions and at different grafting times.

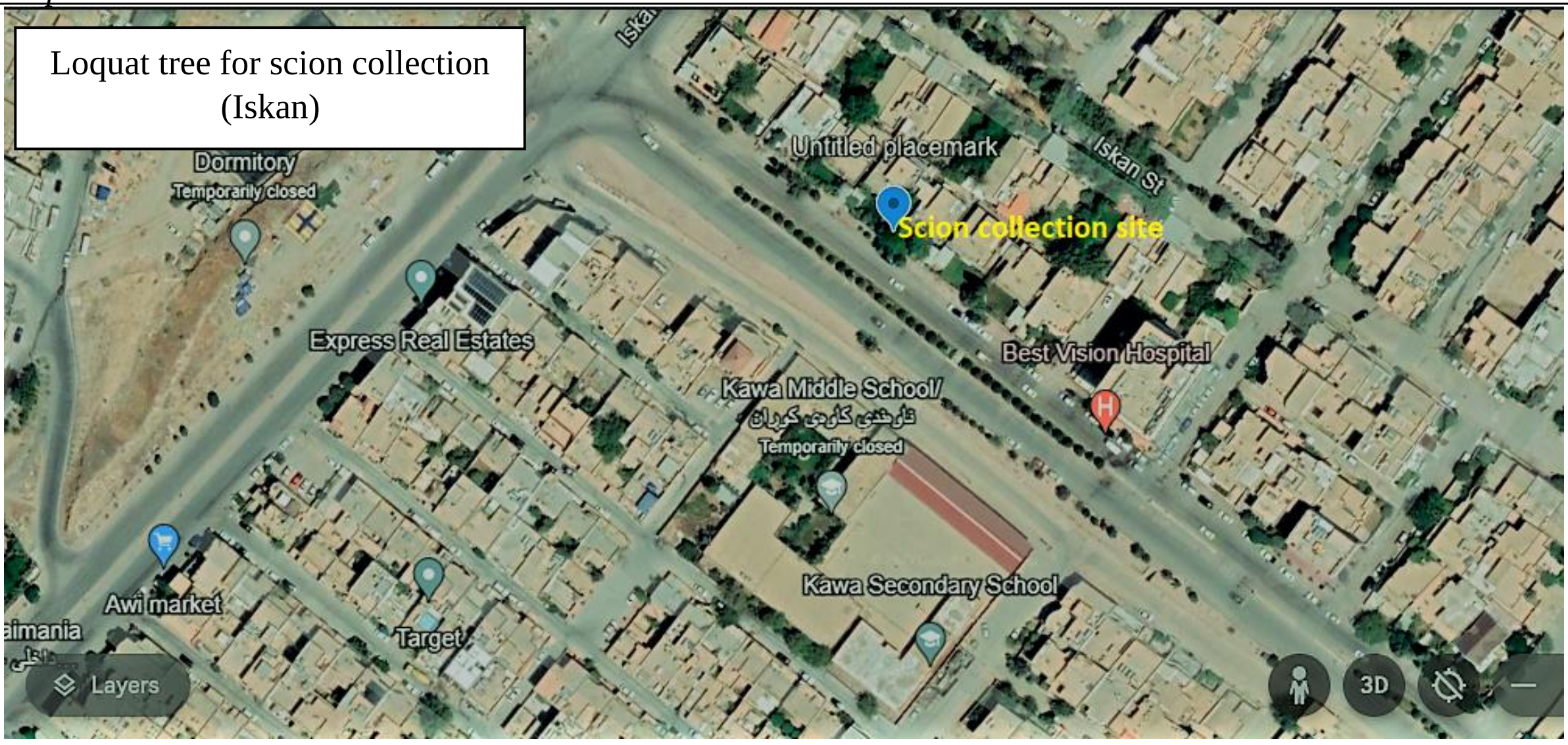

**Figure 3.5 The site map of loquat tree for scion collection.**

**Table 3.5 The number of trees selected for fruit collection used for the experiments.**

| Tree number | Location of loquat trees | Tree number | Location of loquat trees |
|---|---|---|---|
| T1 | Ibrahim Pasha | T11 | Aqari |
| T2 | | T12 | |
| T3 | Ibrahim Ahmad | T13 | Chwarchra |
| T4 | | T14 | |
| T5 | Qrga | T15 | Hawari Shar |
| T6 | | T16 | |
| T7 | Wlluba | T17 | Rapareen |
| T8 | | T18 | |
| T9 | Iskan | T19 | Bakarjo |
| T10 | | T20 | |

## 3.5 Physical Parameters of Loquat Fruits

Mature loquat fruits were collected on May 15, 2021 and subjected to the following measurements:

### 3.5.1 Average fruit weight (g)

It was determined by taking 10 fruits in each replicate and weighing them by electronic balance with an accuracy of 0.0001 g, and then the average weight was calculated (Abdel-Sattar *et al.*, 2021).

### 3.5.2 Average fruit length (mm)

The length of 10 fruits from each replicate was measured using Vernier's caliper, averaged, and expressed in millimeters (mm), with 0.01 mm accuracy (Beyaz, 2020).

### 3.5.3 Average fruit diameter (mm)

The diameter of 10 fruits from each replicate was measured using Vernier's caliper, averaged, and expressed in millimeters (mm), with 0.01 mm accuracy (Beyaz, 2020).

### 3.5.4 Average fruit size ($cm^3$)

The size of loquat fruits was measured using the water displacement method. The fruit was put into water in a large cylinder and carefully recorded to see how much the water level rises. The amount of water increase is equal to the volume of the object (Zare *et al.*, 2022).

### 3.5.5 Fruit stalk length (mm)

The length of 10 ten fruit stalks from each replicate was measured using Vernier's caliper (Model: DMV-SL05, WORKZONE, Germany), averaged and expressed in millimeters (mm), with 0.01 mm accuracy (Ahmad, 2023).

### 3.5.6 Fruit stalk diameter (mm)

The diameter of 10 fruit stalks from each replicate was measured using Vernier electronic caliper (Model: DMV-SL05, WORKZONE, Germany) averaged and expressed in millimeters (mm), with 0.01 mm accuracy (Mahmood, 2020).

### 3.5.7 Average seed length (mm)

The length of 10 loquat seeds from each replicate was measured using Vernier's caliper, averaged and expressed in millimeters (mm), with 0.01 mm accuracy (Şener and Duran, 2020).

### 3.5.8 Average seed diameter (mm)

The diameter of 10 seeds from each replicate was measured using Vernier's caliper averaged and expressed in millimeters (mm), with 0.01 mm accuracy (Şener and Duran, 2020).

### 3.5.9 Number of seeds per fruit

The number of seeds per fruit was calculated by counting the total seeds contained by 10 fruits and then taking the average.

### 3.5.10 Seed weight (g)

It was determined by taking 10 loquat seeds in each replicate from each treatment and weighing them by electronic balance with an accuracy of 0.0001 g, and then the average weight was calculated.

### 3.5.11 Total soluble solids (TSS%)

The total soluble solids were measured by Portable Hand Refractometer, Erma Japan. To measure TSS, a drop of the fruit extract was placed on the prism of the digital refractometer, and the total soluble solids were read in °Brix (Wu *et al*., 2021).

## 3.6 The Data Analyses

All the obtained data of the physical traits and sensory evaluations were organized and statistically analyzed with the analysis of variance (ANOVA) in a randomized complete block design (RCBD) using XLSTAT system (https://www.xlstat.com). and the means were compared according to Duncan's multiple range test ($P \leq 0.05$) **(**Al-Rawi and Khalafullah, 2000).

**Table 3.6 The physical characteristics of loquat fruits grown in Sulaimani city and used in the experiments.**

| Mother tree | Fruit weight (g) | Fruit length (mm) | Fruit diameter (mm) | Fruit size ($cm^3$) | Fruit stalk length (mm) | Fruit stalk diameter (mm) | Seed length (mm) | Seed diameter (mm) | Number of seed/fruit | Seed weight (g) | TSS |
|---|---|---|---|---|---|---|---|---|---|---|---|
| T1 | 19.83 c | 34.80 b-e | 30.12 b-e | 25.50 ab | 25.41 bc | 4.54 b-e | 19.83 a | 12.49 b-d | 2.44 c-f | 1.73 bc | 15.03 cd |
| T2 | 17.75 d | 35.06 b-e | 31.92 a-c | 19.23b-f | 25.33 bc | 4.88 a-e | 16.28 g-i | 12.00 b-d | 2.44 c-f | 1.32 ef | 10.13 hi |
| T3 | 13.23 f | 37.43 b | 33.05 ab | 17.64 d-g | 30.32a-c | 4.95 a-d | 19.85 a | 13.69 ab | 2.44 d-f | 1.45 de | 14.47 c-f |
| T4 | 11.86 g | 29.76 g-i | 26.84d-g | 21.82 b-e | 39.38a | 4.43 b-e | 13.99 j | 12.59 b-d | 1.77 e-h | 1.08 f | 11.70 g-i |
| T5 | 20.89 c | 35.44 b-e | 31.20a-d | 22.56 b-e | 18.94 bc | 4.79 a-e | 19.31a-c | 11.37 cd | 3.66 ab | 1.48 de | 10.50 g-i |
| T6 | 17.36 d | 30.86 f-i | 31.38a-c | 22.11 b-e | 24.69 bc | 4.61 b-e | 17.71c-g | 12.78 b-d | 2.77 b-e | 1.47 de | 12.20 f-i |
| T7 | 20.69 c | 36.33 b-d | 33.32 ab | 24.84 a-c | 22.69 bc | 5.05 a-c | 16.53 e-i | 11.77 b-d | 3.44 bc | 1.43 de | 10.99 g-i |
| T8 | 20.51 c | 33.99 b-f | 32.96 ab | 22.95 b-e | 18.63 bc | 4.88 a-e | 17.16d-h | 12.13 b-d | 2.77 b-e | 1.96 b | 12.37 e-h |
| T9 | 26.44 a | 44.57 a | 33.07 ab | 29.66 a | 19.10 bc | 5.58 a | 19.73 ab | 12.69 b-d | 2.22 e-g | 1.91 bc | 11.37 g-i |
| T10 | 12.88 fg | 37.56 b | 25.50 fg | 17.53 d-g | 19.86 bc | 5.11 a-c | 17.74 c-g | 11.13 c-e | 1.66 f-h | 1.22 ef | 12.76 d-g |
| T11 | 17.44 d | 33.13 d-g | 30.87a-d | 22.23 b-e | 24.19 bc | 5.19 ab | 15.54 h-j | 10.72 de | 1.77 e-h | 1.22 ef | 14.60 c-f |
| T12 | 15.38 e | 28.98 hi | 30.87a-d | 21.36 b-e | 16.07 c | 5.10 a-c | 15.00 ij | 9.13 e | 3.66 ab | 0.83 g | 11.967g-i |
| T13 | 13.12 f | 30.10 g-i | 27.70 c-f | 16.14 e-g | 31.34 ab | 4.53 b-e | 16.30 g-i | 10.78 de | 1.77 e-h | 1.84 bc | 9.77 i |
| T14 | 12.88 fg | 36.86 bc | 26.52e-g | 18.00 c-f | 18.66 bc | 4.06 e | 16.90d-h | 12.13 b-d | 1.66 f-h | 1.48 de | 12.96 d-g |
| T15 | 20.21 c | 29.28 hi | 34.75 a | 21.53 b-e | 24.14 bc | 4.08 e | 16.39 f-i | 11.08 c-e | 4.44 a | 1.24 ef | 14.77 c-e |
| T16 | 9.01 h | 28.74 hi | 23.43 g | 11.28 g | 22.42 bc | 4.74 b-e | 15.65 hi | 13.01 bc | 1.11 h | 1.74 bc | 12.93 d-g |
| T17 | 15.48 e | 28.54 i | 27.79 c-f | 23.66 a-d | 22.31 bc | 4.16 de | 18.09b-f | 13.65 ab | 1.33 gh | 2.19 a | 19.06 a |
| T18 | 13.00 fg | 29.81 g-i | 27.76 c-f | 13.47 fg | 18.87 bc | 4.33 c-e | 15.45 h-j | 14.93 a | 2.00 e-h | 1.38 e | 18.10 ab |
| T19 | 20.98 c | 32.18 e-h | 31.12 a-d | 22.34 b-e | 26.65 a-c | 4.79 a-e | 18.41a-d | 11.01 c-e | 3.66 ab | 1.44 de | 17.91 ab |
| T20 | 22.71 b | 33.68 c-f | 33.39 ab | 24.28 a-d | 24.57 bc | 4.85 a-e | 18.21a-e | 10.93 c-e | 3.33 b-d | 1.68 cd | 16.01 bc |

**Table 3.7 The sensory characteristics of loquat fruits grown in Sulaimani city and used in the experiments.**

| Mother tree | Appearance | Flavor | Soft and Moist | Shape | Exocarp | Colour | Size | Texture | Taste |
|---|---|---|---|---|---|---|---|---|---|
| T1 | 3.89 a-e | 3.55 a | 3.00 fg | 2.22 de | 1.89 bc | 3.00 a-d | 2.00 de | 3.33 ab | 4.44 a-c |
| T2 | 2.89 d-f | 3.00 bc | 3.78 b-e | 3.44 ab | 1.78 c | 3.00 a-d | 2.11 de | 2.44 c | 3.11 fg |
| T3 | 4.11 a-d | 3.11 ab | 4.00 b-d | 3.00 a-c | 2.33 a-c | 3.34 ab | 3.22 bc | 3.00 bc | 4.00 bc |
| T4 | 3.00 c-f | 2.63 cd | 4.00 b-d | 2.56 c-e | 2.11 bc | 3.00 a-d | 2.11 de | 2.44 c | 2.67 f-h |
| T5 | 4.33 ab | 3.11 a-c | 4.33 ab | 2.33 c-e | 1.78 c | 3.44 ab | 3.34 ab | 4.00 a | 3.89 c-e |
| T6 | 3.11 b-f | 2.22 d-f | 3.78 b-e | 2.89 b-d | 1.78 c | 3.00 a-d | 2.00 de | 2.67 bc | 3.34 d-f |
| T7 | 3.00 c-f | 2.22 d-f | 3.89 b-d | 2.33 c-e | 2.00 bc | 3.00 a-d | 2.00 de | 2.78 bc | 3.34 ef |
| T8 | 4.22 a-c | 2.22 d-f | 4.00 b-d | 3.00 a-c | 2.00 bc | 3.11 a-d | 2.00 de | 3.00 bc | 4.00 b-d |
| T9 | 4.89 a | 3.22 ab | 4.78 a | 3.67 a | 2.56 ab | 3.67 a | 4.00 a | 3.89 a | 4.78 a |
| T10 | 3.34 b-f | 2.44 de | 3.55 c-f | 2.78 b-d | 1.89 bc | 2.74 b-d | 2.67 b-d | 2.67 bc | 2.22 h |
| T11 | 3.44 b-f | 2.00 ef | 3.11 e-g | 3.33 ab | 1.78 c | 3.00 a-d | 2.67 b-d | 2.44 c | 2.56 gh |
| T12 | 2.67 ef | 1.89 f | 3.33 d-g | 2.56 c-e | 1.78 c | 2.67 b-d | 2.00 de | 2.44 c | 3.11 fg |
| T13 | 3.56 b-f | 2.22 d-f | 3.67 b-e | 2.89 b-d | 2.00 bc | 3.11 a-d | 2.11 de | 2.44 c | 3.89 c-e |
| T14 | 3.44 b-f | 2.00 ef | 3.44 d-g | 3.00 a-c | 2.44 a-c | 3.00 a-d | 2.11 de | 2.44 c | 3.00 fg |
| T15 | 4.00 a-d | 2.11 d-f | 4.00 b-d | 3.00 a-c | 1.78 c | 3.11 a-d | 2.55 c-e | 3.00 bc | 4.44 a-c |
| T16 | 2.33 f | 2.22 d-f | 2.89 g | 2.00 e | 2.11 bc | 2.33 d | 1.78 e | 3.00 bc | 2.78 f-h |
| T17 | 3.45 b-f | 2.22 d-f | 3.78 b-e | 2.55 c-e | 1.89 bc | 2.45 cd | 1.89 de | 3.33 ab | 4.22 a-c |
| T18 | 3.22 b-f | 1.89 f | 3.67 b-e | 2.89 b-d | 2.00 bc | 3.22 a-c | 2.00 de | 2.89 bc | 4.67 ab |
| T19 | 3.22 b-f | 2.22 d-f | 4.22 a-c | 1.89 e | 2.89 a | 3.00 a-d | 2.67 b-d | 2.67 bc | 4.33 a-c |
| T20 | 4.22 a-c | 2.22 d-f | 4.00 b-d | 3.33 ab | 1.89 bc | 2.33 d | 3.33 ab | 2.78 bc | 4.22 a-c |

### 3.7 Collection of Loquat Scion Shoot Materials

The scion shoots of the loquat tree (*Eriobotrya japonica* Lindl.) were collected from healthy, matured, and disease-free trees. These source trees (T9) of the local variety were aged 10-12 years, known for their heavy fruit load, large fruits, and low number of seeds. The scion shoots were carefully chosen from the terminal shoots of the current season's growths. They were about 7-8 months old at the time of collection. Immediately after separating the bud sticks from the trees, they were wrapped in a moist clean cloth to maintain their freshness and to prevent drying and transported to the grafting site. The scion shoots were then divided into several scions, with each scion containing two active buds. This division of scion shoots allowed for multiple grafting attempts and increased the chances of successful grafting. On average, the scions were 6-8 cm long and had a diameter of 8±2 mm.

### 3.8 Grafting Operation

Cleft grafting was used in this work, in which two smooth slanting cuts, about 3–4 cm long, were made at the proximal end of the scion on both sides, opposite to each other, in such a way that the end portion becomes very thin. This was done by using a sharp knife. The smooth long sloping wedge cuts at the base of the scion cut an appearance of a sharp chisel. The rootstock was first headed back by making a horizontal cut, and then a vertical split cut or cleft was made by using a thin and sharp-bladed grafting knife at the center of the horizontal cut surface of the stock, having a depth of approximately 3–4 cm. Then, the scion was inserted into the cleft of the rootstock through a slight opening of the splits. As a result, both components were brought into close contact, particularly the cambium layers face to face, and were then tied firmly with a polythene strip. After wrapping the graft union, the scion, along with the union portion, was covered with a polythene cap to protect it from losing moisture through transpiration. In this method, both the stock and the scion were of the same thickness to match each other.

### 3.9 Experiments

This study included five experiments, which were conducted as described below:

#### 3.9.1 Experiment 1: Rooting of hardwood cuttings of quince (*Cydonia oblonga* Mill.) as influenced by IBA and rooting substrate

This study was conducted to investigate the effect of rooting substrate and different concentrations of IBA (indole-3- butyric acid) on the rooting of local quince hardwood cuttings in the lathhouse at the College of Agricultural Engineering Sciences, University of Sulaimani, Kurdistan Region, Iraq, in 2021.

### 3.9.1.1 Collecting the quince hardwood cuttings

Plant material of the local variety of quince was taken from 3-year-old quince trees at Kani Panka Nursery Station on February 25, 2021, from the middle part of one-year-old branches with 25±1 cm length and 10±2 mm diameter. The plant wood materials were then brought to the laboratory, where treated with 3 $g.L^{-}$ Captan (50%) fungicide and put in black plastic bags, then sealed and stored at 4±1 °C (Noori and Muhammad, 2020) for 20 days.

### 3.9.1.2 Preparation of rooting substrate

The rooting substrates were prepared from three different rooting media: River sand, river sand + peatmoss (1:1 v/v), and perlite + peatmoss (1:1 v/v), each with unique properties for plant rooting experiments.

### 3.9.1.3 Preparation, IBA treatment, and planting of the hardwood cuttings

The plant wood materials were removed from the cold storage on March 15, 2021, and brought to the laboratory. Then, two-incision cuts of 0.5 cm depth were made on opposite sides of the base of the cuttings (Tworkoski and Takeda, 2007). After that, the cuttings were randomly divided into four lots, each lot included 72 cuttings, then every lot was separately and quickly dipped into the control treatment (ethanol with distilled water), and different IBA concentrations for 10 seconds (Sebastiani and Tognetti, 2004). There were 12 treatments with 3 replications. The experiment included a total number of 288 cuttings. Each treatment involved 24 cuttings which were divided into three replications, 8 cuttings for each replicate (Table 3.8). The cuttings were planted in the three rooting substrates in a black plastic pot with a 43×36 cm. The pots were placed in a lathhouse and arranged in a randomized complete block design (RCBD). The weekly averages of temperature and relative humidity were recorded by a data logger (Model: Perfect prime TH0160) in the lathhouse during the entire period of the experiment (Table 3.9). After 4 months, on July 15, 2021, the experiment was terminated and the cuttings were checked to measure the effect of the treatments.

**Table 3.8 The treatment combinations used for each cutting were planted in rooting substrate in the experiment 1.**

| Treatment number | Treatment combinations | |
|---|---|---|
| | IBA concentration (mg.L-) | Rooting media |
| 1 | 0 | Riversand |
| 2 | | Riversand+ Peatmoss |
| 3 | | Peatmoss + Perlite |
| 4 | 1000 | Riversand |
| 5 | | Riversand+ Peatmoss |
| 6 | | Peatmoss + Perlite |
| 7 | 2000 | Riversand |
| 8 | | Riversand+ Peatmoss |
| 9 | | Peatmoss + Perlite |
| 10 | 3000 | Riversand |
| 11 | | Riversand+ Peatmoss |
| 12 | | Peatmoss + Perlite |

**Table 3.9 The average weekly temperature and the relative humidity inside the lathhouse during the study period in 2021.**

| Date | Weeks | Temperature (°C) | Relative Humidity (%) |
|---|---|---|---|
| April 13-19 | Week 1 | 19.11 | 49.42 |
| April 20-26 | Week 2 | 24.31 | 45.23 |
| April 27-May 3 | Week 3 | 24.16 | 46.29 |
| May 4-10 | Week 4 | 25.39 | 41.36 |
| May 11-17 | Week 5 | 28.10 | 33.37 |
| May 18-24 | Week 6 | 28.08 | 32.72 |
| May 25-31 | Week 7 | 28.14 | 33.76 |
| June 1-7 | Week 8 | 29.13 | 25.96 |
| June 8-14 | Week 9 | 29.08 | 20.92 |
| June 15-21 | Week 10 | 32.59 | 17.68 |
| June 22-28 | Week 11 | 33.34 | 17.70 |
| June29-July5 | Week 12 | 33.37 | 19.09 |
| July 6-12 | Week 13 | 33.51 | 19.26 |
| July 13-15 | Week 14 | 33.95 | 24.57 |

#### 3.9.1.4 The studied parameters

The experiment was terminated on July 15, 2021, by taking the following parameters:

1. Rooting percentage: After separating the shoots and taking all data related to the shoot system, later, the pots were turned over to take the data related to the root system (Kang *et al*., 2005 and Debner, 2016).
2. Root length (cm): The root length of each cutting within a replicate was measured and averaged (Debner, 2016).
3. Root number: Total roots that remain on the cutting were calculated.

4. Root fresh weight (g): The roots were weighed by using a digital balance directly after pulling out the roots from the soil, washing off from any loose soil, and removing any free surface blot moisture.
5. Root dry weight (g): The roots were dried at 60 °C for 72 hours in an oven.
6. Shoot length (cm): The shoot length of each cutting within a replicate was measured with a measuring tape and averaged from the number of sprouted cuttings (Debner, 2016).
7. Shoot diameter (cm): The diameter of the shoot for each cutting was measured at 3 cm from the shoot base using a Vernier calliper, then the recordings were averaged (Şener and Duran, 2020).
8. Shoot fresh weight (g): Weighed by digital balance directly after washing off from any loose dust and air-drying.
9. Shoot dry weight (g): The shoots were dried at 60 °C for 72 hours in an oven.
10. The number of leaves per cutling: The total number of leaves per cutling was counted and averaged from the number of sprouted cuttings (Villa *et al.*, 2003).
11. Leaf area ($cm^2$): Measured by using a software program application (Digimizer image analysis) **(https://www.digimizer.com/)**.
12. Total chlorophyll content of leaves (SPAD units): Determined using a chlorophyll meter (Model OPTI-SCIENCES).

### 3.9.2 Experiment 2: The Impact of grafting dates, the cutting types, and the IBA concentrations on grafting success of bench grafted loquats

This experiment aimed to assess the success of grafting loquat cuttings onto two different loquat stocks (loquat and quince). It involved grafting on various dates (February 10, February 20, March 2, and March 12) and using different concentrations of IBA (0, 1000, 2000, 3000, and 4000 $mg.L^{-}$) (Table 3.10). The study took place in the lathhouse at the College of Agricultural Engineering Sciences, University of Sulaimani, Kurdistan Region, Iraq, throughout the period from February to June 2022.

#### 3.9.2.1 Collecting the hardwood cuttings

The loquat hardwood cuttings were prepared from 10-12-year-old trees grown from seeds on a private orchard cultivated in Qularaisy. In contrast, local quince cuttings were collected from 3-year-old trees grown at Kani Panka Nursery Station. Both loquat and quince hardwood cuttings were taken at different dates (February 10, February 20, March 2, and March 12) in 2022. All the cuttings were obtained from the middle part of one-year-old branches, measuring 25±1 cm in length and 10±2 mm in diameter. After obtaining the cuttings at each given time, they were brought to the laboratory for bench grafting. Subsequently, the cuttings (grafts) were treated with Captan (50%) at a ratio of 3 g.L$^-$ of water as a fungicide.

**Table 3.10 The treatment combinations used for all grafted cuttings in the experiment 2.**

| Treatment combinations | | | | | | | |
|---|---|---|---|---|---|---|---|
| Number | Stock cutting types | Date of grafting | IBA Concentrations (mg.L$^-$) | Number | Stock cutting types | Date of grafting | IBA Concentrations (mg.L$^-$) |
| 1 | Loquat | February 10 | 0 | 21 | Quince | February 10 | 0 |
| 2 | | | 1000 | 22 | | | 1000 |
| 3 | | | 2000 | 23 | | | 2000 |
| 4 | | | 3000 | 24 | | | 3000 |
| 5 | | | 4000 | 25 | | | 4000 |
| 6 | | February 20 | 0 | 26 | | February 20 | 0 |
| 7 | | | 1000 | 27 | | | 1000 |
| 8 | | | 2000 | 28 | | | 2000 |
| 9 | | | 3000 | 29 | | | 3000 |
| 10 | | | 4000 | 30 | | | 4000 |
| 11 | | March 2 | 0 | 31 | | March 2 | 0 |
| 12 | | | 1000 | 32 | | | 1000 |
| 13 | | | 2000 | 33 | | | 2000 |
| 14 | | | 3000 | 34 | | | 3000 |
| 15 | | | 4000 | 35 | | | 4000 |
| 16 | | March 12 | 0 | 36 | | March 12 | 0 |
| 17 | | | 1000 | 37 | | | 1000 |
| 18 | | | 2000 | 38 | | | 2000 |
| 19 | | | 3000 | 39 | | | 3000 |
| 20 | | | 4000 | 40 | | | 4000 |

### 3.9.2.2 Treating loquat scions with BA (benzyl adenine) before grafting

The scions were treated by dipping them in a solution of BA with a concentration of 10 mg.L$^-$ for 10 seconds inside a small glass beaker at a temperature of approximately 17-20 C°, with low light intensity present. The scions were then used immediately after they were removed from the solutions.

### 3.9.2.3 Treating grafted cuttings with IBA (Indole-3-butyric acid) before planting

The loquat grafted on two types of cuttings for each time included 120 grafted cuttings for quince and 120 grafted cuttings for loquat were randomly divided into five lots, each with 24 grafted cuttings for both loquat and quince with 8 grafted cuttings for each replicate. Hence, the experiment included a total number of 960 grafted cuttings. Every lot of the grafted cutting were dipped in control (ethanol and distilled water) and different IBA concentrations for 10 seconds (Sebastiani and Tognetti, 2004). After making the bench grafts, the grafted cuttings were taken in the substrate of peat moss and then transferred to the greenhouse and kept at a temperature of 18±2 ºC till callus formation, and finally the grafted cuttings were planted on March 20 in the rooting substrates prepared from mixing two rooting media; riversand + peatmoss (2:1 v/v), in black plastic pots with height and diameter of 32×26.5 cm. The pots were placed in a lathhouse and arranged in a randomized complete block design (RCBD). The weekly averages of temperature and relative humidity were recorded during the whole period of the experiment (Figure 3.6).

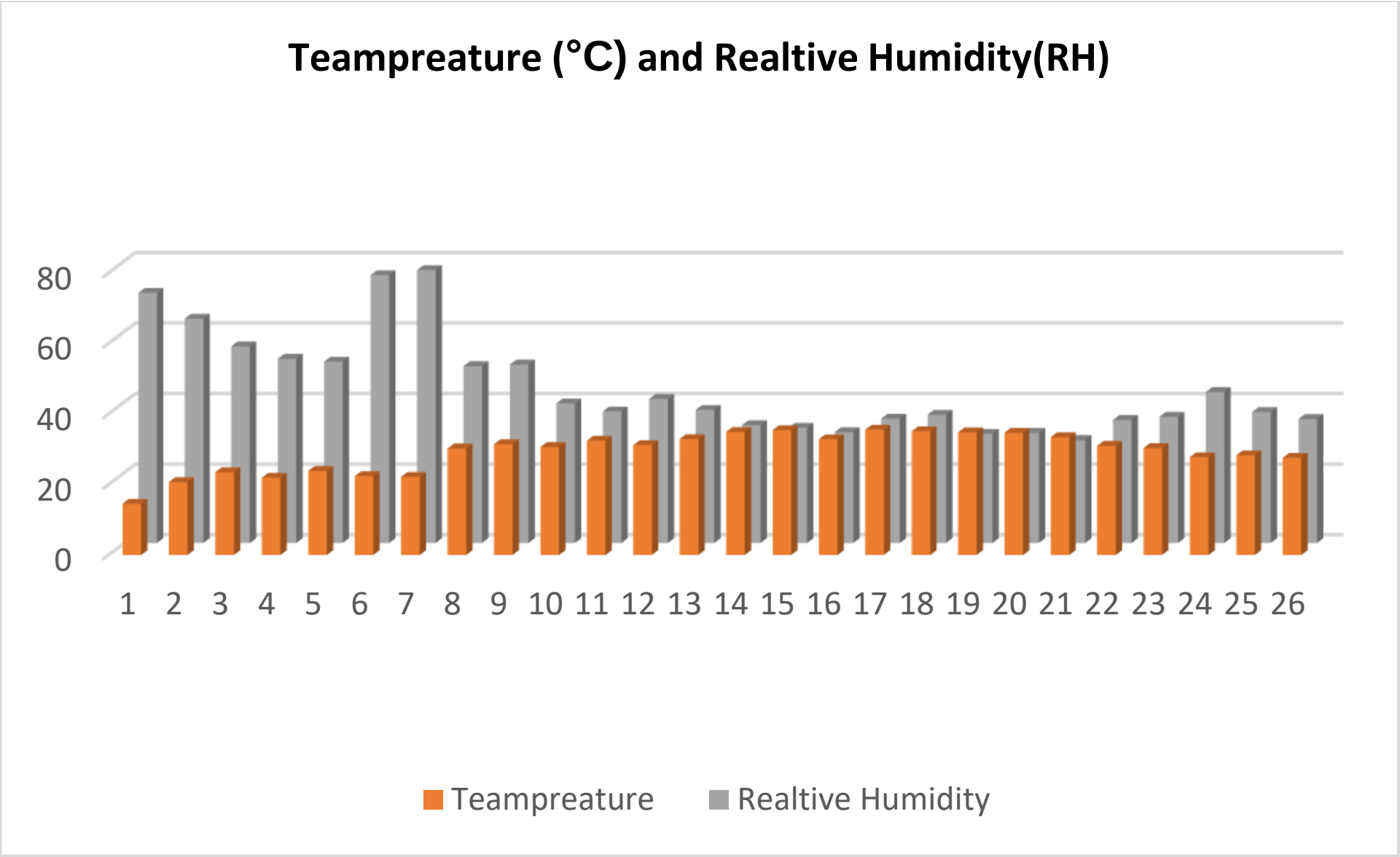

**Figure 3.6 The weekly temperature (°C) and the relative humidity (%) inside the lathhouse starting from March 20, 2022, to October 10, 2022.**

### 3.9.2.4 The studied parameters

The experiment was terminated in 2022, by taking the following parameters:

1. Graft bud sprout percentage on March 30, April 30, and May 30, 2022. The percentage of the successful graft bud sprout was calculated by dividing the sprouted grafts by the total number of grafts per replicate.
2. Rooting percentage. The percentage of successful rooting was calculated by dividing the rooted cuttings by the total number of cuttings per replicate.

### 3.9.3 Experiment 3: The performance of grafting loquats combined onto loquat and quince rootstocks on different dates

This experiment was carried out from February 20 to July 1, 2023, at the lathhouse in the College of Agricultural Engineering Sciences, University of Sulaimani, Kurdistan Region, Iraq. The experiment involved grafting loquats onto two types of rootstocks, namely loquat (*Eriobotrya japonica* Lindl.) and quince (*Cydonia oblonga* Mill), which were performed during the dormant season at three different times (February 20, March 10, and March 30). The experiment consisted of 6 treatments each with 5 grafts and 3 replications, resulting in a total of 90 grafts (Table 3.11).

**Table 3.11 The treatment combinations used for grafting in the experiment 3.**

| Treatment number | Treatment combinations | |
|---|---|---|
| | Type of rootstock | Date of grafting |
| 1 | Loquat | February 20 |
| 2 | | March 10 |
| 3 | | March 30 |
| 4 | Quince | February 20 |
| 5 | | March 10 |
| 6 | | March 30 |

#### 3.9.3.1 Rootstock production

Two types of rootstocks; loquat and quince were used in this experiment. The loquat rootstocks were produced from seeds, the seeds were sown in polyethylene bags measuring 10×30 cm, filled with a mixture of river sand and peat moss (1:1 v/v). The loquat rootstocks were allowed to grow for 1.5 years until they reached a height of 25 cm and 10±2 mm thickness. On the other hand, quince rootstocks were obtained through cuttings from 3-year-old trees and were planted in polyethylene bags measuring 10×30 cm, using the same mixture of river sand and peat moss. The polyethylene bags were placed in a lathhouse and arranged in a randomized complete block design (RCBD), with the same regular maintenance as the loquat rootstocks. The average diameter of both loquat and quince rootstocks was 10±2 mm. The weekly averages of temperature and relative humidity were recorded by a data logger (Model: Perfect prime TH0160) in the lathhouse during the entire period of the experiment (Figure 3.7).

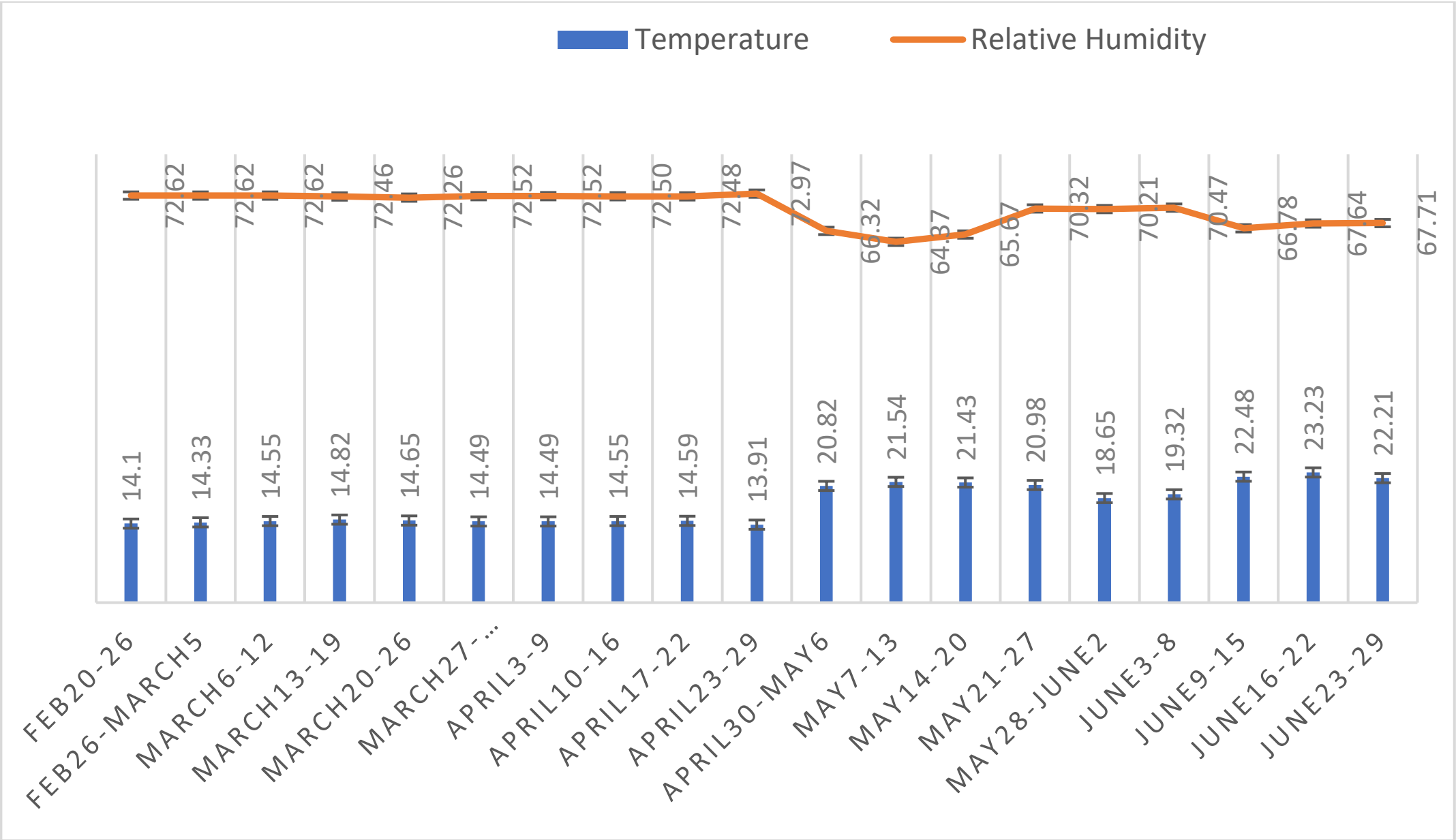

**Figure 3.7 The observed data of mean temperature (°C) and relative humidity (%) in the lathhouse during February-June, 2023.**

### 3.9.3.2 The studied parameters

The experiment was terminated on July 1, 2023, by taking the following parameters:

1. Grafting success (%): The percent graft success was calculated by using the formula given below:

$$\text{Grafting success (\%)} = \frac{\text{No. of successful grafts}}{\text{Total no. of rootstocks grafted}} \times 100 \qquad (3.1)$$

2. Graft shoot length (cm): Graft shoot length was measured with the measuring tape.
3. Graft shoot diameter (mm): This was measured 2 cm above the graft union using the electronic Vernier caliper.
4. Shoot fresh weight (g): Weighted by digital balance directly after washing off from any loose dust and air-drying.
5. Shoot dry weight (g): The shoots were dried at 60 °C for 72 hours in an oven.
6. Number of leaves per budling: The total number of leaves per budling was counted and averaged from the number of graft bud sprouted.
7. Leaf area ($cm^2$): Measured by using a software program application (Digimizer image analysis) **(https://www.digimizer.com/)**.
8. Total chlorophyll content of leaves (SPAD units): Determined using a chlorophyll meter (Model SPAD 502 PLUS).

### 3.9.4 Experiment 4: The effect of grafting dates and stock types on grafting success of loquat and quince treestocks.

This experiment was conducted during February 20 to July 1, 2023, by grafting loquats onto two types of loquat rootstocks; loquat (*Eriobotrya japonica* Lindl.) and quince (*Cydonia oblonga* Mill). The grafting operations were done during the dormant season at three different dates (February 20, March 10, and March 30) at two locations within Sulaimani governorate. The first location was at the College of Agricultural Engineering Sciences, situated in Bakrajo. The meteorological data for this site during the study period are shown in (Table 3.1). On the other hand, at the second location; Kani Waysa village, Sitak, the grafting was performed on local quince trees aged 4-5 years in a private orchard grown, positioned 25 km northeast of Sulaymaniyah city, and at an elevation of 1057 meters above sea level, with coordinates of 35°40′02″ N and 45°34′30″ E. The meteorological data for this location during the study period are presented in (Table 3.2). The experiment comprised of 6 treatments, each involving 5 grafts with 3 replicated, resulting in a total of 90 grafts (Table 3.12).

**Table 3.12 The treatment combinations used for grafting in the experiment 4.**

| Treatment number | Treatment combinations | |
|---|---|---|
| | Type of tree stock | Date of grafting |
| 1 | Loquat | February 20 |
| 2 | | March 10 |
| 3 | | March 30 |
| 4 | Quince | February 20 |
| 5 | | March 10 |
| 6 | | March 30 |

#### 3.9.4.1 Tree stock selection

This study was carried out on 4-5 years of loquat and quince tree stocks, the quince trees were grown on a private orchard in Kani Waysa village. The loquat trees were cultivated from seeds at the College of Agricultural Engineering Sciences. For the study, a total of 9 quince trees and 9 loquat trees with the same trunk and branch diameters ranging from 2.0 to 2.5 centimeters were chosen as tree stocks for grafting. The treatments consisted of two factors; the type of tree stock (loquat and quince) and grafting dates (February 20, March 10, and March 30), using the cleft grafting method on the percentage of survival and growth parameters of loquat scions. The experiment was laid out in a Randomized Complete Block Design (RCBD) with three replications. For each treatment combination, grafting operations were performed on 15 grafts. Thus, a total number of 90 grafts were made for loquat and quince treestocks.

#### 3.9.4.2 The studied parameters

The experiment was terminated on July 1, 2023, by taking the following parameters:

1. Grafting success (%): The percent graft success was calculated by using the formula given below.

$$\text{Grafting success (\%)} = \frac{\text{No. of successful grafts}}{\text{Total no. of tree stocks grafted}} \times 100 \qquad (3.2)$$

2. Graft shoot length (cm): Graft shoot length was measured with the measuring tape.
3. Graft shoot diameter (mm): Was measured 1 cm above the graft union by using the electronic Vernier caliper.
4. Shoot fresh weight (g): Weighted by digital balance directly after washing off from any loose dust and air-drying.
5. Shoot dry weight (g): The shoots were dried at 60 °C for 72 hours in an oven.
6. Number of leaves per budling: The total number of leaves per budling was counted and averaged from the number of graft bud sprouted.
7. Leaf area ($cm^2$): Measured by using a software program application (Digimizer image analysis) (https://www.digimizer.com/).
8. Total chlorophyll content of leaves (SPAD units): Determined using a chlorophyll meter (Model SPAD 502 PLUS).

### 3.9.5 Experiment 5: The Impact of cutting types and grafting dates on graft bud sprout and rooting percentage in loquat bench grafting

This experiment involved loquat bench grafting by using two different types of loquat cuttings (loquat and quince) on various dates, including February 20, March 10, and March 30. and Concentrations of IBA (Indole-3-Butyric Acid) at 4000 $mg.L^-$ were utilized, as detailed in (Table 3.13). The study took place in the lathhouse at the College of Agricultural Engineering Sciences, University of Sulaimani, located 15 km southwest of Sulaymaniyah city in the Kurdistan Region of Iraq. The coordinates for this site are 35°32′15″ N latitude and 45°21′52″ E longitude, with an elevation of 730 meters above sea level. Grafting of the loquat was carried out at this location, and the meteorological data for the lathhouse during the study period, spanning from February to June 2023, are available in Table (3.14).

**Table 3.13 The treatment combinations used for grafting in the experiment 5.**

| Treatment number | Treatment combinations | |
|---|---|---|
| | Type of cuttings | Date of grafting |
| 1 | Loquat | February 20 |
| 2 | | March 10 |
| 3 | | March 30 |
| 4 | Quince | February 20 |
| 5 | | March 10 |
| 6 | | March 30 |

**Table 3.14 The Weekly temperature (°C) and relative humidity (%) inside the lathhouse recorded from March 20 to June 29, 2023.**

| Year 2023 | | |
|---|---|---|
| Date | Temperature (°C) | Relative Humidity (%) |
| February 20-26 | 14.1 | 72.62 |
| February 26-March 5 | 14.33 | 72.62 |
| March 6-12 | 14.55 | 72.62 |
| March 13-19 | 14.82 | 72.46 |
| March 20-26 | 14.65 | 72.26 |
| March 27- April.2 | 14.49 | 72.52 |
| April 3-9 | 14.49 | 72.52 |
| April 10-16 | 14.55 | 72.50 |
| April 17-22 | 14.59 | 72.48 |
| April 23-29 | 13.91 | 72.97 |
| April 30- May.6 | 20.82 | 66.32 |
| May 7-13 | 21.54 | 64.37 |
| May 14-20 | 21.43 | 65.67 |
| May 21-27 | 20.98 | 70.32 |
| May 28- June 2 | 18.65 | 70.21 |
| June 3-8 | 19.32 | 70.47 |
| June 9-15 | 22.48 | 66.78 |
| June 16-22 | 23.23 | 67.64 |
| June 23-29 | 22.21 | 67.71 |

### 3.9.5.1 The preparation of plant growth regulator solutions

The preparation of plant growth regulator solutions, specifically IBA (Indole-3-butyric acid), used for grafted cuttings was carried out as follows: To achieve a concentration of 4000 mg.L$^{-}$ of IBA, 1 g of IBA was fully dissolved in 125 mL of ethanol (96%). This resulting solution was then transferred into a 250 ml volumetric flask and filled to the mark with distilled water (Evert and Smittle, 1990).

### 3.9.5.2 Treating with IBA solutions and planting grafted cuttings

The experiment included six treatments, each with three replications, resulting in a total of 144 grafted cuttings for loquat and quince. These cuttings were randomly divided into two groups, each with 72 grafted cuttings, which were immersed in a solution containing 4000 mg.L$^{-}$ of IBA for 10 seconds, following the method described by Sebastiani and Tognetti (2004). Each treatment involved 24 grafted cuttings, which were further divided into three replicates, with 8 grafted cuttings in each replicate. After bench grafting, the grafted cuttings were planted in rooting substrates prepared by mixing two rooting media; river sand + peatmoss (2:1 v/v), in a black plastic pot with a 32×26.5 cm. The pots were then placed in a lathhouse and arranged in a randomized complete block design (RCBD). Throughout the experiment, weekly records were maintained for temperature and relative humidity (Table 3.14).

### 3.9.5.3 The studied parameters

The experiment was terminated in 2023, by taking the following parameters:

1. Graft bud sprout percentage: The percentage of the successful graft bud sprout was calculated by dividing the sprouted grafts by the total number of grafts per replicate.
2. Leaf number: The total number of leaves per bud sprout counted and averaged.
3. Rooting percentage: The percentage of successful rooting was calculated by dividing the rooted cuttings by the total number of cuttings per replicate.
4. Carbohydrate contents (%): The percentage of total carbohydrates in cuttings was estimated based on dry weight according to the method presented by Kerepesi and Galiba (2014). The total carbohydrate was measured by using a spectrophotometer. Carbohydrate% = 100 - (Moisture +Ash +Protein + Fat + Fiber).
5. Nitrogen (%): The total nitrogen in the plants was estimated according to the micro-kjeldhal method described by Talabani (2020) using concentrated sulfuric acid and hydrogen peroxide with heating for digestion.
6. Total phenols: The concentration of phenolics in plant extracts was determined using spectrophotometric method (Hasperué *et al.*, 2016).

# CHAPTER FOUR
# RESULTS AND DISCUSSIONS

## 4.1 Experiment 1: Rooting of Hardwood Cuttings of Quince (*Cydonia oblonga* Mill.) as Influenced by IBA and Rooting Substrate

### 4.1.1 Rooting success percentage

The data shown in Figure (4.1) demonstrate the effect of IBA concentrations on the rooting percentage of quince hardwood cuttings. It was revealed that using IBA was not significantly effective in improving the rooting percentage of hardwood cuttings of local quince at any concentration compared to control cuttings. Controversially, the highest rooting percentage (62.50 %) was found in control, while 3000 mg.L$^-$ IBA reduced rooting to the lowest percentage (40.97%). The application of exogenous auxins has been used to improve rooting in many species because exogenous auxins increase endogenous auxin, and endogenous auxin which is essential for root formation in cuttings should be at an optimal level to reach the best rooting rate (Hartmann *et al.*, 2002). While the ineffectiveness of IBA in this study could be attributed to that endogenous auxin might not be the only limiting factor for root formation. These results coincided with the results of  Nogueira *et al*. (2007) and Sousa *et al*. (2013) indicating that IBA was not needed for root formation in fig cuttings, and they backed that endogenous auxin may not be the sole factor in inducing rooting in the fig cuttings. They further discussed the factors that were crucial for rooting occurred in the cuttings before treating with exogenous auxin. Besides, the tissue of the quince cuttings, at the time of collection, might not be responsive or sensitive to the applied IBA. The time taken for cuttings is related to the sensitivity of the cutting tissues to the applied exogenous auxins (Mohammed, 2022).

Figure (4.2) shows the effect of rooting media on the rooting percentage of local quince hardwood cuttings. The highest value of rooting (64.58%) was observed from cuttings stuck in the river sand, which was significantly not different from rooting (52.08%) of the cuttings stuck in river sand + peat moss medium, both of them which were significantly higher than those of perlite + peat moss substrate with the lowest value (38.02%) of rooting. The growing or rooting medium is one of the major factors affecting the rooting of stem cuttings. The rooting success of any cutting is affected by the interaction of some factors like water, oxygen, and nutrient availability in the growing media (Bhardwaj, 2013). The results of the current study are contrary to those of Al-Saqri and Alderson (1996) who reported that using perlite with peat moss highly significantly affected the rooting percentage more than vermiculite in *Rosa centifolia* cuttings.

Dvin *et al*. (2011) also reported that using coco peat + perlite medium resulted in a higher root percentage of cuttings. Whereas, the rooting percentage of the quince cuttings in this study agreed with Qadri *et al.* (2018) reported that silt media had recorded a significantly higher survival percentage (83.33±16.33%). Decreasing rooting in perlite + peat moss medium may be related to that this medium could not provide the cuttings with a suitable environment to form roots, such as lacking essential nutrients, ion exchange, beneficial microorganisms, and proper pH (Shaukat *et al.*, 2018).

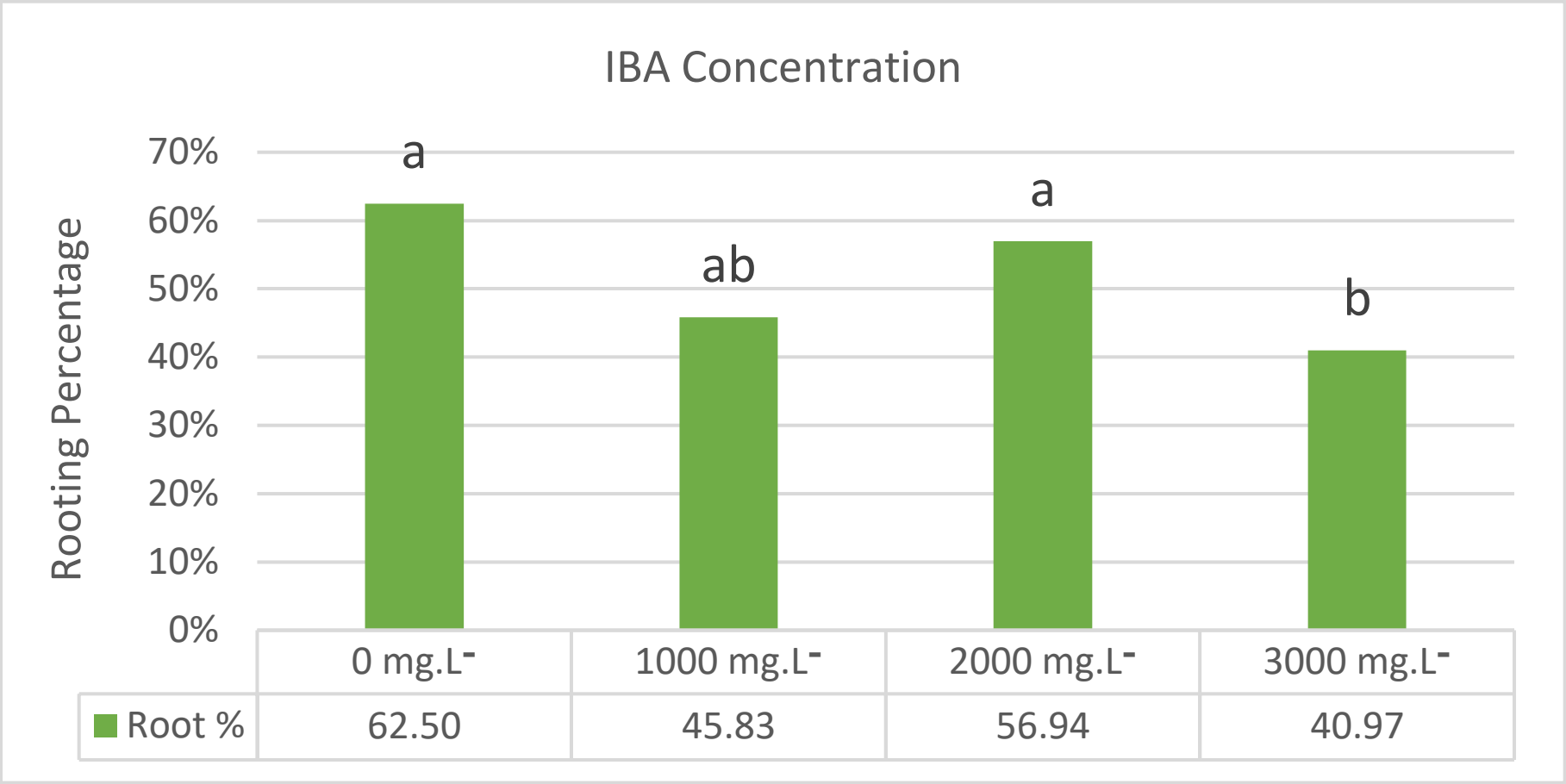

**Figure 4.1 The effect of IBA concentrations (mg.L$^{-}$) on rooting percentage of hardwood cuttings of local quince.** Values that do not share the same superscripts (a, b) differ significantly (P≤0.05).

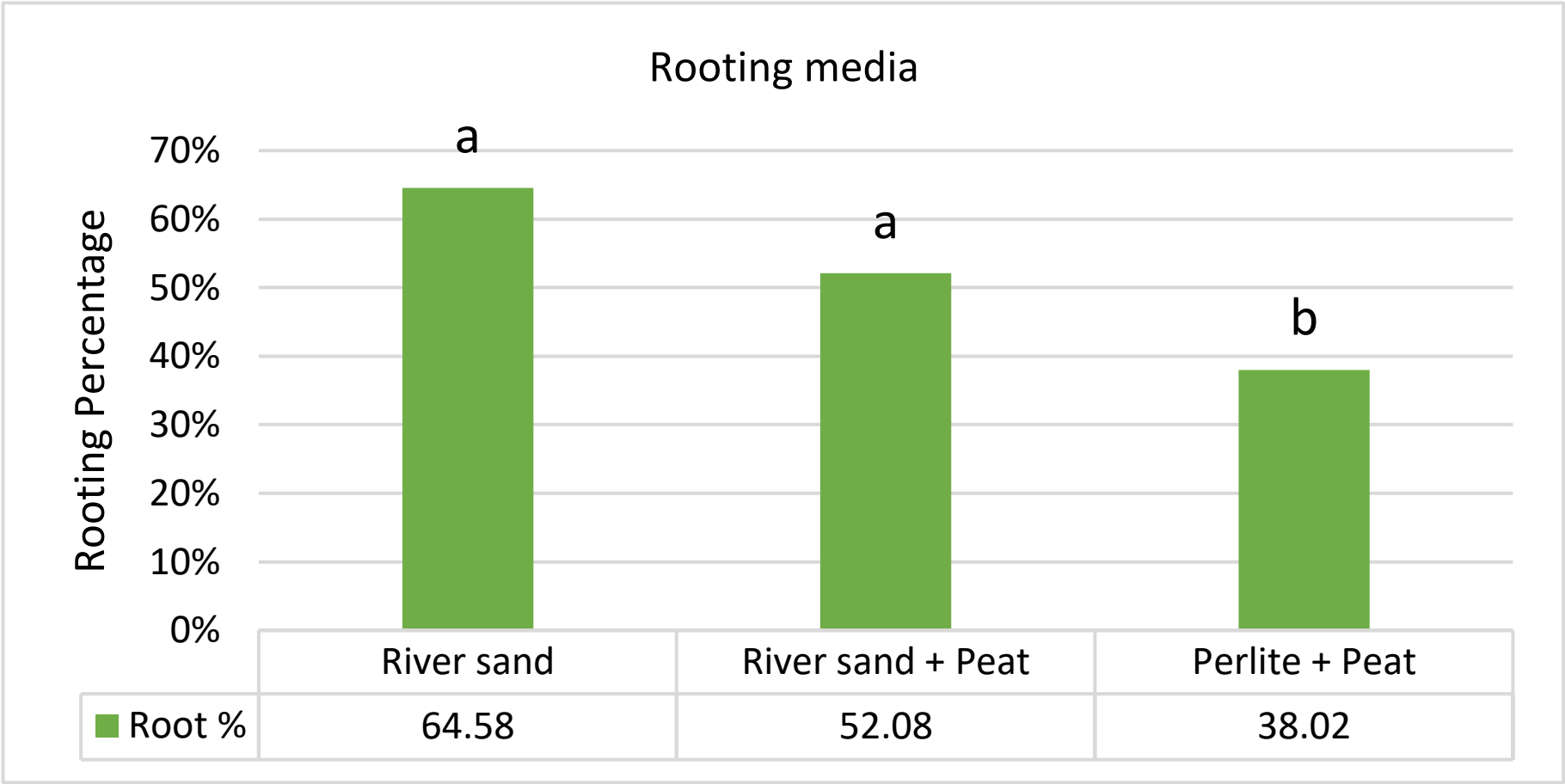

**Figure 4.2 The effect of rooting substrate on rooting percentage of hardwood cuttings of local quince.**
Values that do not share the same superscripts (a, b) differ significantly (P≤0.05).

It can be shown that the interaction of different IBA concentrations and rooting media had a significant connection for enhancing the rooting rate of hardwood cuttings of local quince. The highest rooting percentage (70.83%) was obtained from the combination of river sand and 0 mg.L$^-$ IBA. Besides, the lowest rooting percentage (6.25%) was recorded from the interaction of 3000 mg.L$^-$ IBA and perlite + peat moss medium (Figure 4.3). This might be caused by quince cuttings which are easy to root and do not respond to IBA concentration, and river sand medium afford the cuttings the necessities for better rooting. These results are contrary to those of Sardoei (2014) who reported that a higher rooting percentage (85%) was achieved in perlite/silt (1:1 v/v) medium in guava. Also, Exadaktylou *et al*. (2009) found that a combination of perlite with peat moss is suitable for rooting in hardwood cuttings of the cherry cultivar 'Gisela 5'.

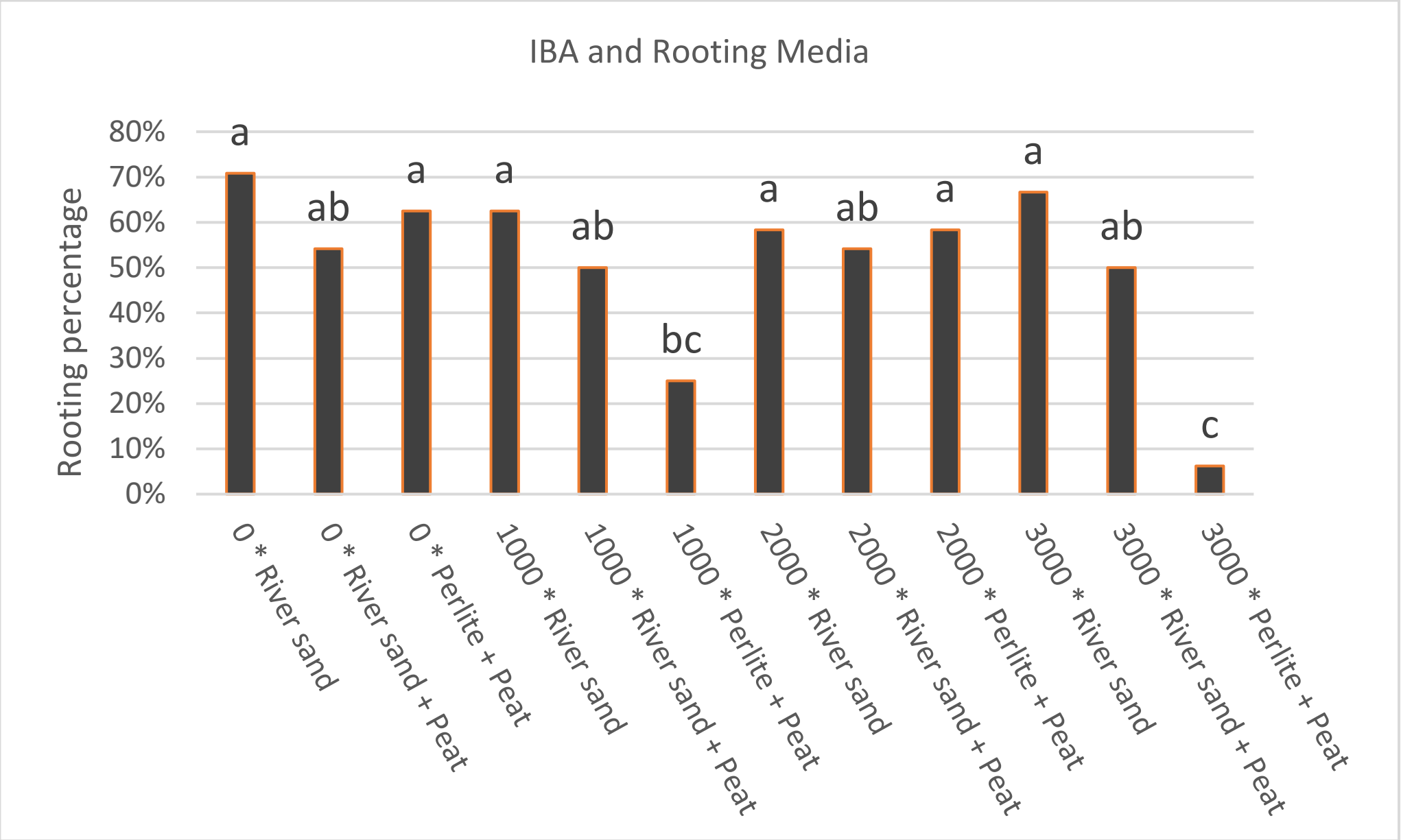

**Figure 4.3 The interaction between IBA concentrations (mg.L$^-$) and rooting substrate on rooting percentage of local quince hardwood cuttings .** Values that do not share the same superscripts (a, b, c) differ significantly (P≤0.05).

### 4.1.2 Root characteristics

Data presented in Table (4.1) show that no significant differences were detected in the root length of local quince hardwood cuttings as a result of the application of IBA at different concentrations. In contrast, the root length was reduced in the cuttings treated with IBA concentrations. So, the longest root (40.85 cm) was observed in control cuttings, but root length was decreased to about half of the control cuttings (22.75 cm) in the ones dipped in 3000 mg.L$^-$ IBA, in the same manner. The high IBA concentration (3000 mg.L$^-$) likely inhibited root growth in local quince hardwood cuttings. This result disagrees with that of Singh *et al*. (2011). The impact of IBA concentrations on root number was not affected, root number was the same in the treated and untreated cuttings, which gave the highest value of root number (13.20) in 2000 mg.L$^{-}$ IBA concentration. This might be due to hormonal effects leading to the accumulation of internal substances and their downward movement and these growth regulators also stimulate cambial activity involved in root initiation in many species. Conversely, the results of root fresh and dry weight proved that IBA concentrations were not significantly sufficient to achieve better root fresh and dry weights than control cuttings. So, root fresh (3.98 g) and dry (1.41 g) weights were superior in control cuttings. On the contrary, 3000 mg.L$^{-}$ IBA decreased fresh and dry weights to the lowest value (1.66 and 0.58 g), respectively. Moreover, IBA concentrations did not affect root length, root fresh weight, and root dry weight. The previous results of Erez and Yablowitz (1981) showed that the adverse influence of IBA was found when higher concentrations were applied. Also, da Costa *et al*. (2013) reported that a comparatively higher auxin concentration is required for adventitious root induction, but its formation was adversely suboptimal auxin concentration. This result disagrees with that of Galavi (2013) who reported that IBA enhanced the maximum number of roots, root length, and root dry weight in grape cuttings.

**Table 4.1 The effect of IBA concentration (mg.L$^{-}$) on root traits of hardwood cuttings of local quince.**

| IBA concentration (mg.L$^{-}$) | Root length (cm) | Root number | Root fresh weight (g) | Root dry weight (g) |
|---|---|---|---|---|
| 0 | 40.85 a | 13.17 a | 3.98 a | 1.41 a |
| 1000 | 34.18 ab | 11.93 a | 3.01 a | 1.11 a |
| 2000 | 35.19 ab | 13.20 a | 2.86 a | 1.09 a |
| 3000 | 22.75 b | 10.59 a | 1.66 b | 0.58 b |

* The values in each column with the same letter do not differ significantly ($P \leq 0.05$) according to Duncan's Multiple Range Test.

The results of the effects of rooting substrate on hardwood cuttings of local quince are shown in Table (4.2), demonstrating that rooting substrates were not effective on root traits, but the maximum value was observed in root length (35.71 cm) for using rooting substrate perlite with peat moss. This may be because perlite has a slow degradation rate, low bulk density, and high porosity, or may be due to the easy translocation of water and minerals to the above-ground parts of the cuttings. Many researchers studied the influence of perlite as a growing substrate. One of the studies found that the highest number of primary roots per plant, the length of the most developed roots, and the number of crowns per runner were obtained from perlite (Ercişli *et al.*, 2002). The minimum value of root length (31.92 cm) was recorded when the rooting substrate, river sand, was used alone. Besides, the maximum root fresh weight (3.07 g) was achieved when planted in perlite with peat moss. In contrast, the minimum root fresh weight (2.70 g) was observed using river sand with peat moss as the substrate. Then the highest values of root number (15.81) and root dry weight (1.11 g) were recorded in river sand. The increase in root weight was due to the greater number of roots, highest root girth, and length of the roots. Whereas, the lowest values of root number (8.90) and root dry weight (1.09 g) were recorded when perlite plus peat moss rooting substrate was used.

**Table 4.2 The effect of rooting substrate on root traits of hardwood cuttings of local quince.**

| Rooting substrate | Root length (cm) | Root number | Root fresh weight (g) | Root dry weight (g) |
|---|---|---|---|---|
| River sand | 31.92 a | 15.81 a | 2.86 a | 1.11 a |
| River sand + Peat moss | 32.11 a | 11.96 ab | 2.70 a | 0.95 a |
| Perlite + Peat moss | 35.71 a | 8.90 b | 3.07 a | 1.09 a |

* The values in each column with the same letter do not differ significantly ($P \leq 0.05$) according to Duncan's Multiple Range Test.

### 4.1.3 Shoot characteristics

Table (4.3) demonstrates the effect of growth regulators on the vegetative traits of local quince hardwood cuttings. IBA concentrations were not effective on vegetative traits compared to control cuttings. The maximum values of shoot length (54.52 cm), shoot diameter (3.19 mm), shoot fresh weight (27.60 g), shoot dry weight (11.27 g), the number of leaves (29.67), and leaf area (26.34 $cm^2$) were recorded in control of IBA concentrations. Whereas, the minimum values of shoot length (32.47 cm), shoot diameter (1.96 mm), shoot fresh weight (12.62 g), shoot dry weight (4.91 g), the number of leaves per seedling (18.02), and leaf area (19.28 $cm^2$) were recorded in (3000 $mg.L^-$) of IBA concentration. These may be caused by hardwood cuttings which contain stored nutrients such as hydrocarbons, nucleic acids, proteins, and natural hormones that can be used for shoot growth and development. Hence, they absorbed more water and nutrients needed for better growth. In this regard, Mohammed (2022) stated that superior root traits are necessary for the best shoot growth because of absorbing high water and nutrients. Moreover, as was seen in the cuttings treated with 3000 $mg.L^-$ IBA in the current study, it has been recorded that high auxin dosage retarded bud sprout and finally vegetative traits, particularly in dormant hardwood cuttings (Hartmann *et al.*, 2014). On the other hand, the maximum value of chlorophyll (21.88 SPAD) was recorded in control of IBA concentration. While the lowest value of chlorophyll (13.68 SPAD) was also found in 3000 $mg.L^-$ IBA concentration. Similarly, Khudhur and Omar (2015) reported that different concentrations and application levels of auxin and their interactions significantly affected chlorophyll, while showing a non-significant effect on chlorophyll b, total chlorophyll, and total carotenoid content of the leaf. On the contrary, Kaur *et al.* (2002) reported that the total chlorophyll content in leaves of grapevine stem cuttings was enhanced after IBA treatment.

**Table 4.3 The effect of IBA concentration (mg.L⁻) on vegetative traits of hardwood cuttings of local quince.**

| IBA concentration (mg.L⁻) | Shoot length (cm) | Shoot diameter (mm) | Shoot fresh weight (g) | Shoot dry weight (g) | Number of leaves | Leaf area (cm²) | Chlorophyll (SPAD) |
|---|---|---|---|---|---|---|---|
| 0 | 54.52 a | 3.19 a | 27.60 a | 11.27 a | 29.67 a | 26.34 a | 21.88 a |
| 1000 | 48.04 a | 2.83 a | 21.92 a | 8.79 a | 25.04 a | 23.61 a | 20.65 a |
| 2000 | 46.78 a | 2.92 a | 22.38 a | 9.93 a | 26.56 a | 26.08 a | 20.60 a |
| 3000 | 32.47 b | 1.96 b | 12.62 b | 4.91 b | 18.02 b | 19.28 a | 13.68 b |

* The values in each column with the same letter do not differ significantly ($P \leq 0.05$) according to Duncan's Multiple Range Test.

It can be seen from the data in Table (4.4), which represents the effect of rooting substrate on vegetative traits of hardwood cuttings of local quince, that these traits were significantly affected by rooting substrate (river sand and river sand with peat moss) above rooting substrate (perlite with peat moss). This could be due to the difference in the level of organic matter content and/or water-holding capacity. While the maximum values of shoot length (55.35 cm), shoot diameter (3.11 mm), shoot fresh weight (23.92 g), shoot dry weight (10.25 g), number of leaves per seedling (28.00), and leaf area (27.41 cm²) were recorded in rooting substrate (river sand with peat moss), but the minimum values of shoot length (32.28 cm), shoot diameter (2.18 mm), shoot fresh weight (16.86 g), shoot dry weight (7.19 g), number of leaves per seedling (19.07) and leaf area (18.27 cm²) were recorded in perlite/peat moss substrate. Superior shoot growth in the river sand with peat moss may be due to improved aeration, drainage, nutrient availability, and pH levels, as opposed to the mixed perlite with peat moss substrate. Similar results were recorded by Qadri *et al.* (2018) who reported that silt substrate had recorded significantly higher shoot length, shoot diameter, and number of leaves. Furthermore, the maximum value of chlorophyll (25.76 SPAD) was recorded in river sand which was significantly above the rooting substrate of perlite with peat moss having (9.85 SPAD) in chlorophyll which was observed as the minimum value. This could be related to the impact of river sand which contains more organic matter than perlite.

**Table 4.4 The effect of rooting substrate on vegetative traits of hardwood cuttings of local quince.**

| Rooting Substrate | Shoot length (cm) | Shoot Diameter (mm) | Shoot fresh weight (g) | Shoot dry weight (g) | Number of leaves | Leaf area (cm²) | Chlorophyll (SPAD) |
|---|---|---|---|---|---|---|---|
| River sand | 48.72 a | 2.88 a | 22.61 a | 8.73 ab | 27.39 a | 25.81 a | 25.76 a |
| River sand + Peat moss | 55.35 a | 3.11 a | 23.92 a | 10.25 a | 28.00 a | 27.41 a | 22.00 a |
| Perlite + Peat moss | 32.28 b | 2.18 b | 16.86 b | 7.19 b | 19.07 b | 18.27 b | 9.85 b |

* The values in each column with the same letter do not differ significantly ($P \leq 0.05$) according to Duncan's Multiple Range Test.

The interaction effects of IBA concentration and rooting substrate (Table 4.5) reveals a significant role in enhancing root measurements; root length, number of roots, root fresh and dry weights. While root length showed no significant differences among IBA concentrations and substrates, exceptions were found for cuttings treated with 3000 mg.L$^-$ IBA and planted in perlite with peat moss. The longest root (51.50 cm) was observed in control cuttings planted in perlite with peat moss, followed by (50.56 cm) in those treated with 2000 mg.L$^-$ IBA in the same substrate. However, the root length decreased to the minimum (7.43 cm) in cuttings treated with the highest IBA dose of 3000 mg.L$^-$ and planted in perlite with peat moss. Moreover, the river sand medium and its interaction with IBA concentrations, or a mixture of river sand with peat moss in combination with IBA concentration levels (1000, 2000, and 3000 mg.L$^-$) had raised to the maximum root number. The highest root number (19.72) was obtained in the control cuttings and planted in the river sand, but the lowest root numbers (6.50,7.78 and 2.22) were found in the cuttings planted in a mixture of river sand with peat moss, perlite, and peat moss substrate after dipping in (0,1000 and 3000) mg.L$^-$ IBA, respectively. Meanwhile, the same cuttings had the highest root dry weight (1.77 g) from the control cuttings planted in a mixture of perlite with peat moss substrate. In contrast, the cuttings dipped in 3000 mg.L$^-$ IBA and planted in a mixture of perlite with peat moss showed the lowest root dry weight (0.23 g). Additionally, the maximum shoot fresh weight (4.97 g) was recorded in control cuttings planted in perlite with peat moss. However, the minimum shoot fresh weight (1.29 g) was obtained when the cuttings were treated with 3000 mg.L$^-$ of IBA and planted in perlite with peat moss.

**Table 4.5 The interaction between IBA concentration and rooting substrate on root traits of hardwood cuttings of local quince.**

| IBA concentration (mg.L$^-$) | Rooting substrate | Root length (cm) | Root number | Root fresh weight (g) | Root dry weight (g) |
|---|---|---|---|---|---|
| 0 | River sand | 35.89 a | 19.72 a | 4.25 ab | 1.77 a |
| | River sand / Peat moss | 35.17 a | 6.50 cd | 2.73 a-d | 0.90 bc |
| | Perlite / Peat moss | 51.50 a | 13.28 a-c | 4.97 a | 1.55 ab |
| 1000 | River sand | 36.61 a | 15.17 a-c | 2.94 a-d | 1.13 a-c |
| | River sand / Peat moss | 32.61 a | 12.83 a-c | 3.36 a-c | 0.98 a-c |
| | Perlite / Peat moss | 33.33 a | 7.78 b-d | 2.73 a-d | 1.22 a-c |
| 2000 | River sand | 25.56 a | 11.72 a-c | 2.23 b-d | 0.68 cd |
| | River sand / Peat moss | 29.44 a | 15.56 a-c | 3.06 a-d | 1.24 a-c |
| | Perlite / Peat moss | 50.56 a | 12.33 a-c | 3.29 a-c | 1.36 a-c |
| 3000 | River sand | 29.61 a | 16.61 ab | 2.03 b-d | 0.83 b-d |
| | River sand / Peat moss | 31.22 a | 12.94 a-c | 1.67 cd | 0.67 cd |
| | Perlite / Peat moss | 7.43 b | 2.22 d | 1.29 d | 0.23 d |

* The values in each column with the same letter(s) do not differ significantly ($P \leq 0.05$) according to Duncan's Multiple Range Test.

Data presented in Table (4.6) illustrate that the combination of the two factors, rooting substrate, and IBA concentration effectively resulted in different shoot characteristics of hardwood cuttings of local quince. The interaction of 1000 mg.L$^-$ IBA with a mixture of river sand plus peat moss substrate differently produced shoot length, shoot diameter, leaf number, and leaf area when compared to the combination of 3000 mg.L$^-$ IBA with a mixture of perlite with peat moss. The cuttings dipped in 1000 mg.L$^-$ IBA and planted in a mixture of river sand with peat moss substrate had the highest values of shoot length (65.78 cm), shoot diameter (3.34 mm), number of leaves per seedling (31.11), and leaf area (29.03 cm$^2$). Whereas, the combination of 3000 mg.L$^-$ IBA with a mixture of perlite with peat moss substrate sharply reduced shoot length (7.24 cm), shoot diameter (0.44 mm), number of leaves (5.18), and leaf area (4.10 cm$^2$).

**Table 4.6 The interaction between IBA concentration and rooting substrate on vegetative traits of hardwood cuttings of local quince.**

| IBA concentration (mg.L$^-$) | Rooting substrate | Shoot length (cm) | Shoot diameter (mm) | Shoot fresh weight (g) | Shoot dry weight (g) | Number of leaves | Leaf area (cm$^2$) | Chlorophyll (SPAD) |
|---|---|---|---|---|---|---|---|---|
| 0 | River sand | 56.11 a | 3.03 a | 32.23 a | 12.64 ab | 30.89 a | 27.70 a | 30.31 a |
| | River sand/ Peat moss | 52.89 a | 3.20 a | 26.23 a-c | 9.91 a-d | 28.78 a | 28.14 a | 25.31 ab |
| | Perlite/ Peat moss | 54.56 a | 3.33 a | 24.35 a-c | 11.24 a-c | 29.33 a | 23.17 a | 10.02 d |
| 1000 | River sand | 55.33 a | 3.23 a | 24.08 a-c | 8.89 b-d | 30.33 a | 24.16 a | 24.10 ab |
| | River sand/ Peat moss | 65.78 a | 3.34 a | 26.49 a-c | 11.04 a-c | 31.11 a | 29.03 a | 23.78 ab |
| | Perlite/ Peat moss | 23.00 bc | 1.91 a | 15.19 c | 6.45 cd | 13.67 bc | 17.64 a | 14.06 cd |
| 2000 | River sand | 40.00 ab | 2.56 a | 15.42 c | 6.07 d | 23.22 ab | 24.33 a | 26.22 ab |
| | River sand/ Peat moss | 56.00 a | 3.19 a | 28.06 ab | 14.17 a | 28.33 a | 25.76 a | 21.18 a-c |
| | Perlite/ Peat moss | 44.33 ab | 3.02 a | 23.65 a-c | 9.55 a-d | 28.11 a | 28.15 a | 14.40 cd |
| 3000 | River sand | 43.44 ab | 2.72 a | 18.71 bc | 7.33 cd | 25.11 ab | 27.03 a | 22.39 a-c |
| | River sand/ Peat moss | 46.72 ab | 2.73 a | 14.91 c | 5.88 d | 23.78 ab | 26.71 a | 17.74 b-d |
| | Perlite/ Peat moss | 7.24 c | 0.44 b | 4.23 d | 1.53 e | 5.18 c | 4.10 b | 0.92 e |

* The values in each column with the same letter do not differ significantly ($P \leq 0.05$) according to Duncan's Multiple Range Test.

Additionally, the highest shoot fresh weight (32.23 g) resulted from 0 mg.L$^-$ IBA with river sand, while the lowest (4.23 g) was observed with 3000 mg.L$^-$ IBA and perlite with peat moss as the substrate. Furthermore, shoot dry weight and leaf area were differently affected by the combinations of the different IBA concentrations and growing substrate. The best shoot dry weight (14.17 g) was observed in the cuttings treated with 2000 mg.L$^-$ IBA and planted in a mixture of river sand with peat moss substrate, followed by (12.64 g) from the control cuttings which were planted in river sand substrate. Additionally, there were significant differences in leaf chlorophyll content of the quince hardwood cuttings as they were dipped in the various doses of IBA and planted in the different rooting substrates. The control cuttings planted in river sand exhibited the greatest chlorophyll content (30.31 SPAD), while chlorophyll content was severely diminished to the lowest level (0.92 SPAD) in the leaves of the cuttings dipped in 3000 $mg.L^{-}$ IBA and planted in a mixture of perlite with peat moss substrate.

## 4.2 Experiment 2: The Impact of Grafting Dates, The Cutting Types, and The IBA Concentrations on Grafting Success of Bench Grafted Loquats

Table (4.7) presents data investigating the influence of grafting dates and stock cutting type on graft bud sprout percentage and rooting in loquat bench grafting. The comparative analysis contrasts the performance of loquat and quince stock cuttings across different dates (March 30, April 30, and May 30). In terms of graft bud sprout percentage, loquat stock cuttings initially showed a higher percentage at (13.13%) on March 30, which slightly decreased to (11.67%) on April 30 and further declined to (10.83%) on May 30. In contrast, the bench grafting of quince stock cuttings started with a lower percentage at (9.17%) on March 30 but showed a consistent increase over time, reaching (13.54%) on April 30 and (14.58%) on May 30. Notably, there was a significant difference between the two stock cutting types in graft bud sprout percentage on May 30. These may be attributed to species-specific growth patterns, genetic differences, environmental preferences, nutrient requirements, cutting quality, and microclimate effects. Factors such as different ecology, rootstock and cultivar characteristics, budding techniques, and budding aftercare can affect budding or grafting success rates (Polat, 2018). Similar results were observed by Abourayya *et al.* (2019) who found that the survival percentage of grafted Thompson Seedless cuttings varied depending on the rootstock used. Romi Red stock yielded the highest survival rate (67.70%), while Salt Creek yielded the lowest (34.08%). In a study carried out by Çelik and Yılmaz (2005), it was found that grafting Foxy and Izabella grapes on different rootstocks, including Kober 5BB, Teleki 5C, 8B, and 140Ru, resulted in the highest success (81.49%) when Izabella was grafted onto the 140Ru rootstock. Additionally, Somkuwar *et al.* (2015) recorded a notably high survival percentage for Superior grape cultivar grafted on the Dog Ridge rootstock, emphasizing the importance of rootstock selection for successful grafting. Each of Lu *et al.* (2008) and Stino *et al.* (2011) reported variations in percentages of success and survival of grape grafting due to the use of different rootstocks. On the other hand, the data presented with (0.00%) values in the rooting percentage for both loquat and quince cutting types indicate a lack of observable instances of successful rooting during the study period. This suggests that the experimental conditions or treatments employed did not lead to any significant rooting for either of the cutting types, loquat, or quince.

**Table 4.7 The effect of cutting types on the percentage of graft bud sprout and rooting in loquat bench grafting.**

| Cutting type | Graft Bud sprout% | | | Rooting% |
|---|---|---|---|---|
| | March 30 | April 30 | May 30 | |
| Loquat | 13.13 a | 11.67 a | 10.83 b | 0.00 |
| Quince | 9.17 a | 13.54 a | 14.58 a | 0.00 |

The values in each column with the same letter do not differ significantly (P≤0.05) according to Duncan's Multiple Range Test.

Table (4.8) demonstrates the impact of loquat bench grafting on different dates (February 10, February 20, March 2, and March 12) on the resulting graft bud sprout percentages for both loquat and quince cuttings. The measurements were recorded on three subsequent dates (March 30, April 30, and May 30). Interestingly, the data indicates that February 20 proved to be the most successful grafting date, with the highest graft bud sprout percentages (17.50%) observed for both loquat and quince cuttings on April 30. Favorable temperature and humidity conditions during this period likely facilitated early contact of the cambium layer, callus formation, and subsequent growth in the lathhouse environment. It is noteworthy that suitable environmental conditions, as discussed by Sharma and Verma (2011), contribute to the rapid flow of plant sap at the stock and scion, leading to the formation of the cambium layer, vascular tissue, and overall graft success. In contrast, February 10 and March 12 were less successful grafting dates, maintaining graft bud sprout percentages at (7.92%) on March 30 for both cuttings. These outcomes are aligned with Singh *et al.* (2019b) in India, emphasizing the influence of grafting timing in walnuts. Likewise, Mir and Kumar (2011) reported earlier bud burst when grafting in late February. Also, Thapa *et al.* (2021) recommended February for successful walnut grafting during dormancy. While, Mehta *et al.* (2018) observed variation in days to bud bursts based on March grafting dates. Furthermore, Thapa *et al.* (2021) concluded that the timing of grafting has a significant effect on graft take and grafting survival. On the other hand, the rooting value of (0.00%) suggests the absence of successful root formation. This indicates that the experimental conditions or treatments did not lead to observable rooting within the specified context of the study.

**Table 4.8 The effect of different dates on graft bud sprout percentage and rooting percentage in loquat bench grafting on loquat and quince cuttings.**

| Date of grafting | Graft bud sprout % | | | Rooting% |
|---|---|---|---|---|
| | March 30 | April 30 | May 30 | |
| February 10 | 7.92 b | 15.83 a | 15.42 a | 0.00 |
| February 20 | 15.83 a | 17.50 a | 16.25 a | 0.00 |
| March 2 | 10.00 b | 9.58 b | 10.42 b | 0.00 |
| March 12 | 7.92 b | 10.42 b | 8.75 b | 0.00 |

The values in each column with the same letter do not differ significantly (P≤0.05) according to Duncan's Multiple Range Test.

The data presented in Table (4.9) provide insights into the effect of various concentrations of IBA on the graft bud sprout percentages of both loquat and quince stock cuttings during the loquat bench grafting process. The measurements were recorded on three different dates (March 30, April 30, and May 30). The data indicate that the highest percentages of graft bud sprout for both loquat and quince stock cuttings reached (17.19%), which were recorded on March 30 without the application of any IBA concentration. In contrast, the lowest percentages of graft bud sprout (4.69%) were observed when the highest IBA concentration of 4000 mg.L$^-$ was applied. This may potentially delay bud sprouting for both loquat and quince stock cuttings, while lower concentrations could yield more favorable outcomes. These results contradict those of Singh *et al.* (2011) who observed that *Bougainvillea* cuttings treated with 3000 mg.L$^-$ IBA had a high sprouting percentage (100%). However, the rooting values consistently remained at (0.00%). This data indicates that the application of IBA did not have a significant impact on the success of grafting or root formation during the observed period. This contrasts with Seyedi *et al.* (2014), who reported that the maximum rooting rate in apple cuttings was achieved with the application of 2400 mg.L$^-$ IBA. Melgarejo *et al.* (2000) also demonstrated that in pomegranates, most clones utilizing modest IBA application concentrations (3000 mg.L$^-$) experienced an increase in the percentage of cuttings that rooted. In a study of Al-Zebari and Al-Brifkany (2015), the application of indole butyric acid (IBA) at concentrations of 500, 1000, and 2000 mg.L$^-$ significantly influenced the rooting percentage in stem cuttings of citron (*Citrus medica* Linnaeus ) Corsian cultivar.

**Table 4.9 The effects of IBA concentrations on graft bud sprout percentage and rooting percentage in loquat bench grafted on loquat and quince cuttings.**

| IBA concentration mg.L$^-$ | Graft bud sprout % | | | Rooting% |
|---|---|---|---|---|
| | March 30 | April 30 | May 30 | |
| 0 | 17.19 a | 16.67 a | 14.58 a | 0.00 |
| 1000 | 10.94 b | 13.54 a | 13.02 ab | 0.00 |
| 2000 | 10.94 b | 14.58 a | 14.06 a | 0.00 |
| 3000 | 8.33 bc | 15.10 a | 13.54 a | 0.00 |
| 4000 | 4.69 c | 6.77 b | 8.33 b | 0.00 |

The values in each column with the same letter do not differ significantly (P≤0.05) according to Duncan's Multiple Range Test.

Table (4.10) illustrates the interaction effects of stock cutting type and different grafting dates (February 10, February 20, March 2, and March 12) of loquat bench grafting on graft bud sprout and rooting percentages which were measured on March 30, April 30, and May 30. the highest graft bud sprout percentage (20.83%) was achieved on April 30 for loquat stock cuttings grafted on February 20. Conversely, the lowest graft bud sprout percentage (4.17%) was observed on May 30, resulting from the interaction between loquat bench grafting on loquat stock cuttings and March 2. Among the grafting of quince stock cuttings, the most favorable graft bud sprout percentage (16.67%) was documented on May 30, specifically when the grafting occurred on March 2. In contrast, the least favorable bud sprout percentage (0.83%) was recorded on March 30, arising from the interaction involving February 10. Furthermore, the maximum graft bud sprout percentage (20.83%) occurred on April 30, arising from the interaction between loquat stock cuttings and February 20. On the other hand, the minimum bud sprout percentage (0.83%) was recorded on March 30 due to the combination of quince cuttings and February 10. Factors influencing these variations include environmental conditions, genetic factors, pruning techniques, nutrient availability, and photoperiod sensitivity. It should be noted that the data for rooting percentage resulted in a value of (0.00%) for all times of grafting.

**Table 4.10 The interaction between cutting types and different dates on graft bud sprout percentage and rooting percentage in loquat bench grafting.**

| Cutting type | Date of grafting | Graft bud sprout % | | | Rooting% |
|---|---|---|---|---|---|
| | | March 30 | April 30 | May 30 | |
| Loquat | February 10 | 15.00 ab | 16.67 ab | 15.00 a | 0.00 |
| | February 20 | 18.33 a | 20.83 a | 18.33 a | 0.00 |
| | March 2 | 5.00 c | 5.83 c | 4.17 c | 0.00 |
| | March 12 | 8.33 bc | 9.17 bc | 5.83 bc | 0.00 |
| Quince | February 10 | 0.83 c | 15.00 ab | 15.83 a | 0.00 |
| | February 20 | 13.33 ab | 14.17 ab | 14.17 a | 0.00 |
| | March 2 | 15.00 ab | 13.33 ab | 16.67 a | 0.00 |
| | March 12 | 7.50 bc | 11.67 bc | 11.67 ab | 0.00 |

* The values in each column with the same letter do not differ significantly (P≤0.05) according to Duncan's Multiple Range Test.

Data presented in Table (4.11) show the interaction effect of stock cutting types (loquat and quince) and IBA concentrations (0, 1000, 2000, 3000, and 4000 mg.L$^{-}$) on graft bud sprout and rooting percentages of loquat bench grafts measured on three different dates (March 30, April 30, and May 30). Consequently, the maximum graft bud sprout percentage (17.71%) for loquat stock cuttings was observed on April 30 when treated with 0 mg.L$^{-}$ IBA. This suggests that loquat cuttings do not require the application of IBA to stimulate graft bud sprout and perform well without it. However, the minimum graft bud sprout percentage (7.29%) was observed on March 30 when treated with 4000 mg.L$^{-}$ IBA. This indicates that the high concentration of IBA (4000 mg.L$^{-}$) is not effective and may even inhibit bud sprout in loquat cuttings. On the other hand, the maximum graft bud sprout percentage (18.75%) for quince cuttings was observed on May 30 when treated with (2000 and 3000) mg.L$^{-}$ IBA. This suggests that 2000 and 3000 mg.L$^{-}$ of IBA are the most effective concentrations for promoting graft bud sprout in quince stock cuttings. However, the minimum graft bud sprout percentage (2.01%) was observed on March 30 when treated with 4000 mg.L$^{-}$ IBA. This indicates that a high concentration of IBA (4000 mg.L$^{-}$) is not effective and may even inhibit bud sprout in quince cuttings. It is worth mentioning that the data for rooting percentage yielded a value of (0.00%) for the interaction between cutting type and different IBA concentrations. On the contrary, Daoud *et al.* (1995) found that applying indole butyric acid (IBA) at different concentrations (0, 1250, 2500, 5000, and 10000 mg.L$^{-}$) to eight citrus rootstock cuttings resulted in diverse rooting percentages. Some exhibited robust responses ranging from 61.1% to 91.7%, while others showed lower percentages ranging from 13.3% to 38.9%.

**Table 4.11 The interaction between cutting types and IBA concentrations on graft bud sprout percentage and rooting percentage in loquat bench grafting.**

| Cutting type | IBA concentration (mg.L$^{-}$) | Graft bud sprout % | | | Rooting% |
|---|---|---|---|---|---|
| | | March 30 | April 30 | May 30 | |
| Loquat | 0 | 16.67 ab | 17.71 a | 14.58 ab | 0.00 |
| | 1000 | 13.54 a-c | 14.58 a | 13.54 ab | 0.00 |
| | 2000 | 9.38 a-d | 11.46 ab | 9.38 b | 0.00 |
| | 3000 | 11.46 a-c | 12.50 ab | 8.33 b | 0.00 |
| | 4000 | 7.29 cd | 9.38 ab | 8.33 b | 0.00 |
| Quince | 0 | 17.71 a | 15.63 a | 14.58 ab | 0.00 |
| | 1000 | 8.33 b-d | 12.50 ab | 12.50 ab | 0.00 |
| | 2000 | 12.50 a-c | 17.71 a | 18.75 a | 0.00 |
| | 3000 | 5.21 cd | 17.71 a | 18.75 a | 0.00 |
| | 4000 | 2.01 d | 4.17 b | 8.33 b | 0.00 |

* The values in each column with the same letter do not differ significantly ($P \leq 0.05$) according to Duncan's Multiple Range Test.

The data presented in Table (4.12) demonstrate the interaction effects of grafting time and IBA concentration on the graft bud sprouting% and rooting% for loquat/loquat and loquat/quince combinations of the combination measured on different dates with monthly intervals (March 30, April 30, and May 30). The data indicates that the maximum percentage of graft bud sprout (31.25%) was achieved on April 30, when the stock cuttings were treated with 3000 mg.L$^{-}$ IBA and grafted on February 20. This may be due to the optimal temperature conditions during this period, promoting enhanced physiological responses to IBA, leading to increased bud sprouting. Conversely, the lowest graft bud sprout percentage (4.17%) was observed on March 30, which may be a result of suboptimal temperatures on February 10 interacting with IBA concentrations of (1000 and 2000 mg.L$^{-}$) thence inhibiting the expected grafting success. Additionally, it is worth noting that the highest observed graft bud sprout percentage, at (29.17%) was recorded on April 30 when the grafting was conducted on February 20 and the stock cuttings were treated with a concentration of 2000 mg.L$^{-}$ IBA in which favorable temperature conditions might be available at this time, supporting successful bud sprouting. Conversely, the lowest graft bud sprout percentage (2.08%) was achieved on both March 30 and April 30. This low percentage resulted from the combination of grafting on February 20 and treatment with a high concentration of 4000 mg.L$^{-}$ IBA which may be due to the inhibitory effect of excessive IBA concentration. Furthermore, the highest value of graft bud sprout percentage (25.00%) occurred on March 30. This result was obtained from the interaction between grafting on March 2 and treating the cuttings with 0 mg.L$^{-}$ IBA. However, the lowest value of graft bud sprout percentage (0.00%) was noted on March 30 when the grafting was conducted on March 2, and the cuttings were treated with a high concentration of 4000 mg.L$^{-}$ of IBA.

**Table 4.12 The interaction between grafting dates and IBA concentrations on graft bud sprout percentage and rooting percentage in loquat bench grafting.**

| Date of Grafting | IBA concentration (mg.L$^{-}$) | Graft Bud sprout % | | | Rooting% |
|---|---|---|---|---|---|
| | | March 30 | April 30 | May 30 | |
| February 10 | 0 | 12.50 a-e | 12.50 b-e | 12.50 bc | 0.00 |
| | 1000 | 4.17 c-e | 8.33 c-e | 10.42 bc | 0.00 |
| | 2000 | 4.17 c-e | 12.50 b-e | 10.42 bc | 0.00 |
| | 3000 | 10.42 b-e | 31.25 a | 29.17 a | 0.00 |
| | 4000 | 8.33 c-e | 14.58 b-e | 14.58 bc | 0.00 |
| February 20 | 0 | 14.58 a-d | 16.67 b-d | 14.58 bc | 0.00 |
| | 1000 | 25.00 a | 25.00 ab | 27.08 a | 0.00 |
| | 2000 | 22.92 ab | 29.17 a | 27.08 a | 0.00 |
| | 3000 | 14.58 a-d | 14.58 b-e | 8.33 bc | 0.00 |
| | 4000 | 2.08 de | 2.08 e | 4.17 c | 0.00 |
| March 2 | 0 | 25.00 a | 20.83 a-c | 16.67 b | 0.00 |
| | 1000 | 10.42 b-e | 10.42 c-e | 6.25 bc | 0.00 |
| | 2000 | 12.50 a-e | 12.50 b-e | 14.58 bc | 0.00 |
| | 3000 | 2.08 de | 2.08 e | 6.25 bc | 0.00 |
| | 4000 | 0.00 e | 2.08 e | 8.33 bc | 0.00 |
| March 12 | 0 | 16.67 a-c | 16.67 b-d | 14.58 bc | 0.00 |
| | 1000 | 4.17 c-e | 10.42 c-e | 8.33 bc | 0.00 |
| | 2000 | 4.17 c-e | 4.17 de | 4.17 c | 0.00 |
| | 3000 | 6.25 c-e | 12.50 b-e | 10.42bc | 0.00 |
| | 4000 | 8.33 c-e | 8.33 c-e | 6.25 bc | 0.00 |

* The values in each column with the same letter do not differ significantly ($P \leq 0.05$) according to Duncan's Multiple Range Test.

In addition, it is important to note that the highest observed graft bud sprout percentage, amounting to (16.67%), was recorded on both March 30 and April 30. This result was obtained when the grafting was conducted on March 12, and the cuttings were treated without IBA, i.e., with 0 mg.L$^{-}$ concentration. On the contrary, the lowest graft bud sprout percentage (4.17%) was consistently achieved on March 30, April 30, and May 30. These low percentages resulted from the combination of grafting on March 12 and the treatment with IBA concentrations of (1000 and 2000 mg.L$^{-}$). It should be noted that the data for rooting percentage also resulted in a value of (0.00%) for all times of grafting and different IBA concentrations, which are controversary to the results obtained by Kaur (2017) who reported that peach cuttings treated with IBA at a concentration of 2400 mg.L$^{-}$ IBA, rooted the best when taken on January 15$^{th}$.

Table (4.13) presents the data of the interaction between cutting type (loquat and quince), different dates of loquat bench grafting (February 10, February 20, March 2, and March 12), and IBA concentrations (0, 1000, 2000, 3000, and 4000 mg.L$^{-}$) on graft bud sprout percentage for both loquat and quince stock cuttings. As a result, the maximum graft bud sprout percentage for loquat stock cuttings reached (37.50%), recorded on May 30 when loquat stock cuttings were grafted on February 20 and treated with (1000 mg.L$^{-}$) IBA. This combination highlights the high efficacy of promoting bud sprouts in loquat stock cuttings. This favorable outcome can be attributed to the synergistic effect of the specific grafting date and a moderate IBA concentration. In contrast, the lowest graft bud sprout percentages (0.00%) for loquat stock cuttings were observed under various combinations, notably on March 2 and March 12, and treated with (2000 and 3000 mg.L$^{-}$) of IBA concentrations. These may be related to factors such as suboptimal grafting dates, inappropriate IBA concentrations, or unfavorable environmental conditions. On the other hand, the highest graft bud sprout percentage for quince stock cuttings (41.67%) was recorded on March 30 when quince stock cuttings were grafted on March 2 and treated with (0 mg.L$^{-}$) of IBA. Similarly, another notable peak of (41.67%) in the graft bud sprout percentage for quince stock cuttings was observed on April 30 and May 30. In these cases, quince cuttings were grafted on February 10 and treated with a substantial IBA concentration of 3000 mg.L$^{-}$, demonstrating the significant role of IBA concentration and grafting timing in stimulating bud sprout in quince cuttings. While, the lowest graft bud sprout percentage (0.00%) was observed for quince stock cuttings, and recorded on April 30 when quince stock cuttings were grafted on both February 10 and February 20 and were treated with either (0 or 4000 mg.L$^{-}$) of IBA. Another instance of the minimum graft bud sprout percentage (0.00%) was observed on May 30 when quince stock cuttings were grafted on February 10 and treated with (0 mg.L$^{-}$) of IBA, underlining the critical role of IBA concentration and grafting timing in achieving optimal bud sprout percentages for quince stock cuttings. Additionally, the rooting percentage resulted in a value of (0.00%) for the interaction between cutting type, different times, and various IBA concentrations. The results might be due to specific combinations of grafting dates and IBA concentrations having separately a significant impact on the bud sprout percentages observed for both quince and loquat cuttings.

**Table 4.13 The interaction between cutting types, different dates, and IBA concentrations on graft bud sprout percentage and rooting percentage in loquat bench grafting.**

| Cutting type | Date of grafting | IBA concentration (mg.L-) | Graft bud sprout % | | | Rooting% |
|---|---|---|---|---|---|---|
| | | | March 30 | April 30 | May 30 | |
| Loquat | February 10 | 0 | 25.00 a-c | 25.00 a-d | 25.00 b-d | 0.00 |
| | | 1000 | 4.17 de | 8.33 d-f | 8.33 d-f | 0.00 |
| | | 2000 | 8.33 c-e | 8.33 d-f | 4.17 ef | 0.00 |
| | | 3000 | 20.83 b-d | 20.83 b-e | 16.67 c-f | 0.00 |
| | | 4000 | 16.67 b-e | 20.83 b-e | 20.83 c-e | 0.00 |
| | February 20 | 0 | 12.50 c-e | 16.67 b-f | 12.50 c-f | 0.00 |
| | | 1000 | 33.33 ab | 33.33 ab | 37.50 a | 0.00 |
| | | 2000 | 20.83 b-d | 29.17 a-c | 25.00 b-d | 0.00 |
| | | 3000 | 20.83 b-d | 20.83 b-e | 12.50 c-f | 0.00 |
| | | 4000 | 4.17 de | 4.17 ef | 4.17 ef | 0.00 |
| | March 2 | 0 | 8.33 c-e | 8.33 d-f | 4.17 ef | 0.00 |
| | | 1000 | 8.33 c-e | 8.33 d-f | 4.17 ef | 0.00 |
| | | 2000 | 8.33 c-e | 8.33 d-f | 8.33 d-f | 0.00 |
| | | 3000 | 0.00 e | 0.00 f | 0.00 f | 0.00 |
| | | 4000 | 0.00 e | 4.17 ef | 4.17 ef | 0.00 |
| | March 12 | 0 | 20.83 b-d | 20.83 b-e | 16.67 c-f | 0.00 |
| | | 1000 | 8.33 c-e | 8.33 d-f | 4.17 ef | 0.00 |
| | | 2000 | 0.00 e | 0.00 f | 0.00 f | 0.00 |
| | | 3000 | 4.17 de | 8.33 d-f | 4.17 ef | 0.00 |
| | | 4000 | 8.33 c-e | 8.33 d-f | 4.17 ef | 0.00 |
| Quince | February 10 | 0 | 0.00 e | 0.00 f | 0.00 f | 0.00 |
| | | 1000 | 4.17 de | 8.33 d-f | 12.50 c-f | 0.00 |
| | | 2000 | 0.00 e | 16.67 b-f | 16.67 c-f | 0.00 |
| | | 3000 | 0.00 e | 41.67 a | 41.67 a | 0.00 |
| | | 4000 | 0.00 e | 8.33 d-f | 8.33 d-f | 0.00 |
| | February 20 | 0 | 16.67 b-e | 16.67 b-f | 16.67 c-f | 0.00 |
| | | 1000 | 16.67 b-e | 16.67 b-f | 16.67 c-f | 0.00 |
| | | 2000 | 25.00 a-c | 29.17 a-c | 29.17 a-c | 0.00 |
| | | 3000 | 8.33 c-e | 8.33 d-f | 4.17 ef | 0.00 |
| | | 4000 | 0.00 e | 0.00 f | 4.17 ef | 0.00 |
| | March 2 | 0 | 41.67 a | 33.33 ab | 29.17 a-c | 0.00 |
| | | 1000 | 12.50 c-e | 12.50 c-f | 8.33 d-f | 0.00 |
| | | 2000 | 16.67 b-e | 16.67 b-f | 20.83 c-e | 0.00 |
| | | 3000 | 4.17 de | 4.17 ef | 12.50 c-f | 0.00 |
| | | 4000 | 0.00 e | 0.00 f | 12.50 c-f | 0.00 |
| | March 12 | 0 | 12.50 c-e | 12.50 c-f | 12.50 c-f | 0.00 |
| | | 1000 | 0.00 e | 12.50 c-f | 12.50 c-f | 0.00 |
| | | 2000 | 8.33 c-e | 8.33 d-f | 8.33 d-f | 0.00 |
| | | 3000 | 8.33 c-e | 16.67 b-f | 16.67 c-f | 0.00 |
| | | 4000 | 8.33 c-e | 8.33 d-f | 8.33 d-f | 0.00 |

* The values in each column with the same letter do not differ significantly ($P \leq 0.05$) according to Duncan's Multiple Range Test.

## 4.3 Experiment 3: The Performance of Grafting Loquats Combined onto Loquat and Quince Rootstocks on Different Dates

### 4.3.1 Grafting success percentage

The results of grafting success percentages of loquat and quince rootstocks showed significant differences between the two types of rootstocks (Figure 4.4). The highest grafting success percentage (97.78%) was recorded on loquat rootstocks, which was significantly higher than that of quince rootstocks which obtained the lowest grafting success percentage (84.44%). This difference may be attributed to genetic compatibility, physiological similarity, and better cambium activity between loquat scions and rootstocks. Other factors, such as environmental adaptation and rootstock vigor could also contribute to the difference between rootstocks. The significant difference between the grafting success percentages of the two rootstocks suggests that loquat rootstock is more favorable for grafting success compared to quince rootstock. Our results conform with the results of Rahman *et al.* (2017) and Öztürk (2021) who observed that rootstocks and varieties have a very important effect on the graft take ratio in the pear.

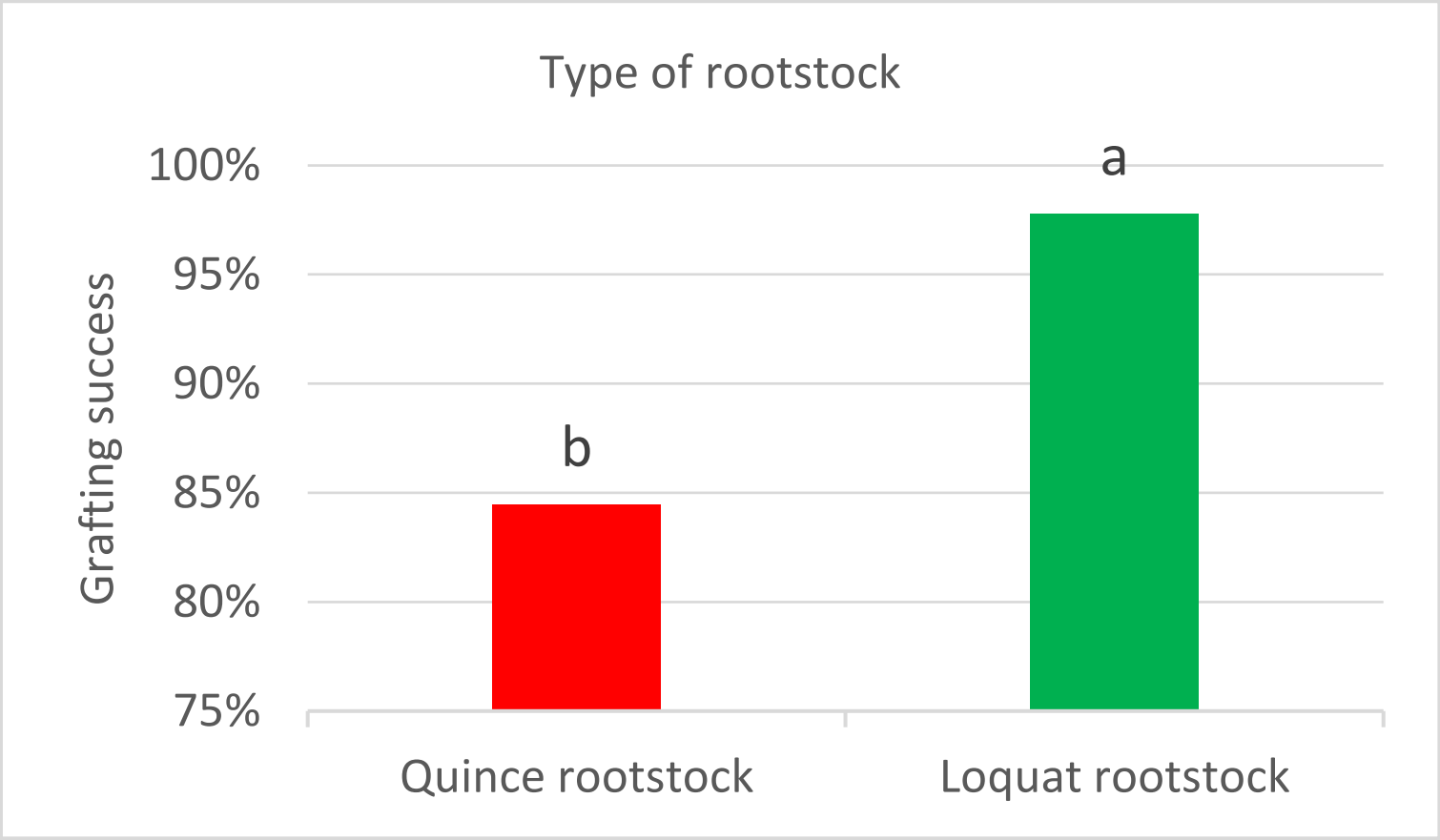

**Figure 4.4 The effect of rootstock type on the grafting success percentage of loquat and quince rootstocks.** Values that do not share the same superscripts (a, b) differ significantly (P≤0.05).

The data presented in Figure (4.5) demonstrate the effect of grafting dates on loquat and quince rootstocks at different periods. It is clear that significant differences were observed between the grafting dates from late February to late March to achieve a successful grafting percentage for the current season on loquat and quince rootstocks. Conversely, the grafting success percentage decreased from February 20 to March 30. Therefore, the highest grafting success rate (96.67%) was recorded on February 20 when grafted on both loquat and quince rootstocks. On the other hand, the lowest grafting success rate (86.67%) was obtained on March 30 when grafted on both loquat and quince rootstocks. These may be due to changing environmental conditions. This result was supported by (Mandal *et al.*, 2011). Significantly, the highest survival percentage at 120 days after grafting was the result. Similar results were also mentioned by (Karna *et al.*, 2018) in mango. Although the grafting process can be performed at any time during the dormant season. However, in the early spring and during sap flow season, the chances for successful healing of the graft union are better.

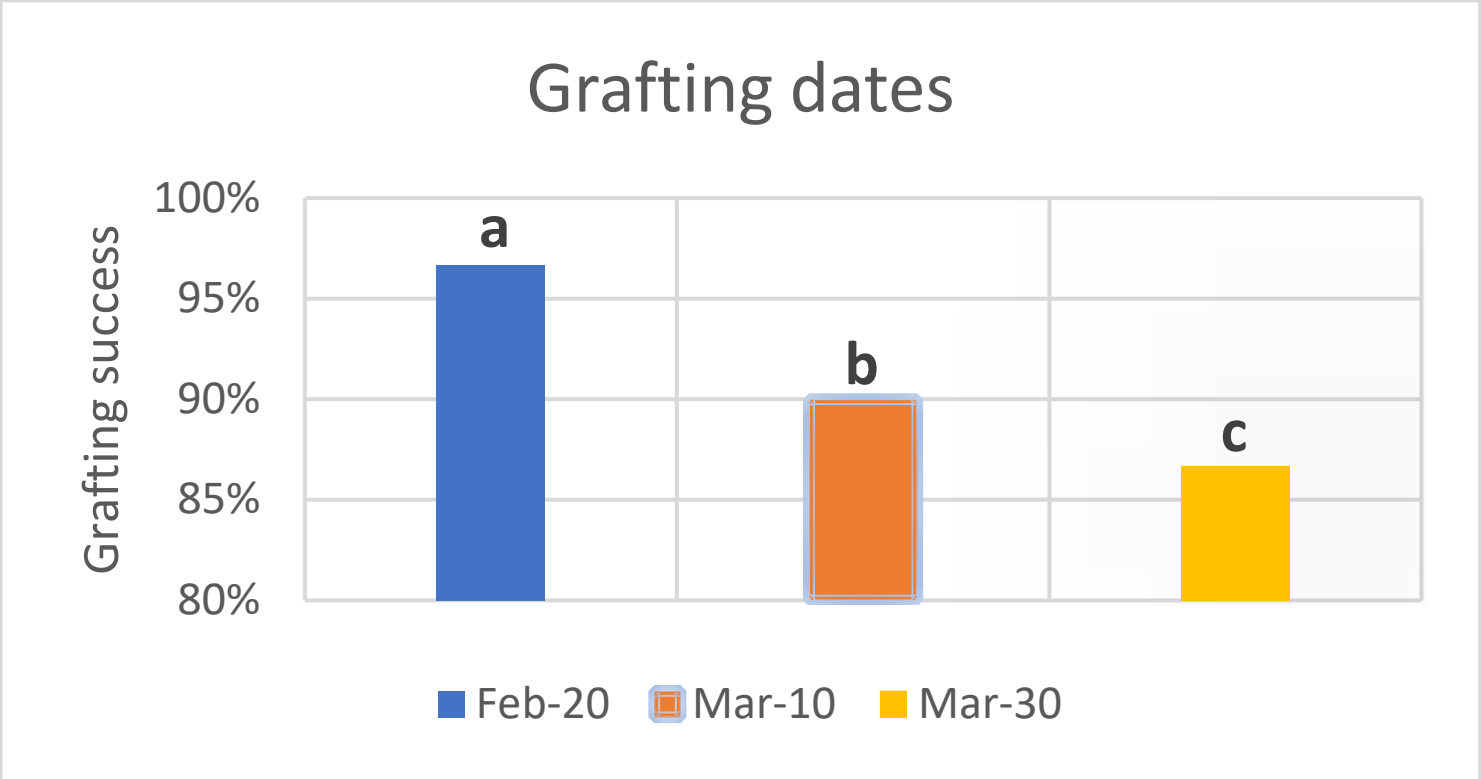

**Figure 4.5 The effect of grafting dates on grafting success percentage of loquat and quince rootstocks.** Values that do not share the same superscripts (a, b, c) differ significantly (P≤0.05).

Figure (4.6) shows the interaction effects of rootstock type and grafting dates on grafting success percentages of loquat grafting onto each loquat and quince rootstock. The combinational effect had a significant connection for enhancing the grafting success percentage of loquat grafting. The highest significant grafting success percentage (100%) was achieved from the combination of loquat rootstock and two grafting dates (February 20 and March 30). This may be attributed to the genetic compatibility and physiological status suitability between loquat scions and rootstocks, resulting in more effective graft unions. This combination likely provided an optimal environment for successful graft union formation and tissue integration. However, the grafting success percentage (93.33%) was recorded from the interaction between quince rootstock on February 20 and loquat rootstock on March 10. Furthermore, the interaction of quince rootstock with March 30 gave rise to the lowest grafting success percentage (73.33%). This might be related to factors such as differences in cambium activity, environmental adaptation, or reduced vigor in the quince rootstock, which hindered the successful integration of the loquat scion.

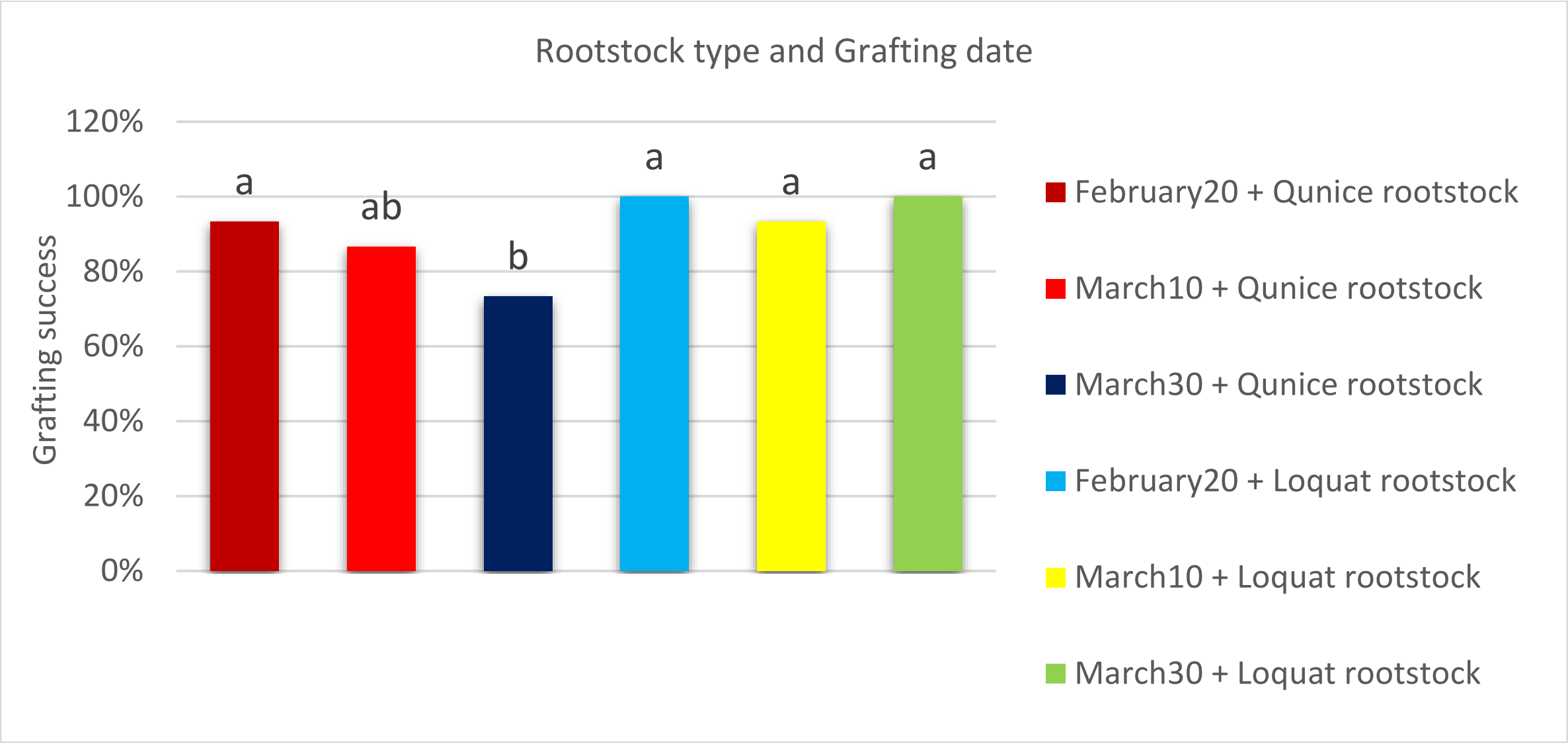

**Figure 4.6 The interaction between rootstock type and grafting dates on the success percentage of loquat grafting, using both loquat and quince rootstocks.** Values that do not share the same superscripts (a, b) differ significantly (P≤0.05).

### 4.3.2 Shoot characteristics

The data presented in Table (4.14) compare the impact of rootstock types, namely loquat and quince on loquat grafting success and subsequent growth parameters on loquat and quince rootstocks. The results show that the loquat rootstock resulted in significantly longer shoots (11.74 cm) compared to the quince rootstock (8.08 cm). This may have a more vigorous growth habit or better nutrient uptake, contributing to longer shoots, or might be more adapted to the specific growing conditions or environmental conditions of the study. Similarly, a budshoot length of 21.23 cm was reported by Chalise *et al.* (2014) in acid lime grafted onto trifoliate orange rootstock at 4 months after grafting. However, there was no significant difference in shoot diameter between loquat (4.06 mm) and quince (4.24 mm) rootstocks. This may be due to similar genetic characteristics that influence shoot diameter. Similarly, Coban and Ozturk (2020) observed a non-significant effect of pear rootstocks and cultivars on graft shoot diameter. Additionally, the study showed that the highest significant shoot fresh and dry weights (10.32 and 6.19 g) were recorded when the loquat was grafted onto loquat rootstocks. However, the lowest shoot fresh and dry weights (5.86 and 3.36 g) were observed when the loquat was grafted onto quince rootstocks. However, the values of leaf area for both loquat and quince rootstocks were statistically similar, with values of (69.98 $cm^2$) for loquat and (69.83 $cm^2$) for quince. Otherwise, quince rootstock had a higher significant leaf number (8.47) and chlorophyll content (43.82 SPAD) compared to loquat rootstock (7.67 and 36.82 SPAD), respectively. These results may be due to genetic variations between loquat and quince rootstocks and their specific responses to the growing conditions, or may be attributed to inherent physiological differences between the two rootstocks.

**Table 4.14 The effect of rootstock type on growth parameters in loquat grafting on loquat and quince rootstocks.**

| Rootstock type | Graft shoot length (cm) | Graft shoot diameter (mm) | Shoot fresh weigh (g) | Shoot dry weight (g) | Number of leaves per budling | Leaf area (cm²) | Leaf chlorophyll content (SPAD) |
|---|---|---|---|---|---|---|---|
| Loquat | 11.74 a | 4.06 a | 10.32 a | 6.19 a | 7.67 b | 69.98 a | 36.82 b |
| Quince | 8.08 b | 4.24 a | 5.86 b | 3.36 b | 8.47 a | 69.83 a | 43.82 a |

* The values in each column with the same letter do not differ significantly (P≤0.05), according to Duncan's Multiple Range Test.

It can be seen from Table (4.15) regarding the effect of loquat grafting dates on vegetative traits on loquat and quince rootstocks that significant effects of grafting dates on shoot length, shoot fresh weight, shoot dry weight, leaf number, leaf area, and leaf chlorophyll content were revealed. The loquat grafted on March 30 had a longer shoot length (11.44 cm) compared to those grafted on March 10 (7.52 cm). This suggests that the timing of grafting might influence shoot elongation, but the grafting of loquat for all three dates had similar results on shoot diameters, ranging from (4.08 mm) to (4.28 mm). This indicates that the grafting date might not have a significant impact on shoot thickness. These results are contrary to those obtained by Onay *et al*. (2003) working with pistachio. Furthermore, regarding shoot fresh weight and dry weight, when grafted on February 20 and March 30 exhibited significantly higher weights (8.55 g and 5.18 g for fresh weight; 8.82 g and 5.32 g for dry weight) than those grafted on March 10 (6.90 g for fresh weight; 3.82 g for dry weight). Additionally, the maximum values of leaf number (9.09), and leaf area (81.26 cm²) were recorded from grafting loquat on February 20, but the minimum values were recorded in leaf number and leaf area (7.39 and 63.12 cm²) respectively. Similar results were observed by Khopade and Jadav (2013) in custard apples. On the contrary with, Panchal *et al*. (2022) recorded significantly higher values of shoot grafts and the leaf area which were noted when grafting was done on 15$^{th}$ April and 15$^{th}$ March. higher leaf area is possibly correlated with the higher number of leaves per graft recorded during these months. Correspondingly, the highest values of the number of leaves and leaf area obtained in the grafts conducted on February 20, might be due to the early bud break resulting from early healing and graft union formation due to favorable environmental conditions of temperature and relative humidity (Figure 3.7). The earlier the healing of graft wounds between scion and rootstock, the more the promotion of bud break resulted from the early and easy availability of raw material for photosynthesis which ultimately increased the growth of the whole plant. On the other hand, the lower number of leaves and leaf area in early grafts might be due to low temperature at the time of graft union formation and leaf emergence and development. The present results were inconsistent with Chalise *et al*. (2014) in the acid lime sapling. Likewise, the maximum value of chlorophyll (41.42 SPAD) was recorded on March 30, but the minimum value (38.40 SPAD) was observed during grafting on March 10. These may be due to variations in environmental conditions, such as sunlight exposure, temperature, and humidity, which can affect chlorophyll production and accumulation in plants. Other factors, such as differences in nutrient availability could also contribute to the variations in chlorophyll content.

**Table 4.15 The effect of grafting dates on growth parameters in loquat grafting on loquat and quince rootstocks.**

| Grafting date | Graft shoot length (cm) | Graft shoot diameter (mm) | Shoot fresh weight (g) | Shoot dry weight (g) | Number of leaves per budling | Leave area (cm²) | Leaf chlorophyll content (SPAD) |
|---|---|---|---|---|---|---|---|
| February 20 | 10.78 a | 4.10 a | 8.55 a | 5.18 a | 9.09 a | 81.26 a | 41.13 a |
| March 10 | 7.52 b | 4.08 a | 6.90 b | 3.82 b | 7.72 b | 65.34 b | 38.40 b |
| March 30 | 11.44 a | 4.28 a | 8.82 a | 5.32 a | 7.39 b | 63.12 b | 41.42 a |

* The values in each column with the same letter do not differ significantly ($P \leq 0.05$), according to Duncan's Multiple Range Test.

Data presented in Table (4.16) demonstrate that the combination of the two factors, rootstock type, and grafting dates effectively resulted in different shoot characteristics of loquat grafting success. The grafting on loquat rootstock on March 30th had the highest significant shoot length (14.78 cm). Whereas, the combination of quince rootstock and March 10th sharply reduced shoot length (6.59 cm). This may be due to the interaction between the genetic compatibility of the rootstock and scion (loquat), and the environmental conditions during the grafting process. However, the study found no significant differences between the interaction of rootstock type and grafting dates on shoot diameter. Although, the highest shoot diameter (4.42 mm) was recorded when grafting on quince rootstock was done on March 30th. Conversely, the lowest shoot diameter (3.77 mm) was observed when using the combination of loquat rootstock and grafting on March 10th. On the other hand, the highest shoot fresh weight recorded was (11.69 g), and the highest shoot dry weight was (6.70 g), observed from loquat grafting on quince rootstocks on March 30th. Conversely, the lowest shoot fresh weight recorded was (4.83 g), and the lowest shoot dry weight was (2.36 g), observed from quince trees grafted on March 10th. This combination resulted in the highest and lowest fresh and dry weights among all other combinations in the table. Additionally, the study found that the highest leaf number (9.11) was observed when using loquat rootstock and grafting on February 20th. But, the lowest value of leaf number (6.89) was recorded from the grafting on loquat rootstock on March 30th. On the contrary, Khushi *et al.* (2019) reported significant relationships between the number of leaves and the timing of grafting in mango. They found that the highest leaf count (22.63) was observed in plants grafted on April 20th, while the lowest count (21.15) occurred for grafting on March 10th. Additionally, there were significant differences in leaf area and leaf chlorophyll content of grafting at different dates on loquat and quince rootstocks. The loquat grafted on quince rootstock on February 20th had the greatest values of leaf area (84.90 cm²), and leaf chlorophyll content (45.51 SPAD), while leaf area was severely diminished to the lowest level (61.29 cm²) in the leaves of the loquat grafted on quince rootstock on March 30th. However, the lowest chlorophyll content (34.37 SPAD) was found from grafting the loquat rootstock on March 10th. observed differences likely arise from environmental conditions, root establishment, nutrient availability, light exposure, genetics, and care practices.

**Table 4.16 The interaction between rootstock type and grafting dates on the growth of loquat and quince rootstocks.**

| Rootstock type | Grafting date | Graft shoot length (cm) | Graft shoot diameter (mm) | Shoot fresh weight (g) | Shoot dry weight (g) | Number of leaves per budling | Leaf area (cm²) | Leaf chlorophyll content (SPAD) |
|---|---|---|---|---|---|---|---|---|
| Loquat | February 20 | 12.00 b | 4.27 a | 10.29 ab | 6.59 a | 9.11 a | 77.61 ab | 36.75 bc |
| | March 10 | 8.44 c | 3.77 a | 8.97 b | 5.28 b | 7.00 b | 67.39 bc | 34.37 c |
| | March 30 | 14.78 a | 4.13 a | 11.69 a | 6.70 a | 6.89 b | 64.95 c | 39.33 b |
| Quince | February 20 | 9.56 c | 3.93 a | 6.80 c | 3.76 c | 9.07 a | 84.90 a | 45.51 a |
| | March 10 | 6.59 d | 4.39 a | 4.83 d | 2.36 d | 8.44 a | 63.30 c | 42.43 a |
| | March 30 | 8.11 cd | 4.42 a | 5.94 cd | 3.94 c | 7.89 ab | 61.29 c | 43.51 a |

* The values in each column with the same letter do not differ significantly ($P \leq 0.05$), according to Duncan's Multiple Range Test.

## 4.4 Experiment 4: The Effect of Grafting Dates and Stock Types on Grafting Success of Loquat and Quince Tree Stocks

### 4.4.1 Grafting success percentage

Figure (4.7) illustrates the effect of loquat grafting on the grafting success percentage of loquat and quince tree stocks. The results indicate that there were no significant differences in grafting success between the two types of treestocks (loquat and quince). It was determined that loquat treestocks had a higher graft success percentage than quince treestocks. The highest graft success percentage was observed in the loquat treestock (73.33%). Whereas, the lowest graft success percentage was obtained in the quince (66.67%) treestock. These may be due to the compatibility between loquat and quince because they are the same. Despite the difference in graft success percentage, the lack of significant differences between the two types of tree stocks suggests that both quince and loquat tree stocks were equally suitable for grafting loquat scions. These results are in line with Somkuwar *et al.* (2015) who reported differences in graft success due to the use of grape rootstocks. The success of grafting is greatly influenced by immediate temperature conditions. Callus tissue formation is also crucial for graft development which relies on suitable environmental factors, especially temperature and humidity (Öztürk, 2021) some pear cultivars make the air temperature in the first 15 days after grafting a direct factor in graft success.

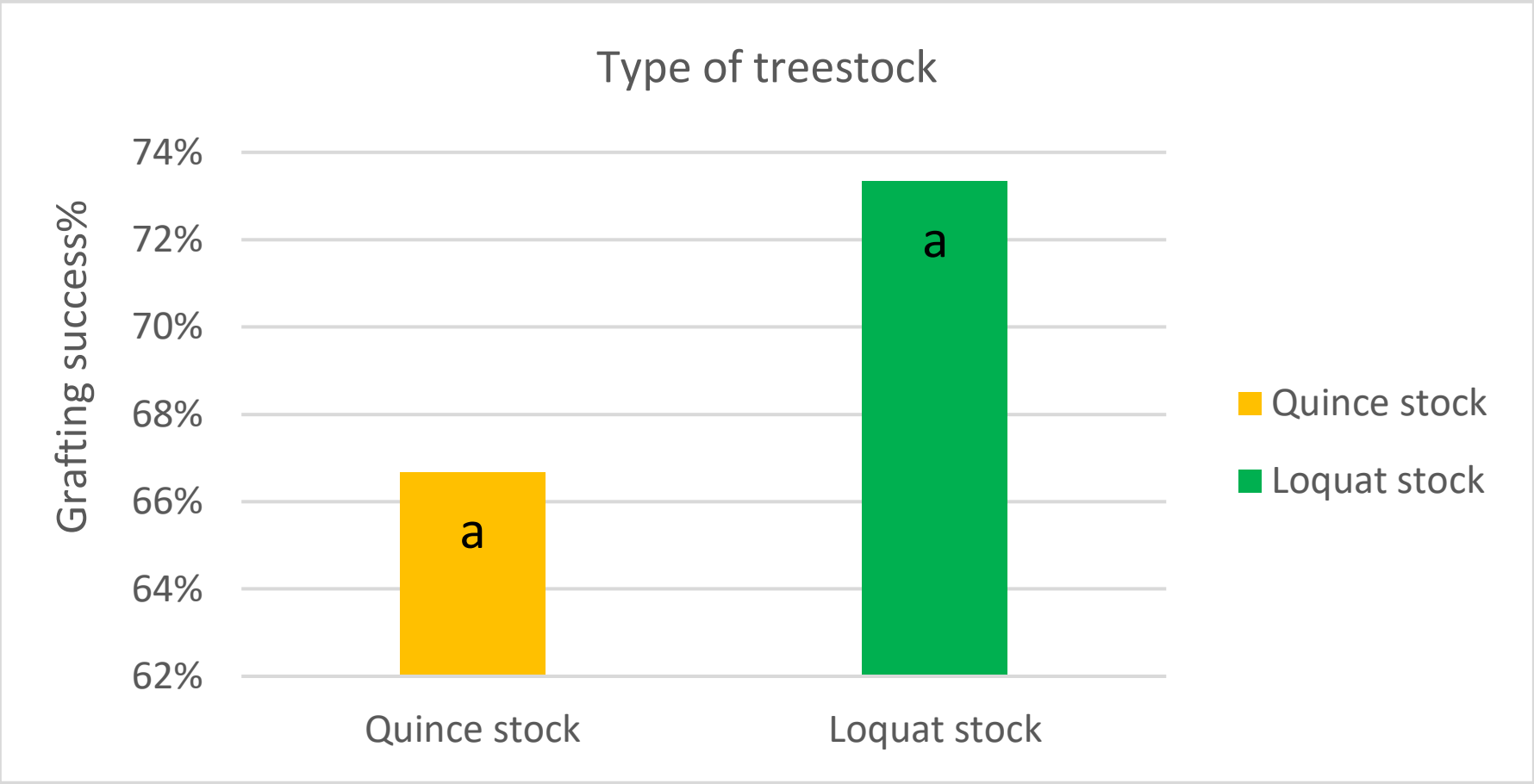

**Figure 4.7 The effect of treestock types on grafting success percentage in loquat grafting.** Values that do not share the same superscripts (a) differ significantly (P≤0.05).

Figure (4.8) demonstrates that the effect of grafting date on grafts for both loquat and quince tree stocks did not result in any significant differences in grafting success percentages. The grafting success percentages for all three grafting dates (February 20, March 10, and March 30) are identical at (70%). The lack of significant differences in grafting success percentage across

all three grafting dates may be due to the experimental conditions being consistent and well-controlled throughout the study. Additionally, the tree stocks used for both loquat and quince may have exhibited similar responsiveness to grafting during these specific periods, resulting in comparable graft success rates. Various factors impact graft success, including ecological, physiological, morphological, and genetic aspects. Temperature, humidity, rootstock growth stage, scion collection time, grafting technique, and expertise also matter. Graft failure or low success can result from graft incompatibility (Hartmann *et al*., 2014). Results obtained by Pektaş *et al.* (2009) who have studied the effects of different grafting techniques on grafting successes in peaches, almonds, apricots, pears, and apples and reported that winter grafts can be very successfully used in apples.

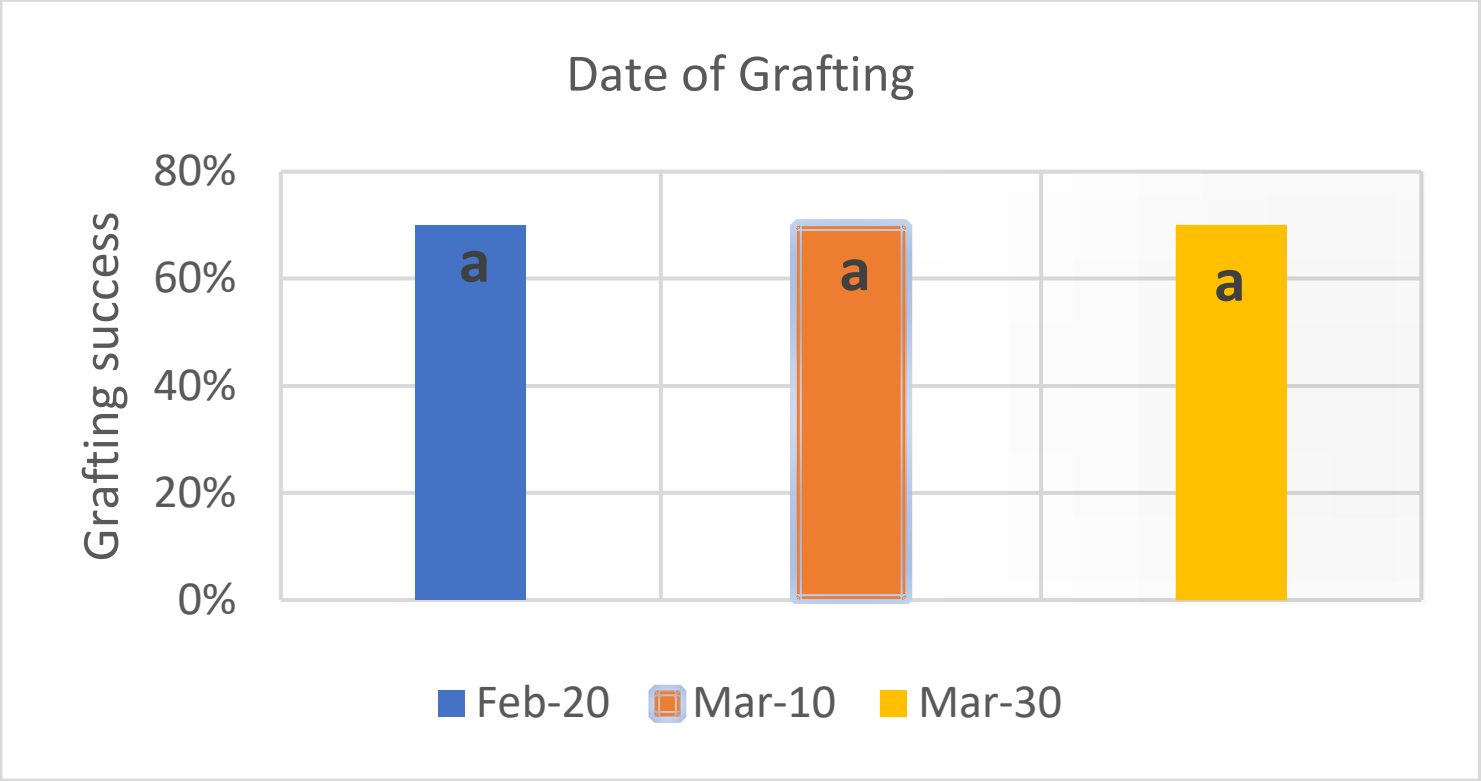

**Figure 4.8 The effect of different grafting dates on grafting success percentage in loquat and quince tree stocks.** Values that do not share the same superscripts (a) differ significantly (P≤0.05).

The study investigated the interaction effects of tree stock type and grafting dates on the grafting success percentage of loquat and quince tree stocks (Figure 4.9). The results showed that there were no significant differences between the two types of tree stocks and grafting dates. Although, the highest grafting success percentage (73.33%) was observed when using the combination of loquat tree stock and three grafting dates (February 20, March 10, and March 30). These findings conform with the findings of Singh *et al.* (2019b) who obtained better success under polyhouse conditions in walnuts. In contrast, the lowest grafting success percentage (66.67%) was achieved when using the interaction of quince tree stock and three grafting dates (February 20, March 10, and March 30). The difference in grafting success rates could stem from tree stock characteristics, physiological responses, or grafting process factors.

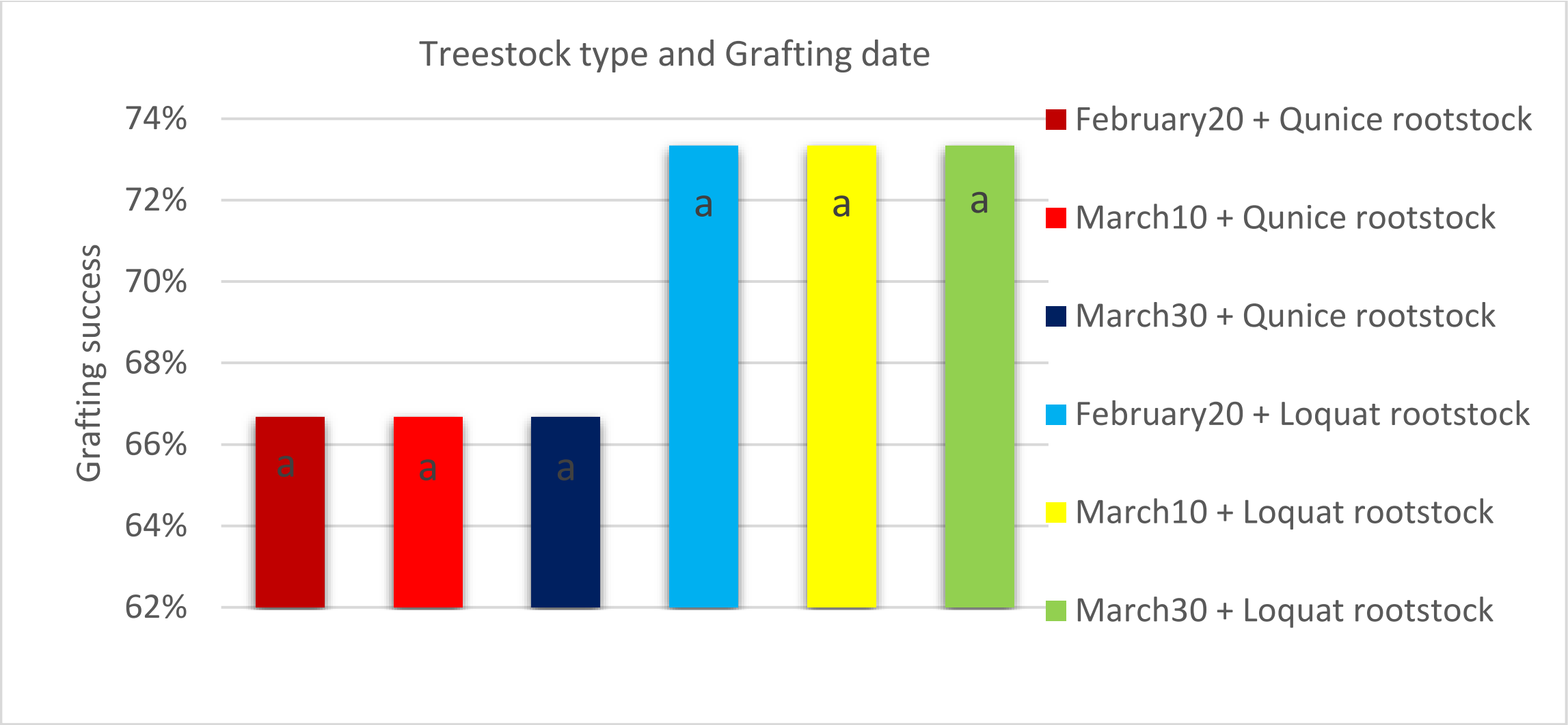

**Figure 4.9 The interaction between treestock types and different grafting dates on grafting success percentage in loquat grafting.** Values that do not share the same superscripts (a,) differ significantly (P≤0.05).

### 4.4.2 Shoot characteristics

Data shown in Table (4.17) compares the effects of two different tree stocks (loquat and quince) on various parameters related to grafting success. The data includes measurements of shoot length, shoot diameter, shoot fresh weight, shoot dry weight, leaf number, leaf area, and chlorophyll content for each tree stock. The data shows that the loquat tree stock generally exhibits better performance compared to the quince tree stock. For most of the parameters like shoot length, shoot diameter, shoot fresh weight, shoot dry weight, leaf number, and leaf area. the loquat tree stock shows higher values, suggesting better growth and development in terms of grafting success. However, interestingly, the quince tree stock has a slightly higher chlorophyll content compared to the loquat tree stock. Typically, the loquat tree stock resulted in longer shoots (16.11 cm) compared to the quince tree stock (11.85 cm). This indicates that the loquat tree stock may promote better shoot growth during grafting which led to a significant increase in shoot fresh weight, shoot dry weight leaf number, and leaf area. Both tree stocks show similar shoot diameters, with loquat tree stock slightly larger (5.26 mm) than quince tree stock (5.09 mm). The difference is not significant, but a larger diameter can contribute to a more robust shoot. Typically, loquat tree stock resulted in heavier shoots (14.85 g) compared to quince tree stock (8.49 g) indicate better overall growth and water uptake capacity. The obtained graft shoot length could be due to the probable differences in genetic differences, ecological and growing conditions (Pektaş *et al.*, 2009 and Hartmann *et al.*, 2014). Correspondingly, the loquat tree stock outperforms the quince tree stock in terms of shoot dry weight (7.20 g vs. 4.33 g). This parameter is essential as it reflects the actual biomass and growth potential of the graft. Also, the number of leaves on the grafted plant is another important indicator of growth. Loquat tree stock produced more leaves (12.33) compared to quince tree stock (9.85), suggesting better leaf development and more efficient photosynthesis. Equally, the leaf area represents the total surface area of the leaves. The loquat tree stock has a larger leaf area (108.38 $cm^2$) compared to quince tree stock (92.69 $cm^2$), which indicates better leaf expansion and potential for higher photosynthetic activity. Similarly, De Souza *et al.* (2015) reported that the leaf area of grapevines, cv. Cabernet Sauvignon was affected by the different rootstocks. Likewise, chlorophyll is a pigment responsible for photosynthesis, and its content is an important indicator of the plant's ability to capture and use light for growth. Surprisingly, the quince tree stock showed a higher chlorophyll content (41.28 SPAD) compared to the loquat tree stock (39.17 SPAD). This could suggest that despite the better growth in other aspects, the loquat tree stock might have slightly lower chlorophyll content, potentially affecting its overall photosynthetic capacity. This suggests that loquat tree stock may be more favorable for achieving better grafting success and overall growth of the grafted plants. These results are contrary to the findings of Çetinbaş *et al.* (2018) who reported the effect of the rootstocks and cultivars on the graft shoot diameter and found no statistical significant differences in pear.

**Table 4.17 The effect of tree stock types on growth parameters in loquat grafting on loquat and quince treestocks.**

| Tree stock type | Graft shoot length (cm) | Graft shoot diameter (mm) | Shoot fresh weigh (g) | Shoot dry weight (g) | Leaf number per budling | Leaf area ($cm^2$) | Leaf chlorophyll content (SPAD) |
|---|---|---|---|---|---|---|---|
| Loquat | 16.11 a | 5.26 a | 14.85 a | 7.20 a | 12.33 a | 108.38 a | 39.17 b |
| Quince | 11.85 b | 5.09 a | 8.49 b | 4.33 b | 9.85 b | 92.69 b | 41.28 a |

* The values in each column with the same letter do not differ significantly (P≤0.05), according to Duncan's Multiple Range Test.

The impact of grafting dates on two types of tree stocks, namely loquat, and quince, was investigated across various growth parameters. The results revealed significant disparities in shoot length, shoot diameter, leaf area, and chlorophyll content. However, no significant differences were observed in shoot fresh weight, shoot dry weight, and leaf number (as indicated in (Table 4.18). This is despite that loquat grafted on tree stock types showed the longest shoot length (16.55 cm) on March 10, followed by those on March 30 (13.89 cm), and the shortest shoot length was observed on February 20 (11.50 cm). These differences may arise from the interplay of grafting timing, physiological responses, and their impact on growth and development. These findings are similar to those of Kudmulwar *et al.* (2008) who reported that in custard apple, the grafting achieved on February 15$^{th}$ recorded the highest length of scion. Mir and Kumar, (2011) reported the highest length of scion when grafting was done on the 3$^{rd}$ week of February which is in contrast with the findings of the present study. Similarly, Wani *et al*. (2017) reported a higher length of shoot when grafting ended on the 20$^{th}$ and 30$^{th}$ of January. Though, the shoot diameter of the grafted trees is quite consistent among the different grafting dates. When grafted on February 20 and March 10 had similar shoot diameters (5.69 and 5.44 mm) respectively, while those on March 30 had a slightly smaller shoot diameter (4.41 mm). The consistent shoot diameters for grafting on February 20 and March 10 may be due to favorable growth conditions. On the other hand, the slightly smaller shoot diameter on March 30 could be attributed to different environmental factors or tree development stages. Similar to our experiment, Mehta *et al.* (2018) reported a higher diameter of the shoot when grafting was done during the first week of March in pecan. Correspondingly, the grafted loquat on March 10 had the highest average shoot fresh and dry weights (14.20 and 6.23 g), followed by March 30 (11.24 and 5.98 g), and February 20 (9.57 and 5.09 g), respectively. The differences in shoot weight among grafting dates may be due to varying environmental conditions and growth stages. Likewise, the loquat grafted on February 20 had the largest average leaf area (111.64 $cm^2$), followed by March 10 (104.91 $cm^2$), and March 30 (85.05 $cm^2$). Also, this indicates that the maximum leaf numbers; (11.33) for February 20, (11.56) for March 10, and a minimum of (10.39) leaves for March 30, maybe due to several factors, including environmental conditions, growth stages, and genetic differences. The rise in leaf numbers during the grafting may be due to the successful grafts, which increased the number of shoots led to an increase number of leaf and area, enhanced photosynthesis efficiency, and nutrient production for grafted seedling growth (Seletsu *et al.*, 2011). However, the highest average chlorophyll content (44.05 SPAD) was recorded on February 20, followed by March 30 (40.08 SPAD), and March 10 (36.56 SPAD). These observed variations might be attributed to a combination of factors related to the interaction between grafting time and the physiological responses of the trees, affecting growth and development.

**Table 4.18 The effect of different grafting dates on growth parameters in loquat grafting on loquat and quince treestocks.**

| Grafting date | Graft shoot length (cm) | Graft shoot diameter (mm) | Shoot fresh weight (g) | Shoot dry weight (g) | Number of leaves per budling | Leaf area (cm²) | Leaf chlorophyll content (SPAD) |
|---|---|---|---|---|---|---|---|
| February 20 | 11.50 c | 5.69 a | 9.57 b | 5.09 a | 11.33 a | 111.64 a | 44.05 a |
| March 10 | 16.55 a | 5.44 a | 14.20 a | 6.23 a | 11.56 a | 104.91 a | 36.56 b |
| March 30 | 13.89 b | 4.41 b | 11.24 ab | 5.98 a | 10.39 a | 85.05 b | 40.08 b |

* The values in each column with the same letter do not differ significantly ($P \leq 0.05$), according to Duncan's Multiple Range Test.

Data presented in Table (4.19) illustrate that the combination of the two factors, tree stock types (loquat and quince), and different grafting dates (February 20, March 10, and March 30), had a significant impact on the various characteristics of shoot growth, in which the successful loquat grafting on loquat tree stock on March 10, the highest significant shoot length (23.22 cm) was resulted. Whereas, the combination of quince tree stock and March 10 sharply reduced shoot length (9.89 cm). This may be due to the interaction between the genetic compatibility of the tree stock and scion (loquat), and the environmental conditions during the grafting process. Also, the study found significant differences between the interaction of tree stock type and grafting dates on shoot diameter, shoot fresh weight, and leaf area. The highest shoot diameter and leaf area were (6.19 mm, and 127.49 cm²) respectively, which were recorded when grafting on loquat tree stock on February 20 was done. Enhanced growth might be due to favorable conditions promoting effective nutrient absorption and photosynthesis, leading to better shoot development. In contrast, the lowest shoot diameter (3.96 mm), and leaf area (79.61 cm²) were noted when using the combination of loquat tree stock and grafting on March 30. This shift could be traced to poor environmental conditions or a growth pattern mismatch between scion and tree stock, possibly hampering nutrient uptake and growth. However, the highest chlorophyll content (44.35 SPAD) was observed on quince trees tock when grafted on February 20, But the lowest chlorophyll content (31.25 SPAD) was recorded from the interaction of loquat tree stock and grafting on March 10. Interestingly, the highest leaf number (13.33) was achieved with loquat tree stock and grafting on March 10, while the lowest number (9.22) occurred when grafting quince tree stock on March 30. Furthermore, there were significant differences in shoot fresh weight and shoot dry weight of grafting from the interaction between tree stock types and different times on loquat and quince tree stocks. The maximum values for shoot fresh weight (19.58 g) and shoot dry weight (8.60 g) were documented when grafting onto loquat tree stock on March 10. Conversely, the minimum shoot fresh weight (6.55 g) and dry weight (3.46 g) were recorded when grafting quince tree stock on February 20.

**Table 4.19 The interaction between treestock types and grafting dates on growth parameters in loquat grafting.**

| Tree stock type | Grafting date | Graft shoot length (cm) | Graft shoot diameter (mm) | Shoot fresh weight (g) | Shoot dry weight (g) | Number of leaves per budling | Leaf area (cm²) | Leaf chlorophyll content (SPAD) |
|---|---|---|---|---|---|---|---|---|
| Loquat | February-20 | 10.34 cd | 6.19 a | 12.59 b | 6.71 ab | 12.11 ab | 127.49 a | 43.75 a |
| | March-10 | 23.22 a | 5.63 ab | 19.58 a | 8.60 a | 13.33 a | 118.02 a | 31.25 c |
| | March-30 | 14.78 b | 3.96 d | 12.39 b | 6.28 ab | 11.56 a-c | 79.61 c | 42.52 ab |
| Quince | February-20 | 12.67b-d | 5.18 bc | 6.55 b | 3.46 b | 10.56 b-c | 95.79 b | 44.35 a |
| | March-10 | 9.89 d | 5.24 bc | 8.82 b | 3.85 b | 9.78 b-c | 91.79 bc | 41.86 ab |
| | March-30 | 13.00 bc | 4.85 c | 10.10 b | 5.68 ab | 9.22 c | 90.48 bc | 37.63 b |

* The values in each column with the same letter do not differ significantly (P≤0.05), according to Duncan's Multiple Range Test.

## 4.5 Experiment 5: The Impact of Cutting Types and Grafting Dates on Graft Bud Sprout and Rooting Percentage in Loquat Bench Grafting

Table (4.20) illustrates the effect of loquat bench grafting on the percentage of graft bud sprouts for two types of loquat stock cuttings namely loquat and quince. It was revealed that the highest value of graft bud sprouts (86.11%) was observed when loquat was grafted onto loquat stock cuttings at 60 days after grafting (Dag). While the lowest value of graft bud sprouts (59.72%) was recorded when loquat was grafted onto quince stock cuttings at the same 60-day interval. This suggests that loquat cuttings consistently exhibit a significant advantage over quince cuttings at all observed time points. As well as, the results indicate that loquat cuttings consistently have higher graft bud sprout percentages compared to quince cuttings across the 20-, 40-, and 60-days periods.

**Table 4.20 The effect of cutting types on graft bud sprout percentage in loquat bench grafting.**

| Cutting stock type | Graft Bud sprout % | | |
|---|---|---|---|
| | 20 Dag | 40 Dag | 60 Dag |
| Loquat | 52.78 a | 75.00 a | 86.11 a |
| Quince | 26.39 b | 43.06 b | 59.72 b |

The values in each column with the same letter do not differ significantly (P≤0.05) according to Duncan's Multiple Range Test.

The study demonstrates the impact of loquat bench grafting on graft bud sprout percentages at different grafting dates February 20, March 10, and March 30 for both loquat and quince stock cuttings (Table 4.21). The results consistently show an increase in grafting success from February 20, with percentages rising from (62.50%) at 20 days to (79.17%) at 40 days and further to (89.58%) at 60 days. This suggests a robust and continuous bud-sprouting pattern, indicating no significant differences between percentages at each time point, but a consistent overall increase. The results are in line with the findings of Sweeti *et al.* (2016) who reported that guava plants grafted on February 20 achieved the highest graft success percentage, as recorded 90 days after grafting, compared to those grafted on March 5. Similarly, Jaipal *et al.* (2021) also observed the early sprouting in peach scions grafted at different time intervals as time progressed from January to February. On the contrary, assessing grafting on March 10 reveals a rising tendency, yet percentages are notably lower than those of February 20. Specifically, at 20 days, the percentage is (33.33%), increasing to (60.42%) at 40 days and (70.83%) at 60 days. The lower values compared to February 20 suggest a slower or less successful bud-sprouting process during this period. By contrast, on March 30, bud sprout percentages were the lowest among the three grafting dates, indicating a less favorable outcome. The data unveils different patterns, with February 20 displaying the highest and most consistent rates, March 10 showing intermediate success, and March 30 yielding comparatively lower and slower bud-sprouting percentages. The earlier bud burst in plants grafted on February 20 could be attributed to the early stimulation of the division of phloem ray cells, xylem ray cells, xylem parenchymal cells, and parenchymal cells between periderm and phloem, as reported by Hartmann *et al.* (2002). This stimulation resulted in earlier callus formation. On the other hand, the delayed bud burst in plants grafted on March 30 could be attributed to poor temperature and humidity conditions, as suggested by (Majd *et al.*, 2019).

**Table 4.21 The effect of different grafting dates on graft bud sprout percentage in loquat bench grafting on loquat and quince cuttings.**

| Date of Grafting | Graft bud sprout % | | |
|---|---|---|---|
| | 20 Dag | 40 Dag | 60 Dag |
| February 20 | 62.50 a | 79.17 a | 89.58 a |
| March 10 | 33.33 b | 60.42 b | 70.83 b |
| March 30 | 22.92 c | 37.50 c | 58.33 c |

The values in each column with the same letter do not differ significantly ($P \leq 0.05$) according to Duncan's Multiple Range Test.

The data shown in Table (4.22) demonstrate the impact of loquat bench grafting on two stock cutting types (loquat and quince) and different dates (February 20, March 10, and March 30) on graft bud sprout percentage. It is clear from the table that loquat grafted from February 20 on loquat stock cuttings showed a consistent increase in graft bud sprout percentage from (75.00%) at 20 days to (91.67%) at 60 days. However, the graft bud sprout percentage ranged from (54.17%) to (95.83%) on March 10, indicating a successful and significant development in graft bud sprout. In contrast, the lower graft bud sprout percentage on loquat stock cuttings was recorded on March 30, which is possibly due to varying environmental conditions. On the other hand, when loquat grafted from February 20, the graft bud sprout percentage on quince stock cuttings increased from (50.00%) at 20 days to (87.50%) at 60 days. The data indicates a vigorous and significant development in graft bud sprout percentage. Conversely, loquat bench grafting on quince stock cuttings on March 10 and March 30 exhibited lower percentages at each time point, hinting at a less successful bud-sprouting process. This might be attributed to temperature differences or other environmental conditions during grafting.

**Table 4.22 The interaction between cutting types and different grafting dates on the graft bud sprout percentage in loquat bench grafting.**

| Cutting type | Date of grafting | Graft bud sprout % | | |
|---|---|---|---|---|
| | | 20 Dag | 40 Dag | 60 Dag |
| Loquat | February 20 | 75.00 a | 87.50 a | 91.67 a |
| | March 10 | 54.17 b | 91.67 a | 95.83 a |
| | March 30 | 29.17 c | 45.83 c | 70.83 b |
| Quince | February 20 | 50.00 b | 70.83 b | 87.50 a |
| | March 10 | 12.50 d | 29.17 d | 45.83 c |
| | March 30 | 16.67 d | 29.17 d | 45.83 c |

The values in each column with the same letter do not differ significantly ($P \leq 0.05$) according to Duncan's Multiple Range Test.

Table (4.23) presents data on the impact of stock cutting type on leaf numbers at different intervals (20, 40, and 60 days) and various parameters during grafting in loquat bench grafting on two stock cutting types namely (loquat and quince). The results showed that loquat scion grafted on quince stock cuttings exhibit a leaf number of (0.96) at 20 days after grafting, slightly lower than loquat scion grafted on loquat stock cuttings (1.15). Both cutting types display an increase in leaf numbers, with loquat reaching (2.00) leaves and quince achieving (1.59) leaves after 40 days of grafting. While after 60 days of grafting, both loquat and quince showed a further increase in leaf numbers, with quince at (2.26) leaves and loquat at the highest count of (2.70) leaves. The data suggests a continuous rise in leaf numbers over time for both cutting types in loquat bench grafting. Notably, loquat consistently demonstrated higher leaf numbers than quince at each observation interval, indicating more vigorous foliage growth. These differences lead to the highest carbohydrate help to a more active metabolic state in loquat, potentially contributing to enhanced foliage growth. Furthermore, the data delves into biochemical parameters such as carbohydrate percentage, nitrogen percentage, and phenol content. The highest significant percentage of carbohydrates (7.44%) was recorded for loquat cuttings, while the lowest (5.38%) was observed for quince cuttings. Conversely, the nitrogen percentage in quince cuttings was slightly higher significant at (1.24%), compared to loquat

cuttings at (1.13%). Nitrogen percentage in quince cuttings was significantly above that of loquat cuttings. Notably, phenol content was markedly higher in quince cuttings at (1737.83 mg.L$^{-}$), contrasting with loquat cuttings at (825.67 mg.L$^{-}$). These results indicate potential differences in the metabolic activities and physiological responses of the two cutting types in the early stages of grafting. The higher carbohydrate percentage and lower phenol content in loquat may suggest a more active metabolic state, while the reversed nitrogen percentage trend may indicate distinct nutrient utilization patterns. The rooting percentage (0.00%) for both types of cuttings suggests that rooting has not been initiated at this early stage. Overall, the data underscores the intricate interplay of various factors influencing the biochemical and physiological aspects of the grafting process in loquat.

**Table 4.23 The effect of loquat bench grafting on leaf number, biochemical parameters, and rooting percentage on loquat and quince stock cuttings.**

| Cutting type | Leaf number | | | Carbohydrate% | Nitrogen% | Phenols (mg.L$^{-}$) | Rooting% |
|---|---|---|---|---|---|---|---|
| | 20 Dag | 40 Dag | 60 Dag | | | | |
| Loquat | 1.15 a | 2.00 a | 2.70 a | 7.44 a | 1.13 b | 825.67 b | 0.00 |
| Quince | 0.96 a | 1.59 b | 2.26 b | 5.38 b | 1.24 a | 1737.83 a | 0.00 |

The values in each column with the same letter do not differ significantly ($P \leq 0.05$) according to Duncan's Multiple Range Test.

Table (4.24) shows the effect of different grafting dates (February 20, March 10, and March 30) on leaf numbers, biochemical parameters, and rooting percentage in loquat bench grafting, considering two stock cutting types (loquat and quince) with observations recorded at (20, 40, and 60) days after grafting. As it is clear from the table, grafting on February 20 resulted in the highest leaf numbers across all intervals. At 20 days after grafting (Dag), the leaf number was (1.78), which increased to (2.56) at 40 days and further to (2.78) at 60 days. Similar results were observed by Thapa *et al.* (2021) who reported that grafting on the 19th of February showed significantly higher results for the number of leaves in Persian walnuts. While, the grafting on March 10, observed leaf numbers were moderate, with values of (0.72) at 20 days, (1.83) at 40 days, and (2.67) at 60 days after grafting. Notably, these leaf numbers were lower than those observed for February 20. Furthermore, grafting on March 30 produced the lowest leaf numbers among the three dates. At 20 days, the leaf number was (0.67), which increased to (1.00) at 40 days and (2.00) at 60 days. This observed variation may be attributed to several factors, including differences in climatic conditions, growth patterns associated with the specific grafting dates, or inherent characteristics of the loquat and quince cutting types.

In addition, the analysis of biochemical parameters across different grafting dates reveals notable trends. The highest significant carbohydrate percentage was consistently observed on both different grafting dates; February 20 and March 30 (6.58%). However, the lowest carbohydrate percentage (6.07%) was recorded when grafting on March 10. Also, the highest nitrogen percentage (1.23%) was achieved when grafting on March 30, while the lowest (1.14%) was recorded on February 20. Furthermore, the maximum phenol content at (1555.75 $mg.L^{-}$), is noted when grafting on March 10 and planted, whereas the minimum, at (1117.80 $mg.L^{-}$), was recorded when grafting on February 20 planted. These trends suggest that grafting timing critically influences the biochemical composition. Consistently high carbohydrate percentages on February 20 and March 30 indicate optimal conditions for synthesis, while variations in nitrogen and phenol content across dates likely result from complex interactions between environmental factors and physiological responses. This emphasizes the importance of selecting appropriate grafting dates to efficiently influence desired biochemical outcomes. The rooting percentages for all three grafting dates February 20, March 10, and March 30—are recorded as (0.00%). The absence of rooting across different grafting dates suggests potential challenges in the root initiation process. It's noteworthy to mention that the initiation of roots can be influenced by a range of endogenous and exogenous factors (Hartmann *et al.*, 2014). Endogenous auxins, the carbohydrate status of cuttings, mineral nutrient content—especially nitrogen—and other biochemical compounds play a crucial role in the formation of adventitious roots (Otiende *et al.*, 2017).

**Table 4.24 The effect of different dates in loquat bench grafting on leaf number, biochemical parameters, and rooting on loquat and quince cuttings.**

| Date of grafting | Leaf number | | | Carbohydrate % | Nitrogen% | Phenol ($mg.L^{-}$) | Rooting% |
|---|---|---|---|---|---|---|---|
| | 20 Dag | 40 Dag | 60 Dag | | | | |
| February 20 | 1.78 a | 2.56 a | 2.78 a | 6.58 a | 1.14 c | 1117.80 c | 0.00 |
| March 10 | 0.72 b | 1.83 b | 2.67 a | 6.07 b | 1.19 b | 1555.75 a | 0.00 |
| March 30 | 0.67 b | 1.00 c | 2.00 b | 6.58 a | 1.23 a | 1171.70 b | 0.00 |

The values in each column with the same letter do not differ significantly ($P \leq 0.05$) according to Duncan's Multiple Range Test.

Table (4.25) presents the interaction between stock cutting types (loquat and quince) and different grafting dates (February 20, March 10, and March 30) on leaf numbers, biochemical parameters, and rooting percentages at different intervals (20, 40, and 60) days after grafting (Dag) in loquat bench grafting. The data reveals distinct patterns in leaf development based on the combination of stock-cutting type and grafting date. For loquat bench grafting on loquat

stock cuttings, the highest leaf number was recorded on March 10 (3.11 at 60 days), while the lowest was on March 30 (0.44 at 20 days). Therefore, grafting the loquat on March 10 appears to be more conducive to achieve a greater leaf number. On the other hand, loquat scions bench grafted on quince stock cuttings, the highest leaf number was observed on February 20 (2.67 at 60 days), while the lowest was recorded on March 10 (0.33 at 20 days). This suggests that grafting quince on February 20 is favorable for achieving a higher leaf count. The variations in leaf numbers indicate the influence of both stock-cutting type and grafting date on the success of loquat bench grafting, emphasizing the importance of timing and selection of stock-cutting type in optimizing leaf development for improved grafting outcomes. Concurrently, the highest carbohydrate percentage recorded in loquat cuttings on February 20 (7.58%) suggesting that grafting on this date promotes a more favorable accumulation of carbohydrates, potentially contributing to enhanced growth and development. Conversely, the lowest carbohydrate percentage in quince cuttings on March 10 (4.91%) indicates a less optimal condition for carbohydrate synthesis and allocation. Nitrogen percentage exhibited a similar pattern, with quince cuttings on March 10 showing the highest nitrogen content (1.25%), potentially indicating improved nutrient assimilation during this grafting period. In contrast, the lowest nitrogen percentage in loquat cuttings on February 20 (1.05%) suggests a less efficient uptake of nitrogen at this specific time. The peak in phenol content in quince cuttings was observed on March 10 (2148.20 $mg.L^{-}$). This may potentially reflect a heightened response to stress or environmental conditions during that period, while the lowest phenol content in loquat cuttings on February 20 (713.20 $mg.L^{-}$) indicates a less pronounced stress response during this grafting timeframe. The consistently low rooting percentage (0.00%) across all conditions suggests a need for further investigation into factors influencing root development in loquat bench grafting. On the contrary, Solgi *et al.* (2022) reported that high amounts of starch, soluble sugars, and C/N ratio (carbohydrate/nitrogen) lead to increased grafting success. Overall, these findings emphasize the importance of considering both cutting type and grafting date for achieving optimal outcomes in loquat grafting, as they influence not only leaf development but also key physiological and biochemical aspects of the grafted plants.

**Table 4.25 The interaction between cutting types and different dates on leaf number, biochemical parameters, and rooting percentage in loquat bench grafting.**

| Cutting type | Date of grafting | Leaf number | | | Carbohydrate % | Nitrogen % | Phenols ($mg.L^{-}$) | Rooting % |
|---|---|---|---|---|---|---|---|---|
| | | 20 Dag | 40 Dag | 60 Dag | | | | |
| Loquat | February-20 | 1.89 a | 2.78 a | 2.89 ab | 7.58 a | 1.05 c | 713.20 f | 0.00 |
| | March-10 | 1.11 b | 2.44 a | 3.11 a | 7.23 b | 1.12 b | 963.30 d | 0.00 |
| | March-30 | 0.44 c | 0.78 b | 2.11 c | 7.49 a | 1.23 a | 800.50 e | 0.00 |
| Quince | February-20 | 1.67 a | 2.33 a | 2.67 b | 5.58 c | 1.23 a | 1522.40c | 0.00 |
| | March-10 | 0.33 c | 1.22 b | 2.22 c | 4.91 d | 1.25 a | 2148.20a | 0.00 |
| | March-30 | 0.89 b | 1.22 b | 1.89 c | 5.66 c | 1.23 a | 1542.90b | 0.00 |

The values in each column with the same letter do not differ significantly ($P \leq 0.05$) according to Duncan's Multiple Range Test.

# CONCLUSIONS AND RECOMMENDATIONS

**A. Conclusions:** Depending on the most important results obtained from all experiments of this study, the following points of view can be mainly expressed as conclusions:

Experiment 1. Rooting of quince hardwood cuttings:

1. Rooting success was influenced by both the concentrations of IBA and the selection of rooting media. However, the control group (without IBA) notably enhanced rooting when compared to the various IBA concentrations.

2. Cuttings in the control group (without IBA) and those planted in river sand exhibited notably high percentages of successful rooting, underscoring the importance of the selected planting medium.

Experiment 2. Bench grafting of loquat:

1. The success of grafting loquat cutting stocks varied based on grafting dates, types of cuttings, and concentrations of IBA. However, IBA at different concentrations did not have a significant impact.

2. Notably, certain interactions such as grafting on February 20 with loquat stock cuttings, yielded higher percentages of successful graft bud sprouting.

Experiment 3. Performance of grafting loquats onto different rootstocks:

1. Grafting success was notably influenced by the selection of rootstock, with loquat rootstock demonstrating superior performance compared to quince.

2. The highest significant levels of successful grafting were attained on February 20, underscoring the crucial role of grafting dates.

Experiment 4. Impact of tree stock types on grafting success:

1. Grafting success percentage was higher in loquat tree stock when compared to quince.

2. The consistency of grafting success percentages across three dates underscores the significant influence of rootstock type.

Experiment 5. Bench grafting of loquat cutting stocks:

1. Graft bud sprout percentages exhibited variations, with loquat stock cuttings surpassing quince.

2. Grafting success demonstrated a consistent increase from February 20 to March 30, underscoring the importance of selecting appropriate grafting dates.

**B. Recommendations:** Based on the comprehensive findings of the research, the following recommendations are proposed:

1. Improvement of Grafting Techniques: Continued efforts should be directed towards refining grafting techniques, taking into account factors such as IBA concentrations, grafting dates, and stock types. Fine-tuning these variables can significantly contribute to enhanced overall success in loquat bench grafting.
2. Optimal Rooting Substrates: Considering the results of Experiment 1, where river sand and river sand/peat moss proved outstanding for quince cuttings, future grafting endeavors should prioritize these substrates to maximize rooting success.
3. Selective IBA Application: As emphasized in Experiment 2, the impact of IBA concentrations on bud sprout percentages varied. Therefore, a more nuanced approach to IBA application is recommended, with careful consideration of grafting dates and stock types.
4. Rootstock Selection: Experiment 3 demonstrated significantly higher grafting success with loquat rootstock. Future grafting projects should prioritize the use of loquat rootstock for improved success rates.
5. Grafting Date Significance: Experiment 5 emphasized the critical role of grafting date selection, with February 20 yielding the highest carbohydrate percentages. Researchers and practitioners should consider this information when planning loquat bench grafting activities.
6. Further Exploration in Quince Grafting: Given the consistently low rooting percentages in quince stocks, additional research is recommended to explore and optimize grafting techniques specifically for quince, addressing the challenges identified in the study.
7. Continuous Monitoring and Adjustment: Grafting success is a dynamic process influenced by various factors. Continuous monitoring of environmental conditions and adapting grafting practices accordingly is crucial for sustained success.
8. Knowledge Sharing: The results of this research provide valuable insights into loquat bench grafting. To advance the field, it is recommended to share these findings with the scientific community through publications and presentations, fostering collaboration and collective advancement in grafting practices.

# APPENDICES

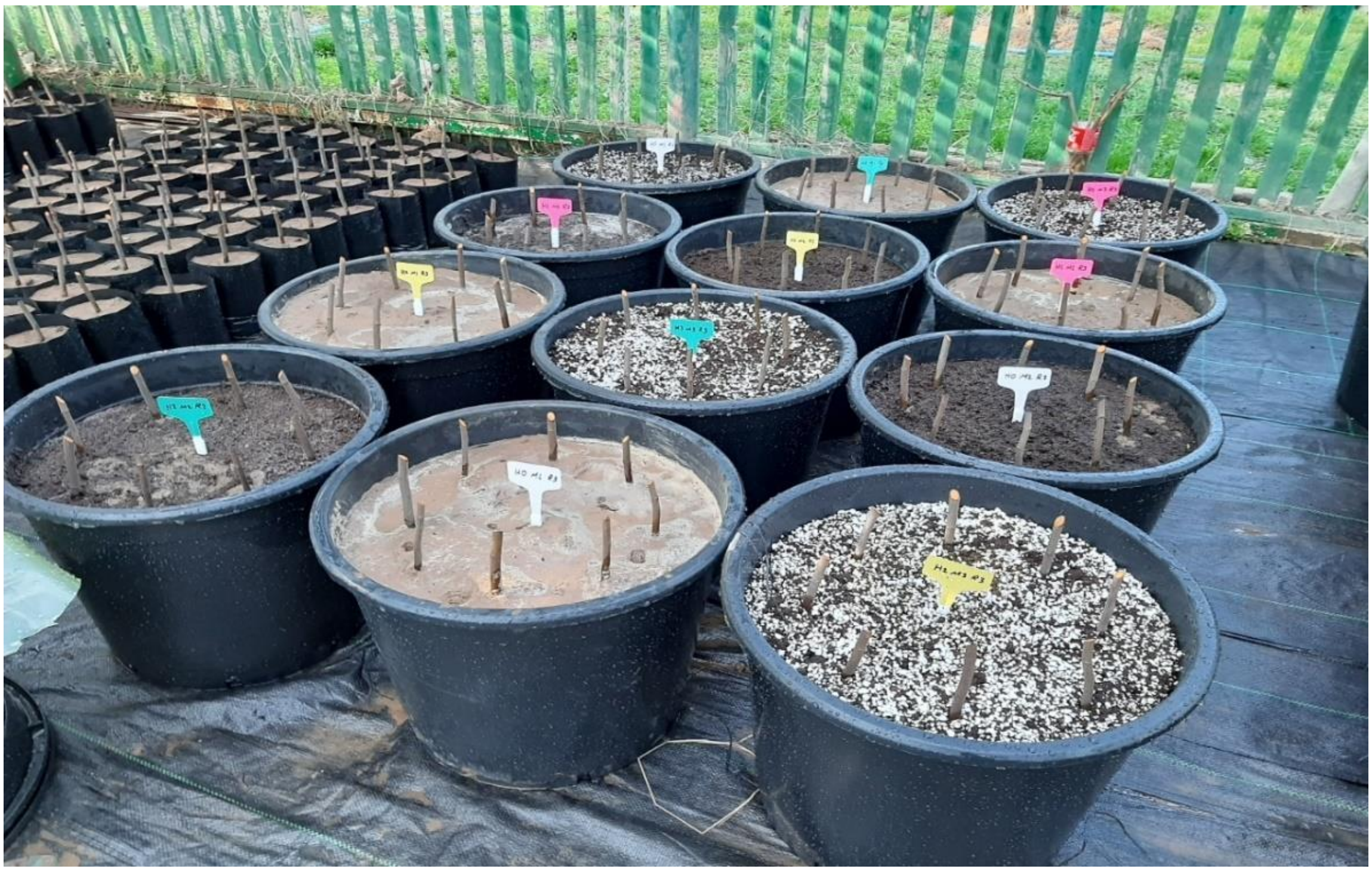

**Appendix 1.** The planting quince cuttings in various rooting media within the lathhouse.

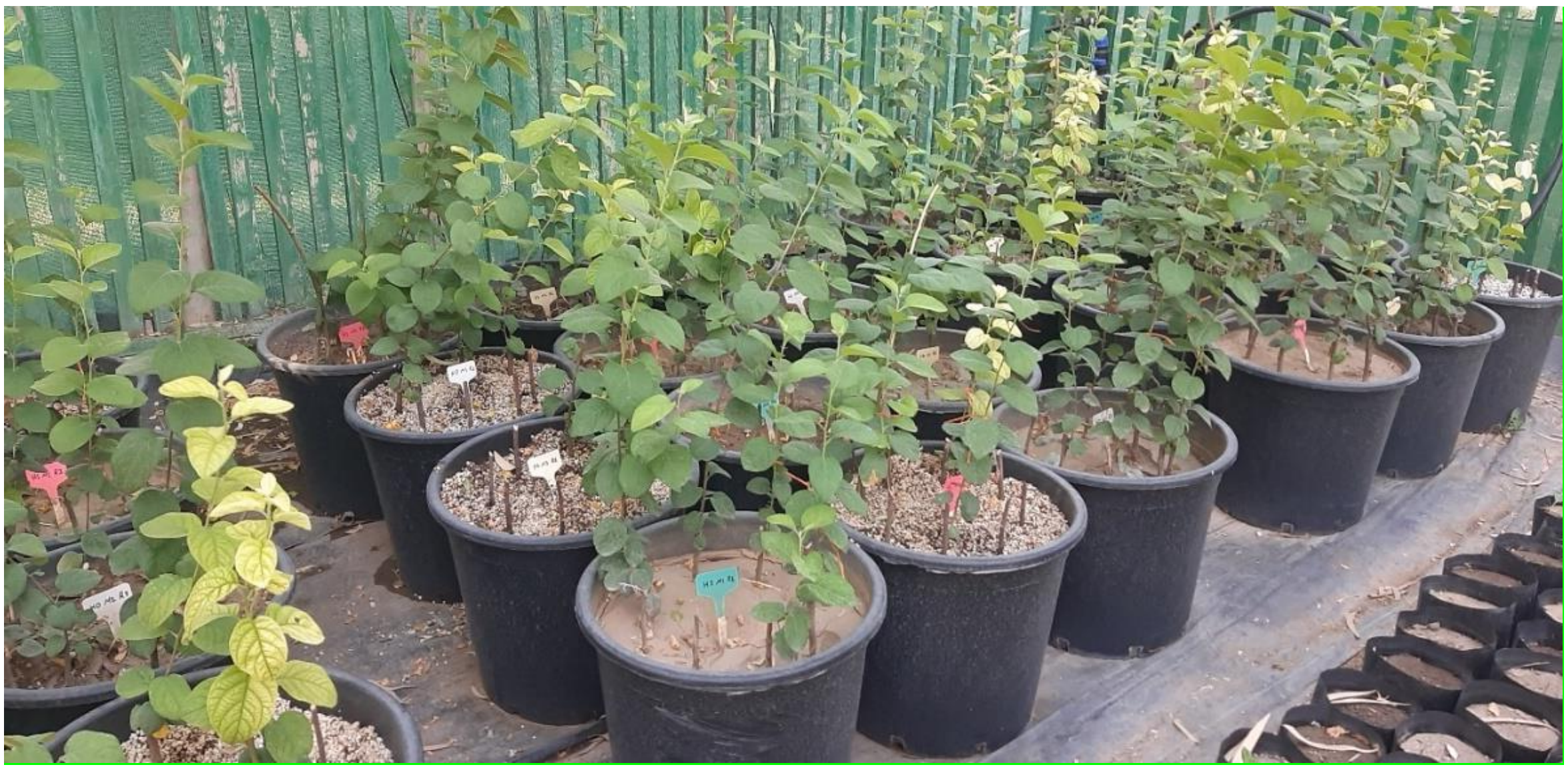

**Appendix 2.** The success of planting quince cuttings in varied rooting media within the lath house.

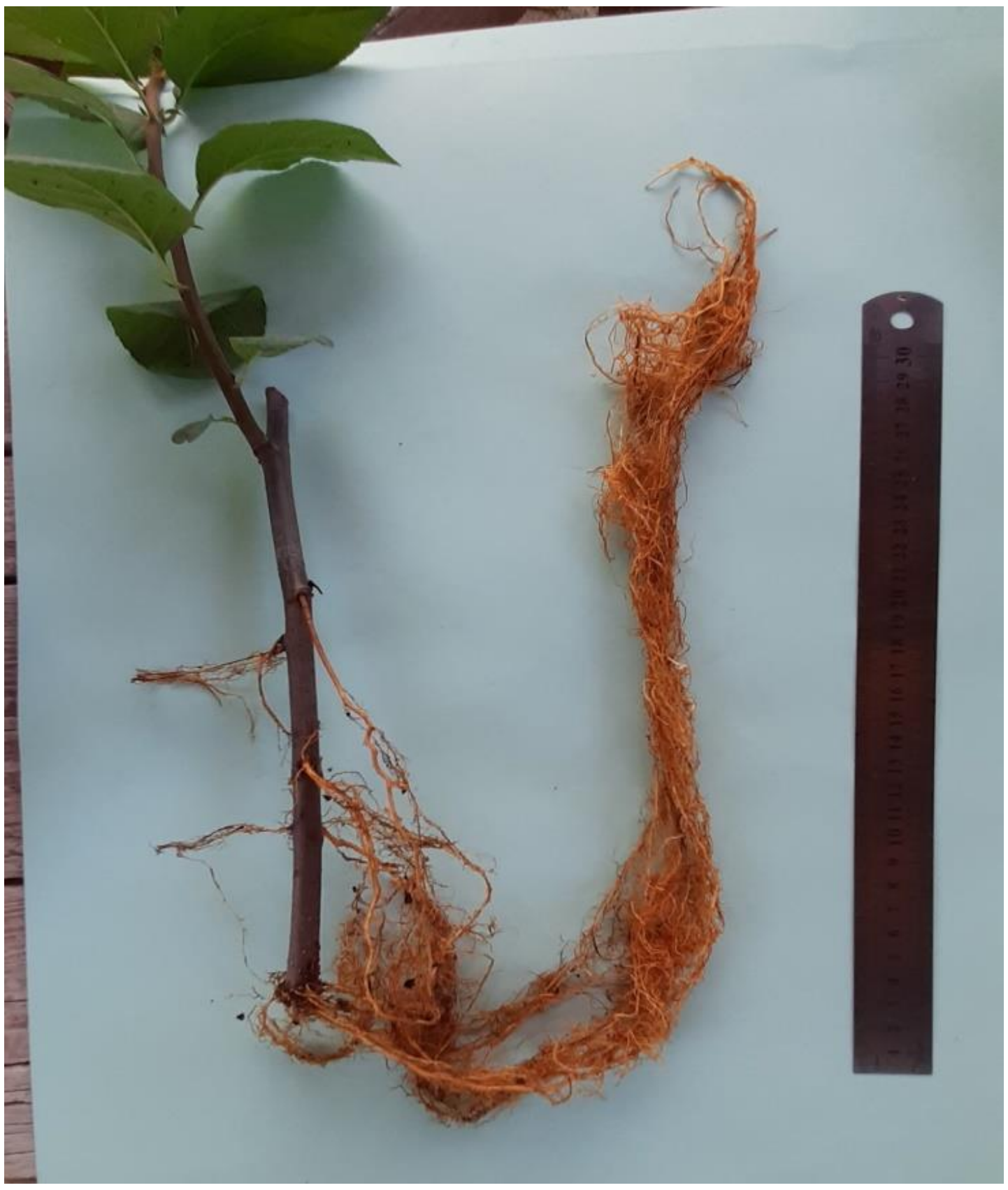

**Appendix 3.** The successful rooting of quince cuttings planted in different rooting substrate within the lath house.

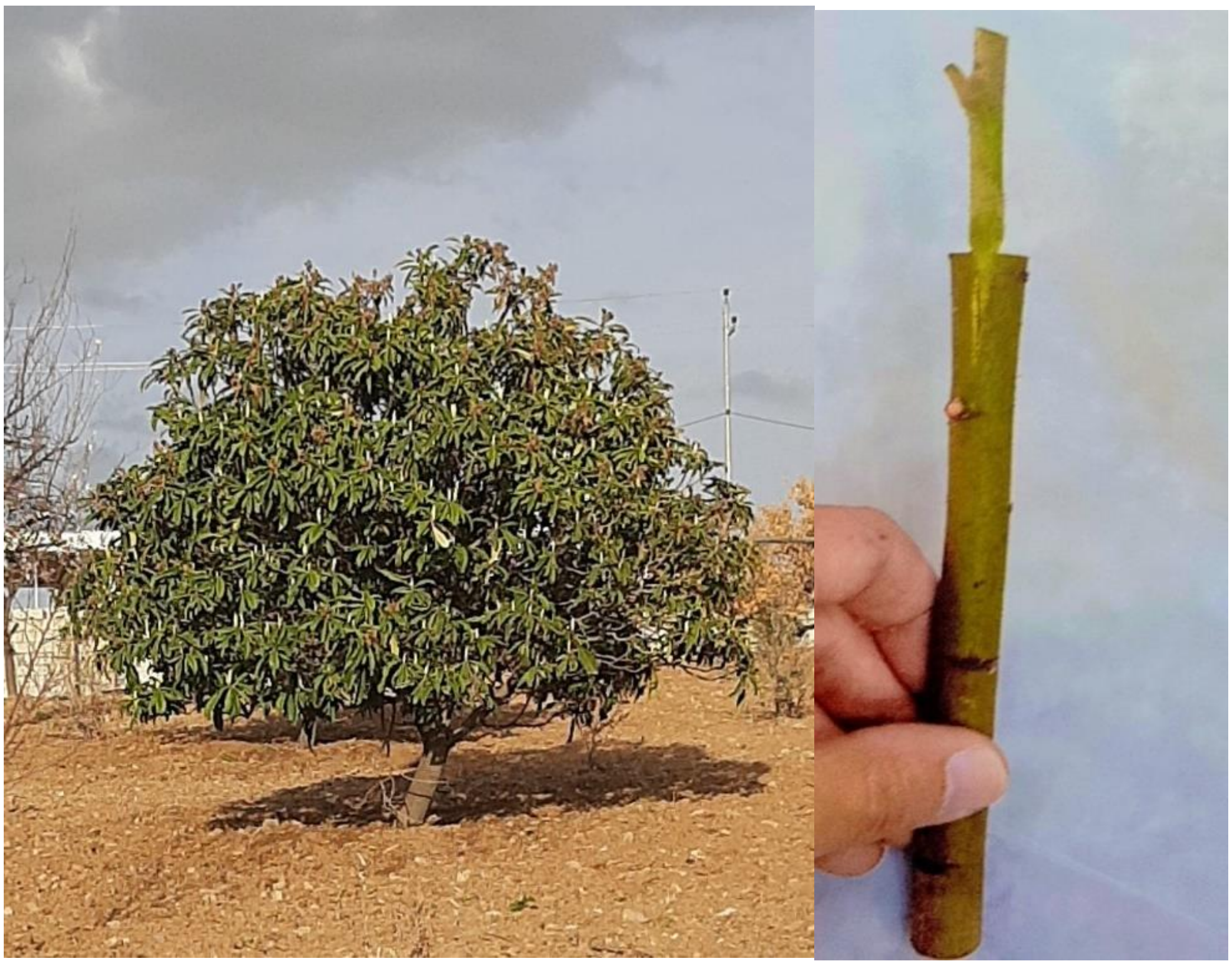

**Appendix 4.** Collecting cuttings from the loquat mother tree at the Qularaisy site.

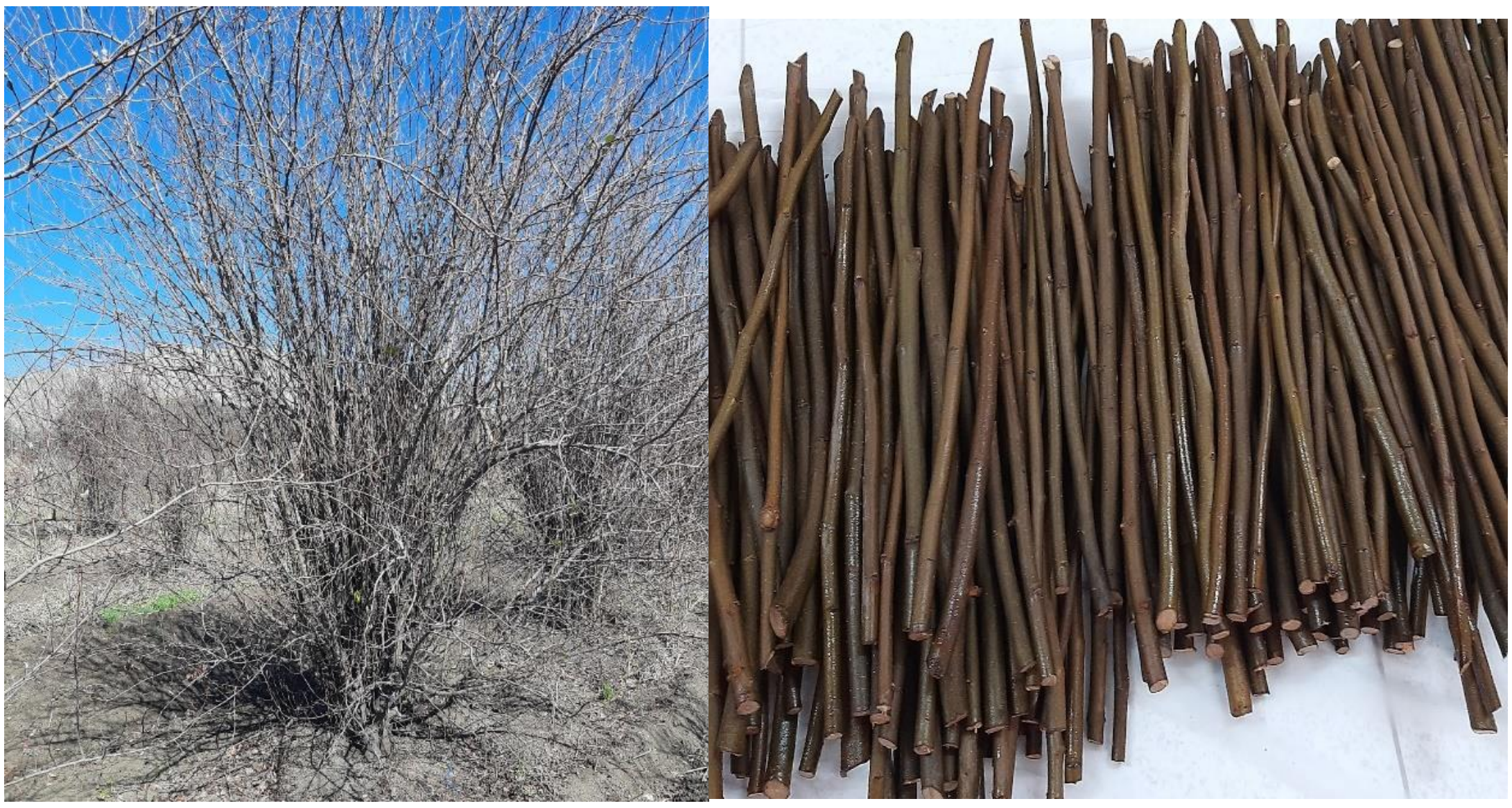

**Appendix 5.** Collecting cuttings from the quince mother tree at the Kani Panka site.

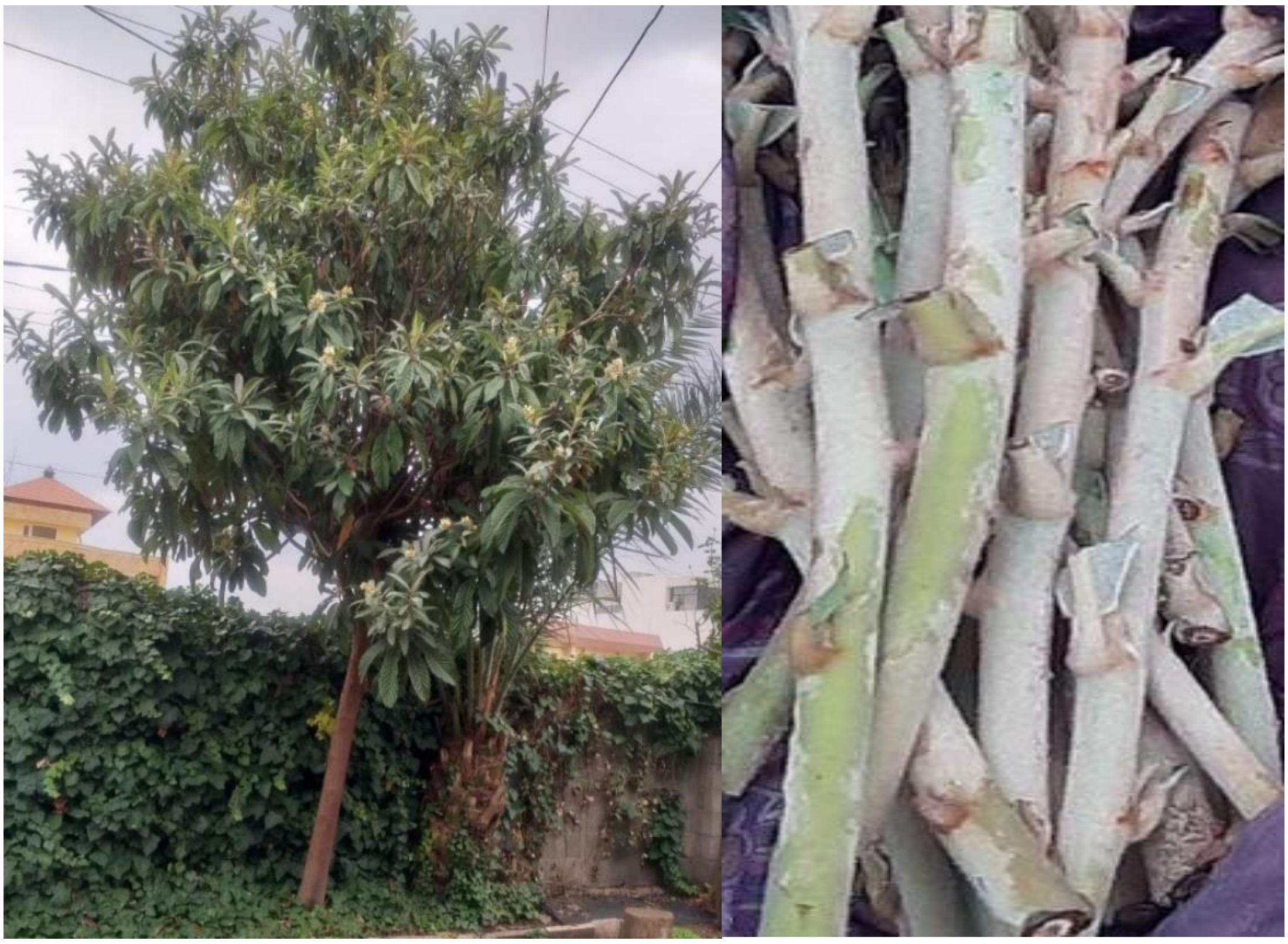

**Appendix 6.** Collecting scions from the loquat mother tree in the Iskan site.

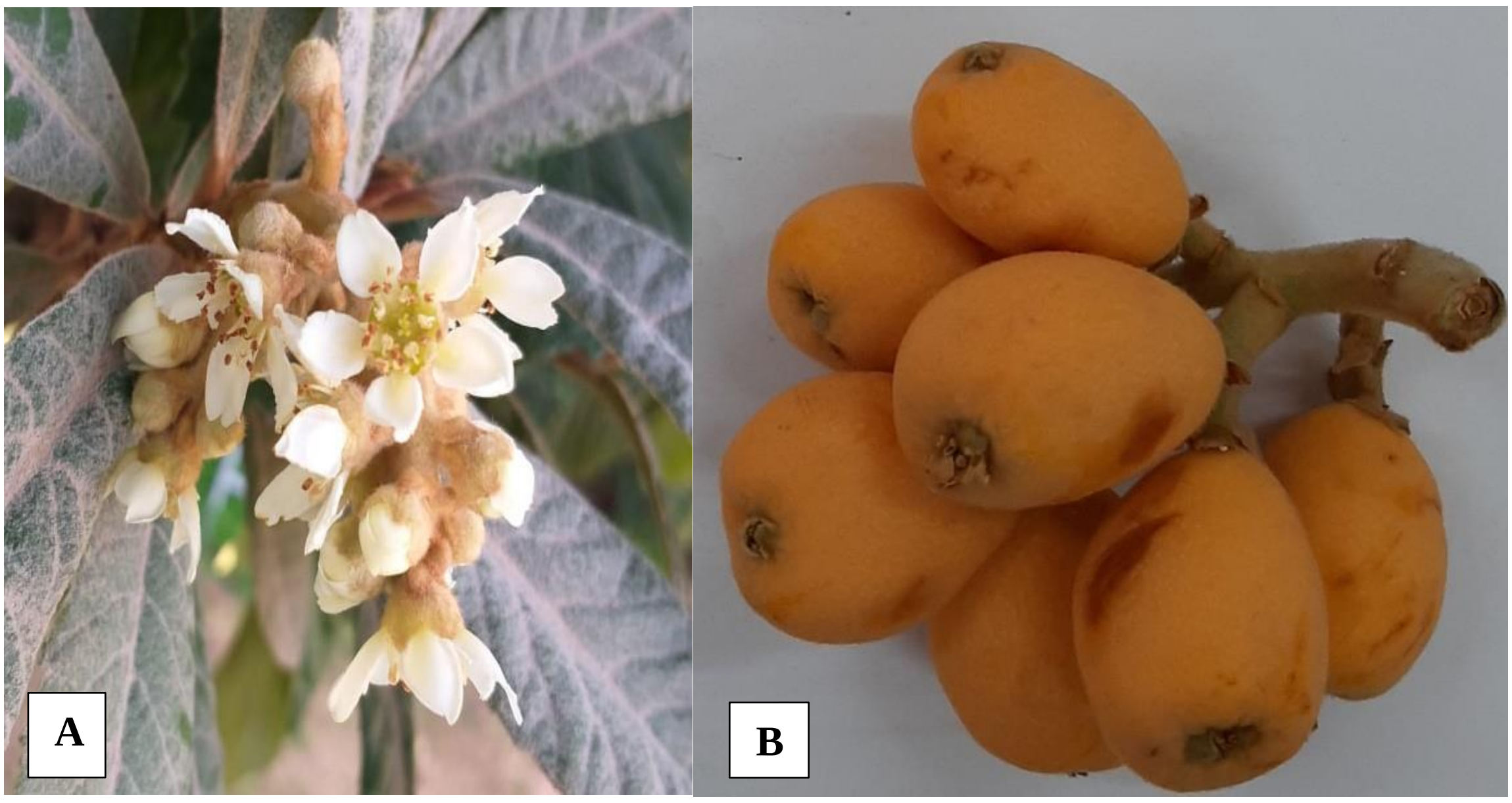

**Appendix 7.** A-B picture of loquat tree flowers and fruits used for scion collection at a site in the Iskan area.

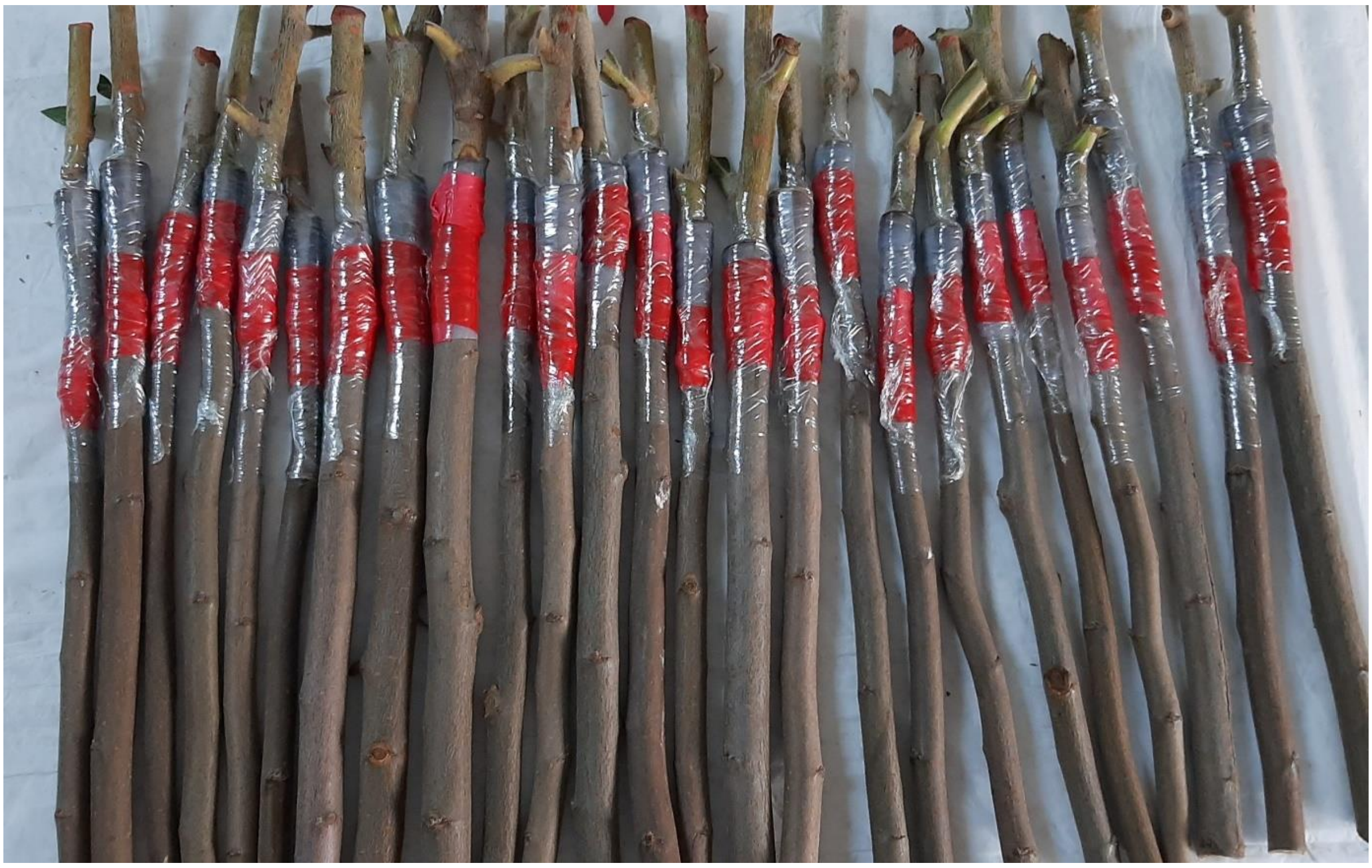

**Appendix 8.** The loquat scion was bench-grafted onto quince cuttings.

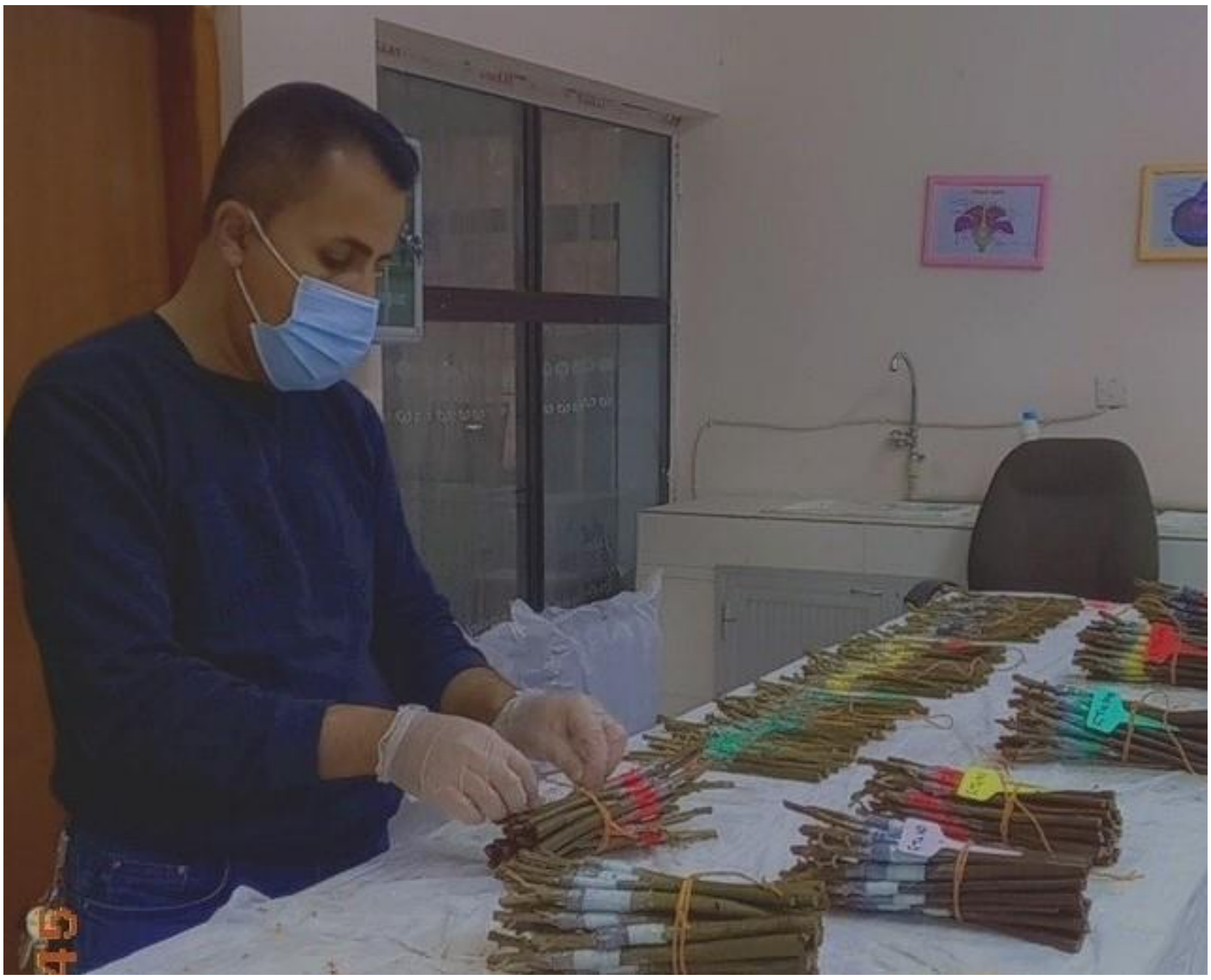

**Appendix 9.** The grafted cuttings before placed in a peat moss substrate for callus formation at a temperature of 18±2 °C.

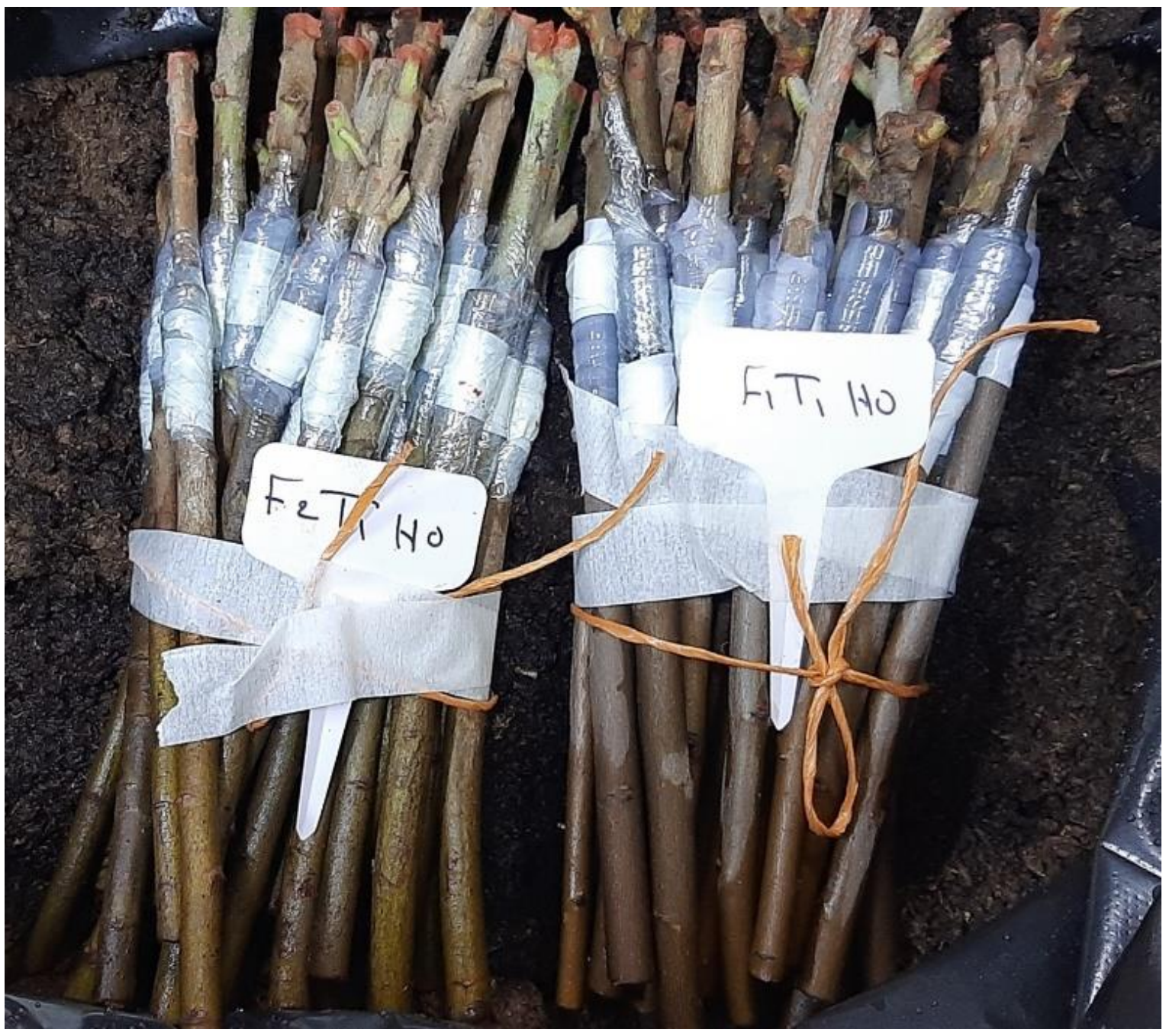

**Appendix 10.** The grafted cuttings were placed in a peat moss substrate for callus formation at a temperature of 18±2 °C.

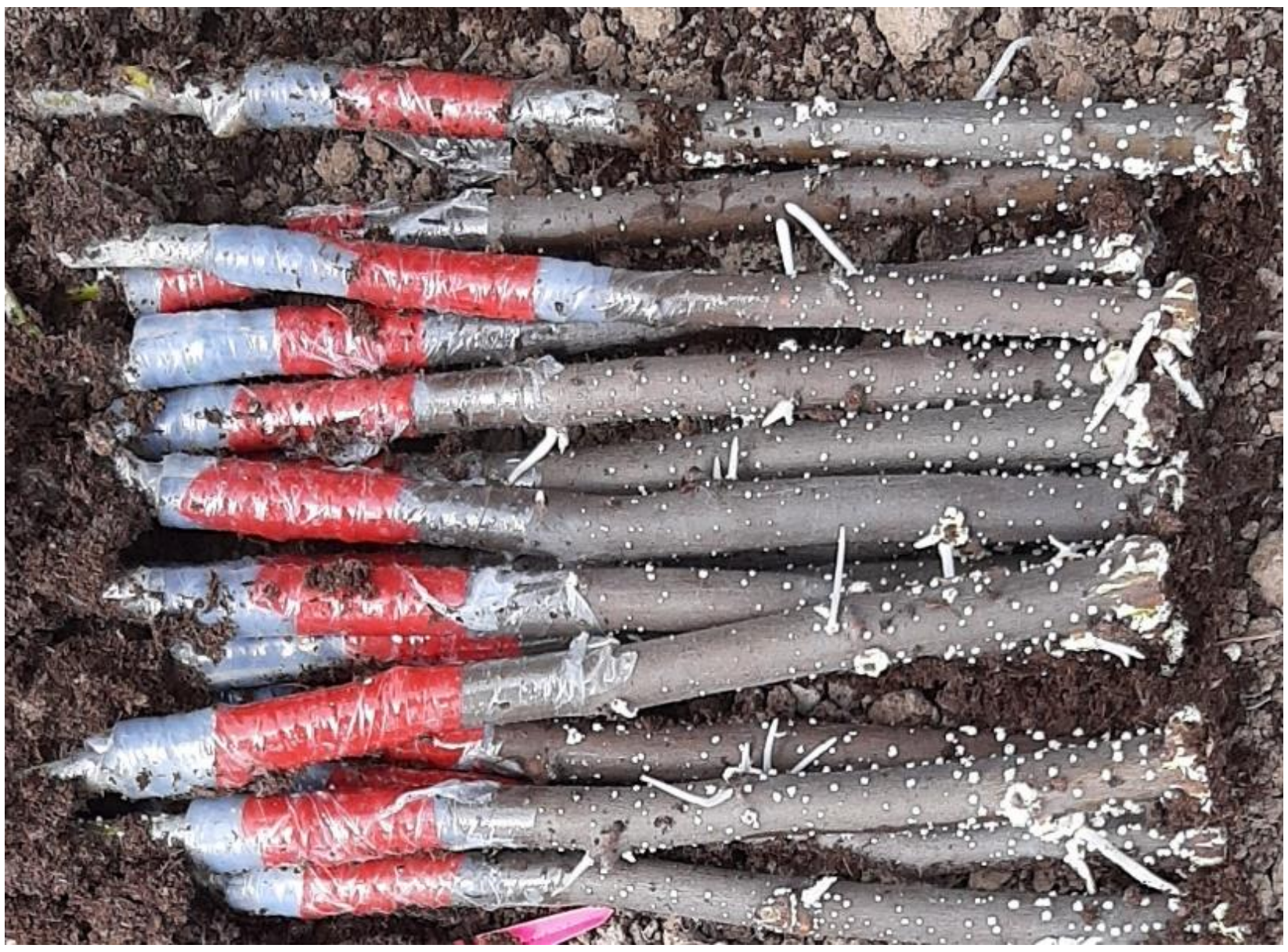

**Appendix 11.** The grafted cuttings after placing in a peat moss substrate for callus formation at a temperature of 18±2 °C.

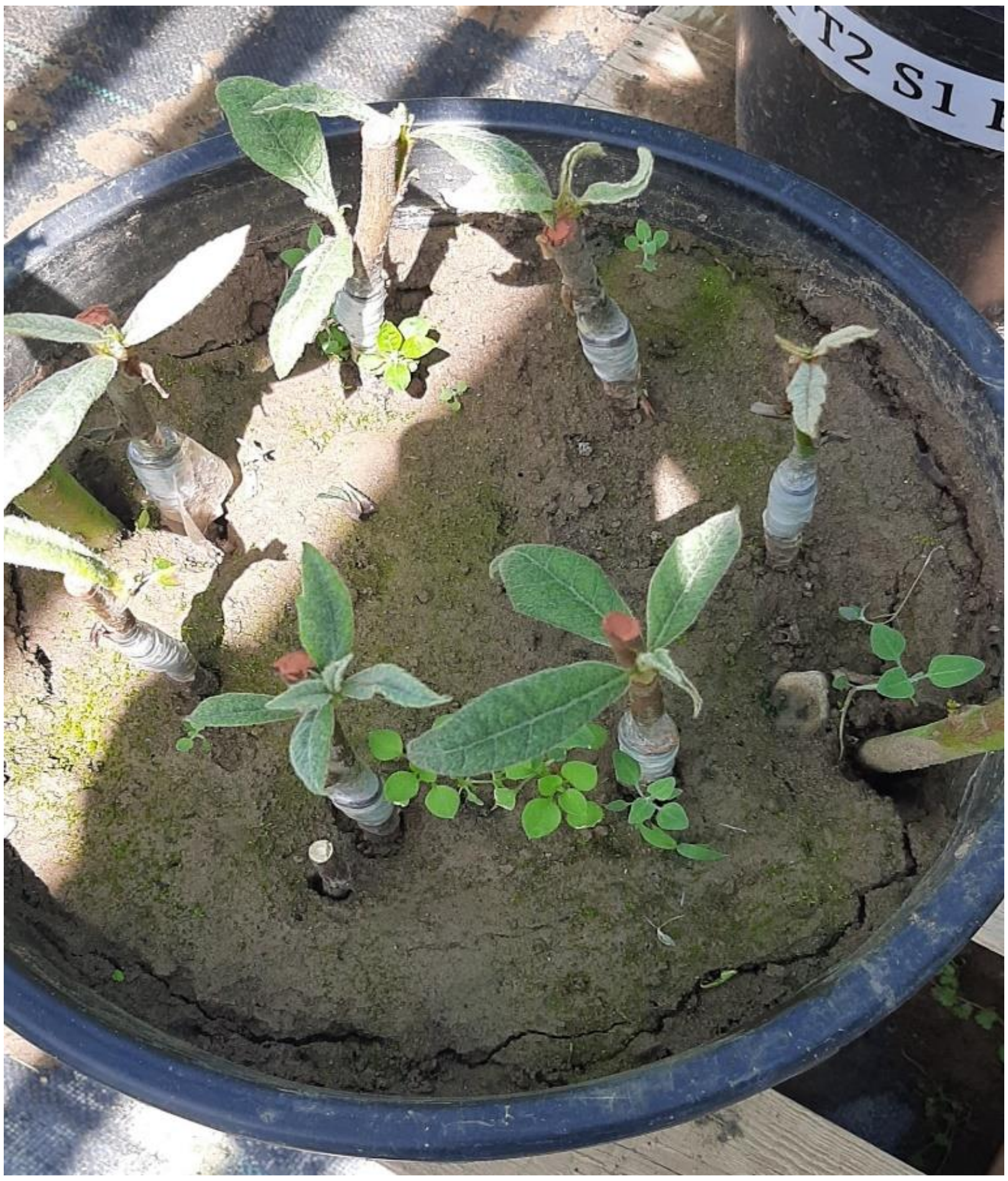

**Appendix 12.** The bud sprouts from the bench grafting loquat on loquat cuttings in the lath house.

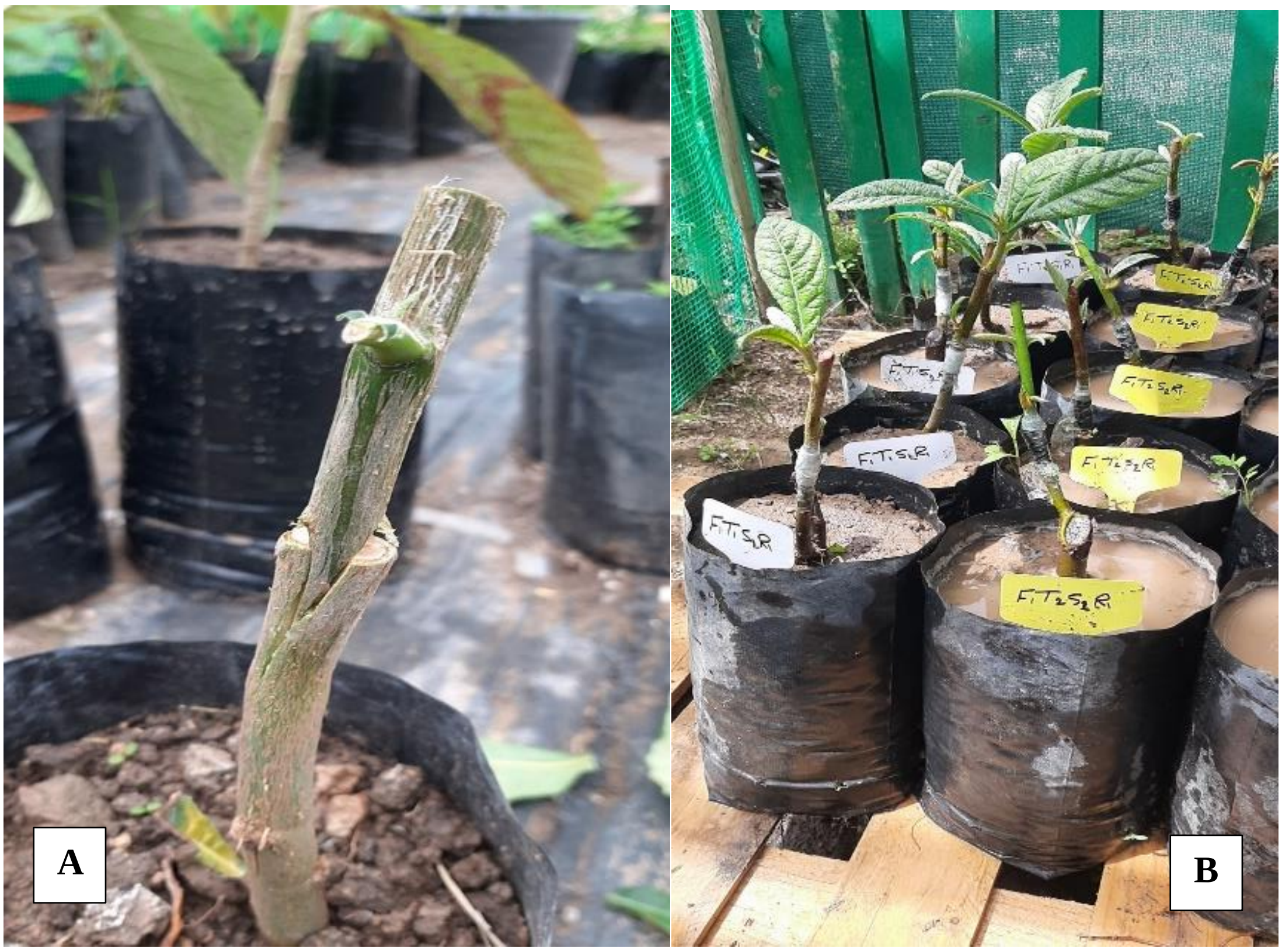

**Appendix 13.** The loquat scion was inserted for grafting onto a loquat rootstock (A), and the successful loquat grafting on both loquat and quince rootstock (B).

حكومة اقليم كردستان

وزارة التعليم العالي و البحث العلمي

جامعة السليمانية

كلية علوم الهندسة الزراعية

# دراسات التوافق فى طعوم الينكى دنيا مع اصول الينكى دنيا والسفرجل

**أطروحة**

**مقدمة الى مجلس كلية علوم الهندسة الزراعية في جامعة السليمانية**

**كجزء من متطلبات نيل شهادة دكتوراه فلسفة في علوم البستنة**

**انتاج فاكهة مستديمة الخضرة**

من قبل

رسول رفيق عزيز

بكالوريوس البستنة (2004)، كلية الزراعة، جامعة السليمانية

ماجستير انتاج الفاكهة (2011)، كلية الزراعة ، جامعة السليمانية

باشراف

د. فخرالدين مصطفى حمه صالح — استاذ مساعد

د. ابراهيم معروف نوري — استاذ مساعد

١٤٤٤هـ

٢٠٢٤ م

## الملخص

أجري هذا البحث خلال الفترة من ٢٠٢١ إلى ٢٠٢٣ في كلية علوم الهندسة الزراعية، جامعة السليمانية، إقليم كردستان-العراق، للتحقيق في نسب نجاح التركيب والتجذير في تداخل العقل المركبة لكل من الينكى دنيا/السفرجل و الينكى دنيا/ الينكى دنيا. فاعتبرت الدراسة عوامل مختلفة، بما في ذلك تراكيز مختلفة من إندول -3- حامض البيوتريك (IBA)، والمواعيد، والأصول، و اوساط التجذير، عبر موقعين. الموقع الاول يقع في الكلية، حيث أجريت أربع تجارب في الظلة الخشبية على مدار ثلاثة مواسم نمو متتالية. ومع ذلك، فإن الموقع الثاني، يتضمن تركيب أشجار السفرجل المزروعة في قرية كاني ويسه بطعوم أشجار الينكى دنيا المزروعة في اسكان. تم استخدام التركيب الشقي في جميع التجارب. صممت التجارب وفق تصميم القطاعات العشوائية الكاملة العاملية بثلاثة مكررات، وتم تحليل البيانات باستخدام تحليل التباين(ANOVA) ، وقورنت المتوسطات باستخدام اختبار دنكن المتعدد المدى (P<0.05) .

**التجربة الاولى: تجذير العقل الخشبية الصلبة للسفرجل *Cydonia oblonga* L. تحت تأثير IBA ووسط التجذير:**

أجريت هذه التجربة في الظلة الخشبية بموقع الكلية، خلال موسم النمو من شباط إلى تموز ٢٠٢١، بهدف دراسة تأثير تراكيز IBA المختلفة (٠، ١٠٠٠، ٢٠٠٠، و ٣٠٠٠ مغ.لتر) واوساط تجذير مختلفة (الرمل النهري، الرمل النهري + خث الطحلب (١:١ حجما)، والبيرلايت + الخث الطحلبي (١:١ حجما) في نجاح تجذير العقل الخشبية الصلبة للسفرجل *Cydonia oblonga* L. . تضمنت المعلمات صفات الجذر والمجموع الخضري ومحتوى الأوراق من الكلوروفيل. أظهر تأثير العوامل الفردية أن التجذير والصفات الأخرى للعقل المجذرة كانت مستقلة عن تأثير IBA. سجلت أعلى نسبة تجذير (٦٢.٥٠%) في عقل المقارنة، مع تحسين الصفات الأخرى. بالإضافة إلى ذلك فإن أفضل تجذير (٦٤.٥٨%) وجد في العقل المزروعة في وسط الرمل النهري. أظهرت تأثيرات التداخل بين العاملين أن عقل المزروعة في الرمل النهري أعطت أعلى نسبة تجذير (٧٠.٨٣%) وأعلى صفات الجذور والمجموع الخضري الأخرى. كان كل من الرمل النهري و الرمل النهري + خث الطحلب رائعًا بالنسبة لعقل السفرجل، ولكن لم تكن هناك حاجة إلى IBA عند التراكيز المستخدمة في هذه الدراسة.

**التجربة الثانية: تأثير مواعيد التركيب وأنواع العقل وتراكيز IBA في نجاح التركيب المنضدي الينكى دنيا.**

أجريت هذه التجربة خلال الفترة من شباط إلى تموز ٢٠٢٢ في الظلة الخشبية بموقع الكلية لتقييم التركيب المنضدي للينكى دنيا على نوعين من العقل؛ الينكى دنيا والسفرجل، تم إجراؤهما في مواعيد مختلفة (١٠ شباط، ٢٠ شباط، ٢ اذار، و١٢ اذار)، وتراكيز IBA (٠، ١٠٠٠، ٢٠٠٠، ٣٠٠٠ و ٤٠٠٠ مغ.لتر). تم تسجيل نسب نجاح التركيب في ٣٠ اذار و٣٠ نيسان و٣٠ مايس بالإضافة إلى نسب التجذير. أظهرت النتائج أن أعلى نسب نجاح التركيب (٣٧.٥٠%) لعقل الينكى دنيا والتي سجلت في ٣٠ مايس، كانت مرتبطة بتركيب ٢٠ شباط بـ ١٠٠٠ مغ.لتر- IBA وعلى العكس من ذلك، فإن أقل نسبة نجاح التركيب (٠.٠٠%) حدث عند إجراء عمليات التركيب

في ٢ و١٢ اذار، ولا سيما بتراكيز ٢٠٠٠ و٣٠٠٠ مغ.لتر⁻ IBA. بلغت نسب نجاح تركيب السفرجل ذروتها عند (٤١.٦٧٪) في ٣٠ اذار (بدون IBA)، مع أدنى مستوى (٠.٠٠٪) في ٣٠ نيسان و٣٠ مايس. أظهرت الاصول المركبة لينكى دنيا في ٢٠ شباط أعلى تداخل في ٣٠ مايس (٣٧.٥٠٪)، بينما سجل ٣٠ اذار أقل النسب (٢.٠٨٪ إلى ٤.١٧٪) لكلا الاصلين على التوالي. حقق التجذير نتائج ثابتة (٠.٠٠٪)، مما يؤكد الحاجة إلى مزيد من الاستكشاف في تركيب الينكى دنيا.

**التجربة الثالثة: أداء تركيب الينكى دنيا المتجمعة على أصول الينكى دنيا والسفرجل في مواعيد مختلفة:**

أجريت هذه الدراسة في الظلة الخشبية بموقع الكلية في الفترة من ٢٠ شباط إلى ١ تموز ٢٠٢٣، بهدف تقييم تأثير الاصول (الينكى دنيا والسفرجل) و مواعيد التركيب (٢٠ شباط، ١٠ اذار، و ٣٠ اذار) في نجاح تركيب الينكى دنيا. تم تسجيل نسب نجاح التركيب والصفات الخضرية ومحتوى الكلوروفيل. أشارت النتائج إلى أن نسبة نجاح التركيب أعلى بشكل ملحوظ (٩٧.٧٨٪) لأصل الينكى دنيا مقارنة بأصل السفرجل (٨٤.٤٤٪). حدث نجاح التركيب الأمثل بنسبة (٩٦.٦٧٪) في ٢٠ شباط عندما تم استخدام كلا من أصلي الينكى دنيا والسفرجل. على العكس من ذلك، لوحظت النتيجة الأقل ملاءمة (٨٦.٦٧٪) في ٣٠ اذار باستخدام نفس الاصلين في التركيب. ومن اللافت للنظر أن الجمع بين اصل الينكى دنيا ومواعيدي التطعيم (٢٠ شباط و٣٠ اذار) أدى إلى أعلى نسبة نجاح التركيب (١٠٠٪) .

**التجربة الرابعة: تأثير مواعيد التركيب وأنواع الاصول على نجاح تركيب اصلي الينكى دنيا والسفرجل:**

أجريت هذه التجربة في موقع الكلية في الظلة الخشبية، وتهدف إلى تقييم تأثير أنواع الاصول (الينكى دنيا والسفرجل) ومواعيد التركيب (٢٠ شباط، ١٠ اذار، و٣٠ اذار) في نجاح تركيب الينكى دنيا. تم قياس نسبة نجاح التركيب وصفات المجموع الخضري ومحتوى الكلوروفيل. كشفت النتائج عن عدم وجود فروق ذات دلالة إحصائية في نجاح التركيب بين النوعين من الاصول. ومع ذلك، أظهر اصل الينكى دنيا نسبة نجاح تركيب أعلى (٧٣.٣٣%) مقارنة بالسفرجل (٦٦.٦٧٪). وكانت نسب نجاح التركيب في المواعيد الثلاثة ثابتة (٧٠٪). حقق تداخل اصل الينكى دنيا مع مواعيد التركيب الثلاثة أعلى نسب نجاح (٧٣.٣٣٪)، بينما أدى تداخل اصل السفرجل مع نفس المواعيد إلى أقل نسبة نجاح للتركيب (٦٦.٦٧٪).

**التجربة الخامسة: تأثير التركيب المنضدي لينكى دنيا ومواعيد مختلفة في نسب نجاح التركيب و التجذير:**

أجريت هذه التجربة في موقع الكلية في الظلة الخشبية، لدراسة تأثير تركيب الينكى دنيا في نسب نجاح التركيب، وتطور الأوراق، والمعلمات البيوكيميائية، وذلك باستخدام نوعين من عقل الاصول (الينكى دنيا والسفرجل) عبر مواعيد تركيب مختلفة ( ٢٠ شباط، ١٠ اذار، ٣٠ اذار). تباينت نسب نجاح التركيب بشكل ملحوظ، حيث بلغت أعلى نسبة (٨٦.١١٪) في عقل الينكى دنيا عند ٦٠ يومًا، في حين سجلت عقل السفرجل أقل نسبة (٥٩.٧٢٪). زادت نسب نجاح التطعيم باستمرار من ٢٠ شباط إلى ٣٠ اذار. فضلت أعداد أوراق الينكى دنيا (٢.٧٠)، مع التطور الأمثل في ٢٠ شباط (٢.٧٨). كشفت المعايير البيوكيميائية عن اختلافات واضحة وأظهرت ارتفاع

محتوى الكربوهيدرات (٧.٤٤٪) وانخفاض مستوى الفينولات (٨٢٥.٦٧٪) في الينكى دنيا. اكدت هذه النتائج على الدور الحاسم لاختيار مواعيد التركيب، حيث حققا موعدى ٢٠ شباط و٣٠ اذار أعلى نسب الكربوهيدرات (٦.٥٨٪).

حکومەتی هەرێمی کوردستان
وەزارەتی خوێندنی باڵا و توێژینەوەی زانستی
زانکۆی سلێمانی
کۆلێجی زانستە ئەندازیارییە کشتوکاڵییەکان

# لێکۆڵینەوەی تەبایی لەخۆراکەی یەنگی دنیا لەگەڵ بنەمای یەنگی دنیا و بەهێ

**تێزی دکتۆرایە**

**پێشکەش کراوە بە ئەنجومەنی کۆلێجی زانستە ئەندازیارییە کشتوکاڵیەکان لە زانکۆی سلێمانی وەك بەشێك لە پێداویستیەکانی بەدەستهێنانی بڕوانامەی دکتۆرای فەلسەفە لە زانستی باخداری**

**بەرهەمهێنانی میوەی هەمیشەسەوز**

لە لایەن

**ڕەسوڵ ڕەفیق عەزیز**

بە کالۆریۆس لە باخداری (2004)، کۆلێجی کشتوکاڵ، زانکۆی سلێمانی

ماستەر لە بەرهەمهێنانی میوە (2011)، کۆلێجی کشتوکاڵ ، زانکۆی سلێمانی

بە سەرپەرشتی

**د. فەخرەددین مستەفا حەمە ساڵح**
پرۆفیسۆری یاریدەدەر

**د. ئیبراهیم مەعروف نوری**
پرۆفیسۆری یاریدەدەر

**2024 ز** **2724 ك**

## پوختە

ئەم توێژینەوەیە لە ماوەی ساڵانی ٢٠٢١ تا ٢٠٢٣ لە کۆلێژی زانستە ئەندازیارییە کشتوکاڵییەکان، زانکۆی سلێمانی، هەرێمی کوردستان-عێراق ئەنجامدرا، بۆ لێکۆڵینەوە لە ڕێژەی سەدی سەرکەوتنی پێوەندکردن و ڕەگگرتنی کۆکراوەی پێوەندکراوی یەنگی دنیا/بەهێ و یەنگی دنیا/یەنگی دنیا. لە توێژینەوەکەدا چەند هۆکارێک لەبەرچاو گیرا، لەوانە چڕیی جیاوازی ئیندۆلی-3-ترشی بوتیریک (IBA) ، بەروار، بنک و تێکەڵەی ڕەگگرتن، لە دوو شوێن. شوێنی یەکەم لە کۆلێژەکەدا هەڵکەوتووە، کە تێیدا چوار تاقیکردنەوە لە کەپری دارێندا ئەنجامدرا لە ماوەی سێ وەرزی گەشەکردندا. بەڵام شوێنی دووەم، بریتی بوو لە پێوەندکردنی یەنگی دنیا لەسەر ئەو دار بەهێیانەی کە لە گوندی کانی وەیسە و دارەکانی یەنگی دنیا کە لە کۆلێژەکەدا چێنراون. بۆ هەموو تاقیکردنەوەکان جۆری پێوەندکردنی بە درز بەکارهات. تاقیکردنەوەکان لە دیزاینێکی بلۆکی تەواو هەڕەمەکی فاکتەریدا داڕێژران لەگەڵ سێ دووبارەکردنەوە، وە داتاکان بە بەکارهێنانی ANOVA شیکرانەوە، وە ناوەندەژمێرەییەکان بە بەکارهێنانی تاقیکردنەوەی فرە مەودای دنکن بەراورد کران ($P \leq 0.05$) .

**تاقیکردنەوەی یەکەم: ڕەگگرتنی قەڵەمی دارینی بەهێ *Cydonia oblonga* L. لەژێر کاریگەریی IBA و تێکەڵەی ڕەگگرتن:**

ئەم تاقیکردنەوەیە لەژێر کەپرەدارینەی کۆلێژەکەدا ئەنجامدرا، لە وەرزی گەشەکردندا، لە شوبات تا تەمموزی ٢٠٢١، وە ئامانجی تاقیکردنەوەکە لێکۆڵینەوە بوو لە کاریگەری خەستیی جیاوازی IBA (٠، ١٠٠٠، ٢٠٠٠، و ٣٠٠٠ مگ.لیتر⁻) وە تێکەڵەی ڕەگگرتن؛ قوم ، قوم + پیتمۆس (١:١) ، و پێرلایت + پیتمۆس (١:١) لە سەرکەوتنی ڕەگگرتنی قەڵەمی دارینی بەهێ *Cydonia oblonga* L. . پێوەرەکان بریتی بوون لە خەسڵەتەکانی ڕەگ وسەوزەگەشە و ڕێژەی کلۆرۆفیلی گەڵا. کاریگەری هۆکارە تاکەکان دەریخست کە ڕەگگرتن و تایبەتمەندییەکانی تری قەڵەمە ڕەگگرتووەکان سەربەخۆن لە کاریگەری IBA . بەرزترین ڕێژەی ڕەگگرتن (٦٢.٥٠٪) لە قەڵەمەکانی کۆنترۆڵدا بەدەست هات، لەگەڵ باشتربوونی تایبەتمەندییەکانی تر. سەرەڕای ئەوەش، باشترین ڕەگگرتن (٦٤.٥٨٪) لەو قەڵەمانەدا دۆزرایەوە کە لە ناوەندی قومدا چێندرابوون. کاریگەرییەکانی کارلێکی ئەو دوو هۆکارە دەریخست کە قەڵەمەکانی کۆنترۆڵ کە لە قومدا چێنراون زۆرترین ڕەگگرتن (٧٠.٨٣٪) و بەرزترین تایبەتمەندی ڕەگ و خەسڵەتەکانی تریان بەخشیوە. هەردوو تێکەڵەی قوم و قوم + پیتمۆس بۆ قەڵەمی بەهێ نایاب بوون، بەڵام IBA پێویست نەبوو بەو چڕییانەی کە لەم لێکۆڵینەوەیەدا بەکارهێنران.

**تاقیکردنەوەی دووەم: کاریگەری بەرواری پێوەندکردن، جۆری قەڵەم، و چڕیی IBA لە سەرکەوتنی پێوەندکردنی سەرمێز بۆ یەنگی دنیا:**

ئەم تاقیکردنەوەیە لە ماوەی شوبات تا حوزەیرانی ٢٠٢٢ دا و لە کەپرەدارینەی کۆلێژەکە ئەنجامدرا، بۆ هەڵسەنگاندنی پێوەندکردنی یەنگی دنیا لەسەر دوو جۆری قەڵەم؛ یەنگی دنیا و بەهێ ، کە لە بەرواری جیاوازدا ئەنجام دران (١٠ی شوبات، ٢٠ی شوبات، ٢ی ئازار، و ١٢ی ئازار)، وە چڕیی IBA (٠ ، ١٠٠٠، ٢٠٠٠، ٣٠٠٠ و ٤٠٠٠ مگ.لیتر⁻). ڕێژەی سەرکەوتنی پێوەندکردن لە ٣٠ی ئازار، ٣٠ی نیسان و ٣٠ی ئایار تۆمارکرا، لەگەڵ ڕێژەی ڕەگگرتندا. ئەنجامەکان دەریانخست کە بەرزترین سەرکەوتنی پێوەندکردن (٣٧.٥٠٪) بۆ قەڵەمی یەنگی دنیا بوو کە لە ٣٠ی ئایاردا تۆمارکرا، کە پەیوەست بوو بە پێوەندکردنی ٢٠ی شوباتەوە بە ١٠٠٠ مگ.لیتر⁻ IBA . بە پێچەوانەوەشەوە، کەمترین ڕێژەی سەرکەوتن (٠.٠٠٪) کاتێک ڕوویدا کە پێوەندکردن لە ٢ و ١٢ی ئازاردا ئەنجامدرا، بەتایبەتی لەگەڵ ٢٠٠٠ و ٣٠٠٠ مگ.لیتر⁻ IBA. بنکی بەهێی پێوەندکراو لە ٣٠ی ئازاردا بەبێ IBA گەیشتە لوتکە (٤١.٦٧٪)، لەگەڵ کەمترین (٠.٠٠٪) لە ٣٠ی نیسان و ٣٠ی ئایاردا. لە کاتێکدا ٣٠ی ئازار کەمترین ڕێژەی هەبوو (٢.٠٨٪ بۆ ٤.١٧٪) بۆ هەردوو بنکەکە، بە ڕێکەوت ڕەگگرتن بە بەردەوامی ڕێژەی (٠.٠٠٪)ی بەدەستهێنا، ئەمەش جەخت لەسەر پێویستی لێکۆڵینەوەی زیاتر لە پێوەندکردنی بنکی یەنگی دنیا دەکاتەوە.

**تاقیکردنەوەی سێیەم: ئەنجامدانی پێوەندکردنی کۆکراوەی یەنگی دنیا لەسەر بنکی یەنگی دنیا و بەهێ لە بەرواری جیاوازدا:**

ئەم توێژینەوەیە لەژێر کەپرەدارینەی کۆلێژەکەدا و لە ٢٠ی شوبات تا ١ی تەمموزی ٢٠٢٣ ئەنجامدرا، بە ئامانجی هەڵسەنگاندنی کاریگەریی جۆری بنک (یەنگی دنیا و بەهێ) و بەرواری پێوەندکردن (٢٠ی شوبات، ١٠ی ئازار، و ٣٠ی ئازار) لە یەنگی دنیادا لەسەر سەرکەوتنی پێوەندکردنەکە. ڕێژەی سەدی سەرکەوتنی پێوەندکردن، تایبەتمەندییەکانی سەوزەگەشە و ڕێژەی کلۆرۆفیل تۆمارکرا. ئەنجامەکان ئاماژەیان بە ڕێژەی سەرکەوتنی پێوەندکردن کرد کە بە شێوەیەکی بەرچاو بەرزتر بوو بۆ بنکی یەنگی دنیا (٩٧.٧٨٪) بە بەراورد بە بنکی بەهێ (٨٤.٤٤٪). سەرکەوتنی گونجاوی پێوەندکردن لە (٩٦.٦٧٪) لە ٢٠ی شوبات ڕوویدا کاتێک هەردوو بنکی یەنگی دنیا و بەهێ بەکارهێنران. بە پێچەوانەوە، کەمترین دەرەنجامی لەبار (٨٦.٦٧٪) لە ٣٠ی ئازاردا بە بەکارهێنانی هەمان بنک بۆ پێوەندکردن بینرا. جێگای سەرنجە، کارلێکی نێوان بنکی یەنگی دنیا لەگەڵ هەردوو بەرواری پێوەندکردن (٢٠ی شوبات و ٣٠ی ئازار) بووە هۆی بەرزترین ڕێژەی سەرکەوتنی بەرچاوی پێوەندکردن (١٠٠٪).

**تاقیکردنەوەی چوارەم: کاریگەریی بەرواری پێوەندکردن و جۆری قەڵەم لە سەرکەوتنی پێوەندکردنی داری یەنگی دنیا و بەهێ:**

ئەم تاقیکردنەوەیە لەژێر کەپرەدارینەی کۆلێژەکەدا ئەنجامدرا، بە ئامانجی هەڵسەنگاندنی کاریگەریی جۆری بنک (یەنگی دنیا و بەهێ) و بەرواری پێوەندکردن (٢٠ی شوبات، ١٠ی ئازار، و ٣٠ی ئازار) بوو لە سەرکەوتنی پێوەندکردنی یەنگی دنیادا. ڕێژەی سەدی سەرکەوتنی پێوەندکردن، تایبەتمەندییەکانی سەوزەگەشە، و ڕێژەی کلۆرۆفیل، پێوانە کران. تێبینییەکان هیچ جیاوازییەکی بەرچاویان لە سەرکەوتنی پێوەندکردن لە نێوان هەردوو جۆری بنکەکاندا دەرنەخست. بەڵام، بنکەداری یەنگی دنیا ڕێژەیەکی بەرزی سەرکەوتنی پەیوەندکردنی نیشان دا (٧٣.٣٣٪) بە بەراورد بە بەهێ (٦٦.٦٧٪). ڕێژەی سەرکەوتنی پێوەندکردن بۆ هەر سێ بەروارەکە جێگیر بوو (٧٠٪). کارلێکی نێوان بنکەدارەکانی یەنگی دنیا لەگەڵ هەرسێ بەروارەکە بەرزترین سەرکەوتنی بەدەستهێنا (٧٣.٣٣٪)، لە کاتێکدا کارلێکی نێوان بنکەداری بەهێ لە هەمان ئەو بەروارانەدا کەمترین سەرکەوتنی پێوەندکردنی لێکەوتەوە (٦٦.٦٧٪).

**تاقیکردنەوەی پێنجەم: کاریگەریی پێوەندکردنی سەرمێز بۆ یەنگی دنیا و بەرواری جیاواز لە ڕێژەی پێوەندکردن و ڕەگکردن:**

ئەم تاقیکردنەوەیە لە ژێر کەپرەدارینەی کۆلێژەکەدا ئەنجامدرا، بۆ لێکۆڵینەوە لە کاریگەریی پێوەندکردنی بنکی یەنگی دنیا لەسەر ڕێژەی سەدی سەرکەوتنی پێوەندکردن، گەشەکردنی گەڵا، و پێوەرە بایۆکیمیاییەکانی، بە بەکارهێنانی دوو جۆری قەڵەمی بنچینە (یەنگی دنیا و بەهێ) لە بەرواری جیاوازی پێوەندکردندا ( ٢٠ی شوبات و ١٠ی ئازار و ٣٠ی ئازار). ڕێژەی سەدی سەرکەوتنی پێوەندکردن جیاوازییەکی بەرچاوی هەبوو، کە گەیشتە بەرزترین (٨٦.١١٪) لەسەرقەڵەمی یەنگی دنیا لە ٦٠ ڕۆژدا، لەکاتێکدا لەسەر قەڵەمی بەهێ کەمترین ڕێژە (٥٩.٧٢٪) تۆمارکرا. سەرکەوتنی پێوەندکردن بە بەردەوامی لە ٢٠ی شوباتەوە تا ٣٠ی ئازار زیادی کرد. ژمارەی گەڵاکان (٢.٧٠) لە بەرژەوەندی یەنگی دنیا بوو، لەگەڵ گەشەکردنی گونجاو لە ٢٠ی شوبات (٢.٧٨). پێوەرە بایۆکیمیاییەکان گۆڕانکاری جیاوازیان ئاشکرا کرد و ڕێژەی کاربۆهیدراتی بەرزتر (٧.٤٤٪) و ئاستی فینۆڵی کەمتریان (٨٢٥.٦٧٪) لە یەنگی دنیادا نیشان دا. ئەم ئەنجامەش جەخت لەسەر ڕۆڵی گرنگی هەڵبژاردنی بەرواری پێوەندکردنکردن دەکاتەوە کە لە ٢٠ی شوبات و ٣٠ی ئازاردا بەرزترین ڕێژەی کاربۆهیدرات (٦.٥٨٪) بەدیکرا.